\documentclass[12pt,a4paper]{article}
\usepackage{jheppub}
\usepackage{bm}
\usepackage{ifsym}
\usepackage{amsthm}
\usepackage{amsfonts}
\usepackage{bbm}
\usepackage{ccaption}
\usepackage{mathrsfs}
\usepackage{booktabs}
\usepackage{color}
\usepackage{wrapfig}
\usepackage[labelfont={sf}]{caption}
\usepackage{multicol}
\usepackage{multirow}
\usepackage{epstopdf}
\usepackage{graphics}
\usepackage{rotating}

\newcommand{\hs}[1]{\hspace{#1 cm}}		
\newcommand{\vs}[1]{\vspace{#1 cm}}
\newcommand{\beq}{\begin{equation}}		
\newcommand{\enq}{\end{equation}}
\renewcommand{\(}{\left(}			
\renewcommand{\)}{\right)}
\renewcommand{\[}{\left[}			
\renewcommand{\]}{\right]}

\newcommand{\mc}[1]{\mathcal{#1}}		
\newcommand{\h}[1]{\hat{#1}}

\newcommand{\parcial}[2][]{\frac{\partial #1}{\partial #2}}
\newcommand{\deriv}[2][]{\frac{\dd #1}{\dd #2}}

\newcommand{\munu}{\mu\nu} 			

\newcommand{\riga}{\rightarrow}

\newcommand{\dd}{\mathrm{d}}		
\newcommand{\nab}{\nabla}
\newcommand{\la}{\langle}		
\newcommand{\ra}{\rangle}

\newcommand{\R}{\mathbb{R}}

\newcommand{\E}{\mathbb{E}}

\newcommand{\uni}{\mathbbm{1}}

  \captionnamefont{\bfseries}
  \captiontitlefont{\small\sffamily}
  \captiondelim{: }
  \hangcaption
  \renewcommand{\figurename}{Fig.}

\def\clock{{\count0=\time
           \divide\count0 60
           \ifnum\count0<10 0\fi\the\count0
           \multiply\count0 -60 \advance\count0 \time
           :\ifnum\count0<10 0\fi \the\count0
         }}
\newcommand{\timestamp}{{\small\vbox{\hbox{\tt\jobname.tex}
\hbox{\the\day/\the\month/\the\year, \clock}}}}

\usepackage{tocloft}

\definecolor{rust}{rgb}{0.8,0.2,0.2}
\definecolor{green}{rgb}{0.1,0.8,0.2}

\numberwithin{equation}{section}
\numberwithin{figure}{section}
\title{Critical lumpy black holes in AdS$_\text{p}\times$S$^\text{q}$}

\author{Biel Cardona}
\author{and Pau Figueras}

\affiliation{School of Mathematical Sciences, \\
Queen Mary University of London, \\
Mile End Road, London E1 4NS, UK.}

\emailAdd{c.gabriel@qmul.ac.uk}
\emailAdd{p.figueras@qmul.ac.uk}

\abstract{In this paper we study lumpy black holes with AdS$_p\times S^q$ asymptotics, where the isometry group coming from the sphere factor is broken down to SO$(q)$. Depending on the values of $p$ and $q$, these are solutions to a certain Supergravity theory with a particular  gauge field. We have considered the values $(p,q) = (5,5)$  and  $(p,q) = (4,7)$, corresponding to type IIB supergravity in ten dimensions and eleven-dimensional supergravity respectively. These theories presumably contain an infinite spectrum of families of lumpy black holes, labeled by a harmonic number $\ell$, whose endpoints in solution space merge with another type of black holes with different horizon topology. We have numerically constructed the first four families of lumpy solutions, corresponding to $\ell = 1,2^+,2^-$ and $3$. 
We show that the geometry of the horizon near the merger is well-described by a  cone over a triple product of spheres, thus extending Kol's local model to the present asymptotics. Interestingly, the presence of non-trivial fluxes in the internal sphere implies that the cone is no longer Ricci flat.  This conical manifold accounts for the geometry and the behavior of the physical quantities of the solutions sufficiently close to the critical point. Additionally, we show that the vacuum expectation values of the dual scalar operators approach their critical values with a power law whose exponents are dictated by the local cone geometry in the bulk. }


\begin{document}

\maketitle

\flushbottom

\newpage

%~~~~~~~~~~~~~~~~~~~~~~~~~~~~~~~~~~~~~~~~~~~~~~~
\section{Introduction}
\label{sec:intro}
%~~~~~~~~~~~~~~~~~~~~~~~~~~~~~~~~~~~~~~~~~~~~~~
The gauge/gravity duality has become an important tool to unravel the physics of certain quantum field theories (QFTs) at strong coupling, where the traditional techniques of perturbation theory fail. With the current understanding of the duality, its power mostly lies on the `gravity side' of the correspondence, which involves (semi-)classical theories of gravity that, in general, are more tractable than their  respective field theory duals. The first example of the gauge/gravity duality was introduced by Maldacena in \cite{Maldacena:97}. This was originally coined as the AdS/CFT correspondence, since it relates a conformal field theory (CFT) such as $\mc{N} = 4$ Super Yang-Mills in $3+1$ spacetime dimensions and gauge group SU$(N)$ to type IIB superstring theory in  AdS$_5\times S^5$. Although the equivalence was conjectured to be true for generic values of $N$ and $g_\text{YM}$ (the gauge theory coupling constant), the duality is better understood in the large-$N$ and large  `t Hooft coupling $\lambda=g_\text{YM}^2N$ limit, where the dual superstring theory reduces to classical supergravity in  AdS$_5\times S^5$ \cite{Hooft:74,Maldacenaetal:00}. By now, the duality has been extended to other dimensions and even to non-conformal settings. The literature on the subject is vast and we will not review it here (see the textbook \cite{ammonbook} for a review and extensive references).

In this paper we are interested in equilibrium static black holes in asymptotically global AdS$_p\times S^q$ spacetimes. By the gauge/gravity duality, these black holes are dual to thermal phases of certain gauge theories.  The field theory is defined at the boundary, $B_{p-1}$, of the $p$-dimensional Anti-de Sitter (AdS) space, which enjoys a timelike conformal structure at infinity.\footnote{The internal manifold, $S^q$ in our case, is reflected in the field content of the dual gauge theory.} For instance, if $B_{p-1}$ is taken to be  Minkowski space, then the bulk spacetime is asymptotically  AdS$_p$ in Poincar\'e coordinates. In the Poincar\'e patch there exist two homogeneous solutions of the Einstein's equations, namely pure AdS$_p$ and the planar Schwarzschild-AdS$_p$ black hole. The latter always dominates the canonical ensemble and hence there are no phase transitions. This is consistent with the fact that a CFT cannot have phase transitions on a scale invariant background such as  Minkowski space.  On the other hand, considering $B_{p-1}$ to be a space with non-zero curvature \cite{marolf:13} can lead to new and interesting physics. In the simplest case, one can take $B_{p-1}$ to be the Einstein static universe (ESU), $\mathbbm{R}\times S^{p-2}$, which corresponds to considering asymptotically global AdS$_p$ spacetimes in the bulk.  In this case, there exists the well-known  Hawking-Page phase transition \cite{Hawking:83} between global AdS$_p$ and (large) Schwarzschild black holes in that background. This phase transition was later re-interpreted by Witten \cite{witten:98} (see also Ref.~\cite{witten:981}) as a confinement/deconfinement phase transition in the dual CFT. 

So far the discussion has ignored the structure of the internal manifold, namely the $S^q$ in our case. There are situations where the internal geometry leads to new and interesting dynamics. This is famously the case for the  Gregory-Laflamme (GL)  instability of black branes \cite{Gregory:1993vy}, whose endpoint has led to counterexamples of the weak cosmic censorship conjecture in higher dimensions \cite{Lehner:2010}. In fact, soon after the discovery of the GL instability, it was suggested that black holes in global AdS would undergo a similar dynamical instability  that should lead to localization in the internal space \cite{Banks:98,Peet:98}. 
The instability of black holes in global AdS was addressed in \cite{Hubeny:02}, who found the threshold  zero (i.e.,~time-independent) mode(s) for the Schwarzschild-AdS$_5\times S^5$ black hole. In that paper it was shown that small (compared to the AdS radius) Schwarzschild-AdS$_5\times S^5$ black holes have zero modes (preserving a SO(5) isometry) when the negative modes of the Schwarzschild-AdS black hole \cite{Prestidge:1999uq}  coincided with the momentum along the internal space. For the latter, being  a $S^5$, modes are expanded in terms of spherical harmonics, which are labeled by the harmonic number $\ell$. Then there exists an infinite number of zero modes, one for each (discrete) value of $\ell$ and the corresponding specific horizon radius.

The existence of zero modes suggests two things: First, the presence of a dynamical instability and, second, the existence of new branch of static black holes. Indeed, in the case of the GL instability of black strings, \cite{Gubser:2001ac,Wiseman:2002zc} first showed that there exists a new branch of non-uniform black strings that emanates from the uniform string branch precisely at the onset of the GL instability. Similarly,  \cite{Dias:2015pda} constructed asymptotically global AdS$_5\times S^5$ black holes with non-uniform horizons along the $S^5$ that  emanate from the Schwarzschild-AdS$_5\times S^5$ black holes at the onset of GL instability for the $\ell = 1$ and $\ell=2$ modes respectively. These so-called lumpy black holes have the same horizon topology as their Schwarzschild-AdS$_5\times S^5$ counterparts, but no longer preserve the full isometry group of the internal space. In the examples of \cite{Dias:2015pda}, only a round $S^4$ inside the internal $S^5$ is preserved. The role of the lumpy black holes in the space of black hole solutions in AdS$_5\times S^5$  is analogous to that of the non-uniform black strings in the context of black holes in standard Kaluza-Klein (KK) theory. Similar to the fact that there exist localized black holes in KK theory \cite{Sorkin:2003ka}, there also exist black holes in AdS$_5\times S^5$ that are localized on the $S^5$. The latter were numerically constructed in  \cite{Dias:2016eto}. The localized black holes of  \cite{Dias:2016eto} should presumably merge with the $\ell = 1$ lumpy black holes, but  the  branches of solutions constructed  in \cite{Dias:2015pda} and \cite{Dias:2016eto} are still quite far from the merger point. One of the goals of this paper is to get close to this point from the lumpy black holes side.

Mergers in the space of static black hole solutions in general relativity have been thoroughly studied in the context of KK theory. Kol  \cite{Kol:top} proposed that the horizon topology change between the two branches is locally described by conifold-type-of transition controlled by a (singular) self-similar double-cone geometry. This double-cone model has been confirmed for the localized black hole/non-uniform black string transition   \cite{Kol:2003ja} (see also the recent studies \cite{Kalisch:2016fkm,Kalisch:2017bin,Cardona:2018shd,Ammon:2018sin,Emparan:top2}) and in  asymptotically flat space \cite{Emparan:top,Figueras:bumpy}.
One  non-trivial prediction of the double-cone model is that the behavior of any physical quantity $Q$ near the critical point is given by
\beq\label{coneQ}
Q = Q_{\text{\tiny c}} + C_+\rho_0^{s_+} + C_-\rho_0^{s_-},
\enq
where $Q_{\text{\tiny c}}$ is the value of $Q$ at the critical point, $\rho_0$ is any length that measures the deviation from the cone, and  $s_+$ and $s_-$ are the critical exponents that govern the perturbations away from the double-cone geometry. These critical exponents are dimension dependent, and they are complex for $D<10$ and purely real for $D>10$; $D=10$ is a degenerate case that leads to a power law with a logarithmic correction. This behavior of the physical quantities near the critical point  has been beautifully confirmed for $D<10$ \cite{Kalisch:2016fkm} and $D=10$ \cite{Kalisch:2017bin,Cardona:2018shd}. Whilst physical quantities may be defined on the whole horizon, eq.~\eqref{coneQ} indicates that sufficiently close to the critical point in a topology-changing transition, they are actually governed by the regions of the horizon that pinch-off. 

So far no topology change transitions analogous to the localized black hole/non-uniform black string have been studied in AdS. In this paper we fill this gap by numerically constructing critical lumpy black holes. We review, correct and extend the work of \cite{Dias:2015pda} by constructing lumpy black holes in AdS$_5\times S^5$  and AdS$_4\times S^7$ with non-uniformity parameters (see Section \ref{sec:lumpy} for the definition) $\lambda\sim 25$ and $\lambda\sim 104$ (for $\ell = 1$) respectively.\footnote{Following \cite{Gubser:2001ac}, from now on we will exclusively use $\lambda$ to denote the non-uniformity parameter, as is customary in the field.  This should not be confused with the 't Hooft coupling. } Ref. \cite{Dias:2015pda} only considered the AdS$_5\times S^5$ case, and from their plots we have estimated that their most critical solutions have $\lambda\sim 2$. For the $\ell = 1$ lumpy family, the phase diagrams that we obtain exhibit some differences compared to those reported in \cite{Dias:2015pda}, with our branches being shorter (see Section \ref{ssec:thermo}). This may appear to be a little surprising since we have managed to construct solutions with significantly larger values of $\lambda$, and hence closer to the critical point;  these discrepancies can be attributed to the different quality of our numerical solutions and theirs. To get close to the critical point, we employ the sophisticated methods that we successfully used in \cite{Cardona:2018shd} (see also \cite{Kalisch:2016fkm,Kalisch:2017bin} and Section \ref{sssec:boundarylumpy2}). As we will show in Section \ref{ssec:crit}, in the context of asymptotically AdS$_p\times S^q$ black holes, the critical geometry is controlled by a triple cone which is not Ricci flat anymore. We have checked that, to leading order, the behavior near the critical point of physical quantities of the different families of lumpy black holes that we have constructed (more specifically $\ell=1,2,3$) is given by \eqref{coneQ}. Furthermore, the critical exponents obtained from our numerical solutions (see Tables~\ref{tablecrit10} and \ref{tablecrit11}) match the prediction of the triple-cone model. Therefore, we are confident of the correctness and accuracy of our numerical solutions.  

Another novelty of our work is that we use the holographic dictionary to extract how the physics of the topology change is imprinted in the dual field theory observables. Using the tools of KK holography \cite{Skenderis:2006kkh}, we extract the vacuum expectation values (vev's) of the boundary stress-energy tensor and certain scalar operators (see Section \ref{sec:results}).\footnote{This is only possible when the SO$(q+1)$ isometry of the internal $S^q$ is broken, as it occurs at the threshold point of the GL instability of small Schwarzschild-AdS black holes \cite{Hubeny:02}, and persists along the lumpy branches. Holographically, this is interpreted as a spontaneous symmetry breaking in the dual gauge theory.} The latter parametrize the deformations of the internal space and, unsurprisingly, near the critical point they exhibit the behavior predicted by the triple-cone model \eqref{coneQ}. This result can be interpreted as a prediction from holography of how the dual field theory may detect a topology change in the bulk.  We should point out that our results from KK holography differ from those of \cite{Dias:2015pda}. In particular, the expressions for the holographic stress-energy tensor and the scalar vev's are different, and we also find that $\langle \mathcal{O}_{\mathcal{T}^4}\rangle=0$. To correct possible errors and typos in the literature, we felt the need to provide an extensive review of KK holography and the details of our calculations in Appendices \ref{app:KKhol} and \ref{applyHolo}. 

So far we have focused on  asymptotically AdS$_5\times S^5$ lumpy black hole solutions since these are the relevant ones for type IIB supergravity in ten spacetime dimensions. In this paper we also consider lumpy black holes in asymptotically AdS$_4\times S^7$ spacetimes, hence they are solutions to 11-dimensional supergravity/M-theory.\footnote{We could also have considered $(p,q) = (7,4)$ within $11D$ supergravity; other possibilities are, for instance, type IIB supergravity with a 3-form field strength, corresponding to $(p,q) = (7,3)$, or type IIA  supergravity with a Ramond-Ramond (RR)  2- or 4-form. In this case one can presumably construct black hole solutions for $(p,q) = (4,6)$, $(8,2)$ and $(6,4)$, respectively.} In this case, the only bosonic fields are the metric and the 4-form field strength. The equations of motion for the bosonic fields in this theory are simpler than in the IIB case (because there is no self-duality condition for the field strength), which allows us to get much closer to the merger point and test the triple-cone model with exquisite detail. Lumpy black holes in this theory should be dual to certain thermal phases of ABJM theory \cite{Aharony:ABJM}.

%~~~~~~~~~~~~~~~~~~~~~~~~~~~~~~~~~~~~~~~~~~~~~~~
\subsubsection*{The gravity problem}
%~~~~~~~~~~~~~~~~~~~~~~~~~~~~~~~~~~~~~~~~~~~~~~
Our  problem  consists of finding static black holes that are solutions to certain supergravities and are asymptotically AdS$_p\times S^q$  (so $D = p + q$ is the total number of spacetime dimensions) such that the $S^{p-2}$ inside AdS and an $S^{q-1}$ inside the $S^q$ are round. Asymptotically, the solutions must tend to the metric: \beq\label{asymp}
\dd s^2_{\text{\tiny AdS$\times$S}} = \dd s^2_{(p)}(\text{AdS}) + R^2\dd\Omega_{(q)}^2,
\enq
where $R$ denotes the radius of the $S^q$. $R$ is fixed by supersymmetry  in terms of the radius $L$ of the AdS factor. For $(p,q) = (5,5)$ and $(4,7)$, the background metric (\ref{asymp}) is a solution to the bosonic sector of type IIB SUGRA with only a RR self-dual 5-form field strength or 11$D$ SUGRA respectively. 

The actions for the theories considered in this paper can be derived requiring local supersymmetry and we truncate them to the desired field content \cite{freedmanbook}. We have considered the following normalizations for the field strengths: \beq\begin{split}\label{actions}
S_{\text{IIB}} &= \frac{1}{16\pi G_{10}}\int\(R\star\uni - \frac{1}{4}\star F_{(5)}\wedge F_{(5)}\), \\
S_{11D} &= \frac{1}{16\pi G_{11}}\int\(R\star\uni - \frac{1}{2}\star F_{(4)}\wedge F_{(4)}\),
\end{split}\enq
where $F_{(n)} = \dd A_{(n-1)}$. In type IIB case, one has to additionally impose the self-duality condition $F_{(5)} = \star F_{(5)}$ after deriving the equations of motion. 

A few comments are in order. First, notice that in $D = 10$, this is the minimal field content consistent with the symmetries. Since the $S^3$ within the AdS$_5$  factor is assumed to be round, a 3-form field strength would also be compatible with the symmetry, but it is not necessary. For simplicity, in this paper we have  turned it off. We leave for future work the problem of exploring for phase diagram of asymptotically AdS$_5\times S^5$ black hole solutions with an additional 3-form field strength. Second, in $D = 11$ the action contains a Chern-Simons term, $\sim\int F_{(4)}\wedge F_{(4)}\wedge A_{(3)}$, whose coefficient is fixed by supersymmetry. However, for static configurations and with the symmetry assumptions of this paper,  the Chern-Simons term does not contribute and we omit it. 

The equations of motion that one obtains from (\ref{actions}) can be written as follows: \beq\begin{split}\label{eom}
R_{MN} - \frac{1}{96}F_{(5)MPQRS}F_{(5) N}^{\phantom{(5)N}PQRS} &= 0, \hs{1.1} \dd F_{(5)} = 0, \hs{0.75} F_{(5)} = \star F_{(5)}, \\
R_{MN} - \frac{1}{12}\(F_{(4)MOPQ}F_{(4)N}^{\phantom{(4)N}OPQ} - \frac{1}{12}g_{MN}|F_{(4)}|^2\) &= 0, \hs{0.75} \dd\star F_{(4)} = 0.
\end{split}\enq
The indices run over the total number of spacetime dimensions. None of the equations for the metric has a `bare' cosmological constant. Instead, this emerges from the flux of the gauge field which allows the background (\ref{asymp}) to be a solution of (\ref{eom}) with: \beq\begin{split}\label{gaugeSchw}
F_{(5)} &= \frac{4}{L}\(\text{vol(AdS$_5$)} + \text{vol($S_5$)}\), \hs{0.75} R = L, \\
F_{(4)} &= \frac{3}{L}\text{vol(AdS$_4$)}, \hs{2.99} R = 2L.
\end{split}\enq

\subsubsection*{Numerical methods}
We use the Einstein-DeTurck method to numerically integrate the equations of motion. This was originally introduced in \cite{Headrick:10} to solve static problems in vacuum Einstein's gravity, but it can be straightforwardly generalized to include matter. In fact, matter does not modify the principle part of Einstein's equations and therefore the non-vacuum Einstein-DeTurck equation: \beq\label{EdT}
R_{MN}^H \equiv R_{MN}  - \nab_{(M}\xi_{N)} = 8\pi\,G_D\(T_{MN}-\frac{1}{D-2}g_{MN}T\), \quad \xi^M = g^{NO}\(\Gamma^M_{NO} - \bar{\Gamma}^M_{NO}\),
\enq
is manifestly elliptic on static metrics. $\xi_M$ is the so-called DeTurck vector and contains the usual Levi-Civita connection $\Gamma$ compatible with the spacetime metric $g$, and a Levi-Civita connection $\bar{\Gamma}$ compatible with some reference metric $\bar{g}$ that we are free to prescribe. Additionally, there are also the matter field equations which will not be elliptic due to the underlying local gauge symmetry \beq
A_{(n-1)} \riga A_{(n-1)}+\dd\Lambda_{(n-2)},
\enq
where $\Lambda_{(n-2)}$ is an arbitrary ($n-2$)-form. The matter equations can also be modified by adding a DeTurck-like term, but in our case it is easier to fix the gauge algebraically with a suitable ansatz for the gauge field.

For static spacetimes that are either asymptotically flat, KK or AdS, and whose boundary conditions are compatible with $\xi_M$ vanishing at the boundaries of the manifold (if any exist), \cite{Figueras:Riccisol} showed that all solutions to $R_{MN}^H = 0$ are necessarily Einstein. Even though the proof of \cite{Figueras:Riccisol} is for the pure gravity case, it should be straightforward to generalize to the present case. In fact, as we shall see below, for our particular problem, the boundary conditions that we impose fall within the class considered in \cite{Figueras:Riccisol}, and hence solving (\ref{EdT}) should be equivalent to solving the Einstein equations. In practice, we use the square of the DeTurck vector, $\xi^2\equiv \xi_M\xi^M$, to quantify the validity of our numerical solutions and to perform convergence tests. More details about the DeTurck method can be found in \cite{Headrick:10} and in  the reviews \cite{Wiseman:2011by,Dias:revst}.

From the technical point of view, the equations are always discretized using pseudo-spectral methods on a Chebyshev grid, where it has been proven that exponential convergence is achieved for sufficiently smooth solutions. Then we solve them by a standard iterative Newton-Raphson method; the resulting linear system of equations is solved using LU decomposition (implemented by subroutine \texttt{LinearSolve} in Mathematica.)

\subsubsection*{Organization of the paper}
The rest of the paper is organized as follows. In Section \ref{sec:schwads} we review the simplest black hole solution with the desired asymptotics, namely the Schwarzschild-AdS$_p\times S^q$ black hole. We do so for generic values of $p$ and $q$, and we present the expressions for the physical quantities. In Section \ref{sec:lumpy} we present the metric and gauge field ansatz, again for generic values of $p$ and $q$, that we use for the numerical construction of the lumpy black holes. This section also includes a discussion about the boundary conditions that yield a well-posed boundary value problem. The one-parameter family that we use as ansatz captures, by construction, the threshold GL  zero mode for any $\ell$. We use this mode as the initial seed to find the actual non-linear solutions. The results are presented in Section \ref{sec:results}. This section constitutes the core of the paper, where the phase diagrams are shown and discussed. Special emphasis is put on the isometric embeddings, the topology changes  and the critical behavior of the solutions close enough to the merger point. In particular, we compute the critical exponents and find the critical values of various thermodynamic quantities. We close in Section \ref{disc} with a discussion and outlook of our results. Finally, we include a few appendices with technical details. KK holography in AdS$_5\times S^5$ is reviewed thoroughly in Appendix \ref{app:KKhol}, and detailed application of this procedure  to our solutions can be found in Appendix \ref{applyHolo}. This requires a basis of spherical harmonics that preserve an SO(5) subgroup of the full SO(6) rotational group of the $S^5$; this is presented for a generic value of $q$ in Appendix \ref{harm}. In Appendix \ref{appEmb10d} we show the embedding plots in $D = 10$, which are not included in the main body of the paper. In Appendix \ref{app:conve} we include further details of the numerical construction, especially of the grids employed, and we perform some convergence tests. 

%~~~~~~~~~~~~~~~~~~~~~~~~~~~~~~~~~~~~~~~~~~~~~~~
\section{Schwarzschild-AdS$_{\text{p}}\times$S$^{\text{q}}$ black hole}
\label{sec:schwads}
%~~~~~~~~~~~~~~~~~~~~~~~~~~~~~~~~~~~~~~~~~~~~~~
The simplest black hole solution with the desired asymptotics is the Schwarzschild-AdS$_p\times S^q$ black hole, which is known explicitly.  In this section we review those aspects of this solution that are relevant for our analysis for generic values of $p$ and $q$. Most of the material in this section is well-known and can be skipped by the expert reader.

The horizon of Schwarszchild-AdS$_p \times S^q$ black hole  is uniformly smeared along the internal $S^q$ and it thus respects the full SO$(q+1)$ symmetry; the horizon topology is $S^{p-2}\times S^q$. The metric reads: \beq\label{schwadsp}
\dd s^2 = -f(r)\dd t^2 + \frac{1}{f(r)}\dd r^2 + r^2\dd\Omega_{(p-2)}^2 + R^2\dd\Omega_{(q)}^2, \\
\enq
where $\dd\Omega^2_{(p-2)}$ 
and $\dd\Omega^2_{(q)} = \dd\theta^2 + \sin^2\theta\,\dd\Omega_{(q-1)}^2$ are the line elements of the round spheres appearing in the AdS$_p$ and $S^q$ factors respectively, and  
\beq
f(r) = 1+\frac{r^2}{L^2}-\(\frac{r_+}{r}\)^{p-3}\(1+\frac{r_+^2}{L^2}\), \hs{0.75} (p > 3).
\enq
The field strength and radius of the $S^q$ coincide with those for the background AdS$_p \times S^q$ and are given in \eqref{gaugeSchw}. 

It is straightforward to compute the temperature and entropy of this black hole. The energy may be found by integrating the first law of black hole mechanics and then the free energy is given by $F = E-TS$. We find: 
\beq\begin{split}\label{EEE}
T &= \frac{p-3 + (p-1)(r_+/L)^2}{4\pi r_+}, \\
S &= \frac{1}{4G_D}r_+^{p-2}\Omega_{(p-2)}R^q\Omega_{(q)}, \\
E &= E_{\text{AdS}_p} + \frac{p-2}{16\pi G_D}\(1+\frac{r_+^2}{L^2}\)r_+^{p-3}R^q\Omega_{(p-2)}\Omega_{(q)}, \\
F &= F_{\text{AdS}_p} + \frac{1}{16\pi G_D}\(1-\frac{r_+^2}{L^2}\)r_+^{p-3}R^q\Omega_{(p-2)}\Omega_{(q)}.
\end{split}\enq
$E_{\text{AdS}_p}$ ($=F_{\text{AdS}_p}$) is the vacuum energy of Anti-de Sitter space, which is non-vanishing for $p = 3,5$ (it is zero for $p = 4$), and whose precise expression cannot be found from the first law, as it appears as an integration constant. One can find those `Casimir' energies, for instance, with the prescription of \cite{Balasubramanian:1999re} for defining the stress tensor of asymptotically AdS spacetimes. Here we instead use the results that follow from the KK holography analysis (see Appendix \ref{applyHolo} and in particular equation \eqref{thefinalTmunu}). 

Letting $L$ be the AdS radius, Schwarzschild-AdS$_p\times S^q$ black holes are called small if $r_+\lesssim L$ or large  if $r_+ \gtrsim L$. We may invert the relation between the temperature $T$ and the horizon radius $r_+$ in \eqref{EEE} to find the two branches of black holes at the same temperature $T$: 
\beq
r_+ = \frac{L}{p-1}\(2\pi TL \pm \sqrt{(2\pi TL)^2-3-p(p-4)}\).
\enq
The positive branch describes large AdS black holes, while the negative branch describes small AdS black holes. The temperature has a minimum \beq
T_{\text{min}} = \frac{\sqrt{(p-3)(p-1)}}{2\pi\,L},
\enq
at \beq\label{rmin}
r_+^{\text{min}}=L\, \sqrt{\frac{p-3}{p-1}},
\enq
and diverges in the limits of either large or small $r_+$. This minimum temperature separates the two  branches of black holes and there are no  homogeneous static black holes in AdS$_p\times S^q$ (for fixed values of $p$ and $q$)  for $T<T_\text{min}$. This is summarized in Fig.~\ref{fig:Th}.

With the expressions for the physical quantities in \eqref{EEE}, we can now review the phase diagram of the homogenous asymptotically AdS$_p\times S^q$ solutions. These include the background AdS$_p \times S^q$ itself and the two branches of Schwarzschild-AdS$_p \times S^q$ black holes. Since these spacetimes are static,  we can trivially consider their Euclidean sections. There are two relevant thermodynamic ensembles: the microcanonical and the canonical ensembles. In the former,  we compare solutions with the same energy and those with the largest entropy dominate; in the latter,  the temperature is fixed and the solutions with the lowest free energy are preferred. 

In the microcanonical ensemble the phase diagram is trivial: the background has no horizon and hence vanishing entropy; consequently, there is only one solution at a given energy, namely the corresponding Schwarzschild-AdS$_p \times S^q$ black hole. On the other hand, the phase diagram in the canonical ensemble is much richer. First notice that global AdS$_p\times S^q$ exists at any temperature: the period $\beta$ of the Euclidean time circle can be chosen to have any value. The  geometry has no horizon and therefore vanishing entropy. Consequently, $F_{\text{AdS}_p} = E_{\text{AdS}_p}$. From now on we subtract this contribution in (\ref{EEE}), and consider the difference $\Delta F = F-F_{\text{AdS}_p}$. For the Schwarzschild-AdS$_p\times S^q$, regularity of the Euclidean section requires that the period of the Euclidean time circle is given by $\beta = 1/T$ with $T$ given in \eqref{EEE}.  Furthermore, from \eqref{EEE}  we see that $\Delta F<0$ for $r_+ > L$, and hence large Schwarzschild-AdS$_p\times S^q$ black holes above this threshold dominate the canonical ensemble. On the other hand, $\Delta F>0$  for $r_+ < L$, and thus thermal AdS$_p\times S^q$ is preferred. At $r_+ = L$, corresponding to Hawking-Page (HP) temperature, 
\beq
T_{\text{HP}} = \frac{p-2}{2\pi\,L},
\enq
there is a phase transition between two distinct equilibrium states: large Schwarzschild-AdS$_p\times S^q$ black holes and thermal AdS$_p\times S^q$.  This is summarized in Fig. \ref{fig:Th}.  The energy is discontinuous across the phase transition, $E_{\text{HP}}^+ \neq E_{\text{HP}}^-$, with 
\beq
E_{\text{HP}}^\pm\equiv \lim_{T\riga T_{\text{HP}}^\pm}E(T),
\enq
yielding a non-vanishing latent heat $\Delta E$, and the phase transition is of first order. For temperatures $T < T_{\text{min}}$, there only exists thermal AdS. For $T_{\text{min}} < T < T_{\text{HP}}$, homogenous black hole solutions do exist but do not dominate the canonical ensemble; for  $T>T_{\text{HP}}$, large  Schwarzschild-AdS$_p\times S^q$  black holes are the preferred phase, with both phases co-existing at $T_{\text{HP}}$. Small Schwarzschild-AdS$_p\times S^q$ black holes are always subdominant.\footnote{Notice that planar-AdS$_p(\times S^q)$ black holes are infinitely extended since $B_{p-1} = \R^{1,p-2}$ and can be recovered in the limit $r_+\gg L$. In this limit the free energy is always negative and therefore there are no phase transitions. As we anticipated earlier, these black holes always dominate the canonical ensemble.}
 
\begin{figure}[t] 
\begin{center}
\begin{minipage}{\textwidth}
\begin{center}
\input{./figures/TL.tex}
\end{center}
\end{minipage}
\captionsetup{width=0.9\textwidth}
\captionof{figure}{\textsl{Dimensionless temperature vs.~dimensionless control parameter for the relevant values $(p,q)$. The value at which the HP phase transition takes place is shown.}}
\label{fig:Th}
\end{center}

\vs{0.1}\begin{center}
\begin{tabular}{|r||r|r|r|r|r|r|}
\cline{1-7}
\multirow{2}{*}{$(p,q)$} & \multirow{2}{*}{ \hs{0.2} $r_+^{\text{min}}/L$  } & \multirow{2}{*}{$T_{\text{min}}L$} & \multirow{2}{*}{$T_{\text{HP}}L$} & \multirow{2}{*}{$E_{\text{HP}}^-$} & \multirow{2}{*}{$E_{\text{HP}}^+$} & \multirow{2}{*}{$\Delta E$} \\
 & & & & & & \\ \hline\hline
\multirow{2}{*}{$(5,5)$} & \multirow{2}{*}{$\frac{1}{\sqrt{2}}$} & \multirow{2}{*}{$\frac{\sqrt{2}}{\pi}$} & \multirow{2}{*}{$\frac{3}{2\pi}$} & \multirow{2}{*}{$\frac{3N^2}{16L}$} & \multirow{2}{*}{$\frac{27N^2}{16L}$} & \multirow{2}{*}{$\frac{3N^2}{2L}$} \\ 
 & & & & & & \\ \hline\hline
\multirow{2}{*}{$(4,7)$} & \multirow{2}{*}{$\frac{1}{\sqrt{3}}$} & \multirow{2}{*}{$\frac{\sqrt{3}}{2\pi}$} & \multirow{2}{*}{$\frac{1}{\pi}$} & \multirow{2}{*}{$0$} & \multirow{2}{*}{$\frac{128\pi^4L^8}{3}$} & \multirow{2}{*}{$\frac{128\pi^4L^8}{3}$} \\ 
 & & & & & & \\ \hline
\end{tabular}
\vs{0.25}
\captionsetup{width=0.9\textwidth}
\captionof{table}{\textsl{Relevant thermodynamic values in the phase diagram of the solutions preserving the background's isometries. In $D = 10$, the $G_{10}$ has been expressed in terms of the dual gauge theory parameter $N$ using (\ref{G10G5N}). In $D = 11$, $G_{11}$ is a fundamental constant and can be taken to be one.}}
\label{tab:pqE}
\end{center}
\end{figure}

So far, the physics only presents dependence on $p$. For the values ($p,q$) of interest, these quantities are shown in Table \ref{tab:pqE}. Schwarzschild-AdS$_p\times S^q$ describes the universal sector to which Ref.~\cite{marolf:13} refers, only involving gravity in AdS and with the internal space playing no role. This is indeed the case  if we only consider large black holes and the background.  On the other hand, small enough Schwarzschild-AdS$_p\times S^q$ black holes are dynamically unstable under perturbations that break the isometries of the internal space \cite{Hubeny:02}. Whilst the thermodynamic stability of the small black hole branch changes at
\beq
r_+ = r^{\text{min}}_+,
\enq
becoming unstable for $r_+ < r^\text{min}_+$, this does not necessarily signal the onset of a dynamical instability. Indeed, the onset of the dynamic instability occurs when the Euclidean eigenmode $\lambda_E$ of the Schwarzschild-AdS$_p$ black hole coincides exactly with the `momentum' in the transverse space. In the present case, where we take the internal space to be a $S^q$, the threshold unstable mode will occur when $\lambda_E = -\ell(\ell+q-1)$, where $\ell$ is labels the spherical harmonics on the $S^q$. This is how the zero mode that signals the instability was first obtained. In turn, it suggests the existence of new branches of black holes that emerge from this point and that have deformed horizons along $S^q$. These new branches of black holes are completely analogous to the non-uniform strings in the context of the GL instability of black strings \cite{Gubser:2001ac,Wiseman:2002zc}.  From the $D$-dimensional point of view, for each $\ell$ there is a GL  zero mode that breaks the SO$(q+1)$ isometry of $S^q$ down to SO$(q)$. These zero modes can be uplifted to fully non-linear solutions, leading to the so-called lumpy black holes, first constructed in \cite{Dias:2015pda} in AdS$_5\times S^5$. The aim of this paper is to study in detail several branches of lumpy black holes in AdS$_5\times S^5$ and AdS$_4\times S^7$.

Linear perturbations about the Schwarzchild-AdS$_{p}\times S^q$ spacetime can be decomposed into a sum of scalar, vector and tensor perturbations, depending on how they transform under coordinate transformations on the $S^q$. The GL zero modes are in the scalar sector and are classified in terms of $\ell$. Here, instead of proceeding as in \cite{Hubeny:02} and expand the AdS sector of the metric perturbations in spherical harmonics, \beq\label{zeromodeV}
g_{\munu} \riga g_{\munu} + h_{\munu}^\ell Y_\ell(\theta), \hs{0.75} (\mu,\nu = 0,\dots, p-1),
\enq
we will linearize the full $D$-dimensional general ansatz for the lumpy black holes (see the next section) around Schwarzchild-AdS$_{p}\times S^q$. The problem of  numerically finding the lumpy black holes requires to construct the linearized Einstein-DeTurck operator,\footnote{Around an Einstein metric, this operator coincides with the Lichnerowicz operator.} and hence we can easily adapt our code to find the spectrum of time independent perturbations of Schwarzschild-AdS$_{p}\times S^q$ black holes. Since the latter turn out to be thermodynamically unstable for $r_+ < r_+^{\text{min}}$, we started our search for zero modes at $r_+ = r_+^{\text{min}}$ and gradually decreased this value until a zero mode is found. The first three zero modes, corresponding to $\ell = 1,2,3$, occur at 
\beq\begin{split}\label{GLvalues}
(p,q) &= (5,5): \hs{0.5} y_0^{\ell = 1} = 0.44023414, \hs{0.75} y_0^{\ell = 2} = 0.32388984, \hs{0.75} y_0^{\ell = 3} = 0.25704192, \\
(p,q) &= (4,7): \hs{0.5} y_0^{\ell = 1} = 0.28898162, \hs{0.75} y_0^{\ell = 2} = 0.24819894, \hs{0.75} y_0^{\ell = 3} = 0.21751714,
\end{split}\enq
where $y_0 = r_+/L$ . In practice we use these zero modes as initial seeds for finding the corresponding non-linear families of lumpy black holes. We note that in  the $D = 10$ case, our values coincide with those found in \cite{Dias:2015pda} for $\ell = 1,2$. In the $D = 11$ case, as far as we know, our results are new. The spectrum of negative modes of the Schwarzschild-AdS$_4$ black hole was computed in \cite{Prestidge:1999uq}. The threshold radii for the zero modes that we have found agree, within numerical error, with the values of the radii  where these negative modes coincide with the eigenvalues $-\ell(\ell+6)$  of the scalar harmonics on the $S^7$.

%~~~~~~~~~~~~~~~~~~~~~~~~~~~~~~~~~~~~~~~~~~~~~~~
\section{Lumpy black holes in AdS$_{\text{p}}\times$S$^{\text{q}}$}
\label{sec:lumpy}
%~~~~~~~~~~~~~~~~~~~~~~~~~~~~~~~~~~~~~~~~~~~~~~
In this section we present the numerical construction of lumpy black holes in AdS$_{p}\times S^{q}$. The different cases that we explore differ by the equations of motion and by the $n$-form gauge field, but the ansatz for the metric is general and can be written down for any values of $p$ and $q$.

Lumpy black holes in AdS$_p\times S^q$ only retain a SO($q$) subgroup of the full isometry group of the internal $S^q$. In the space of static solutions, they emanate  from the small Schwarzschild-AdS$_p\times S^q$ black hole branch at the GL threshold point, and therefore they have the same horizon topology: $S^{p-2}\times S^q$. They are classified by $\ell$, which is the harmonic number corresponding to the scalar harmonic on the $S^q$  that labels the zero mode. Each lumpy solution of the $\ell$th family depends on two coordinates: the radial coordinante $r$ in the AdS$_p$ factor, and the polar angle $\theta$ on the $S^q$; using standard spherical coordinates, the polar angle ranges from 0 to $\pi$, with the `north' pole of $S^q$ at $\theta = 0$, the equator at $\theta = \pi/2$ and the `south' pole at $\theta = \pi$.   

Uniform horizons on $S^q$ trivially possess a reflection symmetry $\theta \leftrightarrow \pi-\theta$ with respect to the equator of this $S^q$. However, under this parity transformation  scalar harmonics pick up a factor of $(-1)^\ell$. Recall that  the various families of lumpy black holes are non-linear solutions of the field equations that are continuously connected (in the space of solutions) to the linear deformations of the uniform  black holes parametrized by the scalar harmonics. Therefore, the lumpy families labelled by the harmonic  $\ell$ transform in the same way as the scalar harmonics under reflections about the equator of the internal $S^q$. Consequently, lumpy black holes labelled by an odd $\ell$ are not symmetric with respect to reflections about the equator, and one has to consider the whole range of the polar angle $\theta$ on the $S^q$ to construct them. On the other hand, for even $\ell$ lumpy black holes enjoy a reflection symmetry about the equator and it  suffices to consider $\theta\in\[0,\pi/2\]$.

%~~~~~~~~~~~~~~~~~~~~~~~~~~~~~~~~~~~~~~~~~~~~~~~
\subsection{Metric ansatz}
\label{lumpl1}
%~~~~~~~~~~~~~~~~~~~~~~~~~~~~~~~~~~~~~~~~~~~~~~
Given the previous considerations, a suitable ansatz for the metric of the lumpy black holes consists of  a general enough deformation of the Schwarzschild-AdS$_p\times S^q$ black hole that breaks the SO$(q+1)$ symmetry down to SO$(q)$. To solve the equations of motion numerically, it turns out to be convenient to compactify the radial coordinate and to redefine the polar angle as follows: 
\beq\label{changelu}
r(y) = \frac{r_+}{1-y^2}, \hs{0.75} \theta(a) = \arcsin(1-a^2).
\enq
In these coordinates, the horizon is located at $y = 0$ and the conformal boundary of AdS$_p$ is at $y = 1$, while the north and south poles of the $S^q$ are at $a = \pm 1$ respectively, and  the equator is at $a = 0$. Then for odd/even $\ell$ the integration domain is simply $[0,1]\times[-1,1]$ / $[0,1]\times[0,1]$. We will be parametrize the ansatz in terms of the dimensionless parameter $y_0 = r_+/L$. If we further rescale the time coordinate $t \riga Lt$, then the AdS radius $L$ drops out of the equations of motion, leaving $y_0$ as the only control parameter. This parameter effectively sets the temperature of the black hole. Therefore, for any given harmonic $\ell$, the lumpy black holes form a one-parameter family of solutions specified by the temperature. 

Applying these changes of coordinates and redefinitions to (\ref{schwadsp}), the metric reads: 
\beq\begin{split}\label{schwxa}
\dd s^2 &= \frac{L^2}{(1-y^2)^2}\(-\frac{G_p(y)}{1-y^2}\dd t^2 + \frac{4y^2(1-y^2)y_0^2}{G_p(y)}\dd y^2 + y_0^2\dd\Omega_{(p-2)}^2\) \\
&\hs{0.45}+R^2\(\frac{4\dd a^2}{2-a^2}+ (1-a^2)^2\dd\Omega_{(q-1)}^2\),
\end{split}\enq
with \beq
G_p(y) = 3y^4 - y^6 - (1+y_0^2)\((1 - y^2)^p-1\) - y^2\(3+y_0^2\).
\enq
This is the metric that we will use as the reference metric in the Einstein-DeTurck equations \eqref{EdT}. We can now present the ansatz for the lumpy black holes. For any value of $p$ and $q$, we consider: 
\beq
\begin{split}
\label{ansatzlumpy}
\dd s^2 = &~\frac{L^2}{(1-y^2)^2}\bigg(-\frac{G_p(y)}{1-y^2}Q_1\dd t^2 + \frac{4y_0^2y^2(1-y^2)}{G_p(y)}Q_2\(\dd y - (1-y^2)^2Q_3\dd a\)^2 + y_0^2Q_5\dd\Omega_{(p-2)}^2\bigg)\\
& + R^2\(\frac{4Q_4}{2-a^2}\dd a^2 + (1-a^2)^2Q_6\dd\Omega_{(q-1)}^2\),
\end{split}\enq
where $Q(y,a)\equiv\{Q_1,\dots, Q_6\}(y,a)$ are the unknowns for the metric field. For $Q_i(y,a) = 1$, $\forall i \neq 3$ and $Q_3(y,a) = 0$, we recover (\ref{schwxa}). These are actually the Dirichlet boundary conditions imposed at $y = 1$, as required by the asymptotics. The ansatz (\ref{ansatzlumpy}) preserves the full SO$(p-1)$ symmetry of the $S^{p-2}$ within the AdS$_p$ factor and the unbroken SO$(q)$ symmetry of the $S^{q-1}$ inside the $S^q$.

%~~~~~~~~~~~~~~~~~~~~~~~~~~~~~~~~~~~~~~~~~~~~~~~
\subsection{Gauge field}
\label{sssec:gaugefield1}
%~~~~~~~~~~~~~~~~~~~~~~~~~~~~~~~~~~~~~~~~~~~~~~
The different theories that admit AdS$_p\times S^q$ asymptotic solutions also contain a non-trivial $n$-form gauge field strength; $n = 5$ in $D = 10$ and $n = 4$ in $D = 11$. In order to solve the equations of motion numerically, we need to provide a suitable ansatz for this field. 

Instead of considering the form field strength $F_{(n)}$, we will work with the potential form field $A_{(n-1)}$, such that $F_{(n)} = \dd A_{(n-1)}$. The reason is that the equations of motion for $A_{(n-1)}$ are of 2nd order and elliptic. To find  a suitable ansatz, we consider a deformation of the gauge field $A_{(n-1)}$ corresponding to the Schwarzschild-AdS$_p\times S^q$ black hole (\ref{gaugeSchw}) written  in terms of the coordinates $(y,a)$; see Eq.~\eqref{changelu}. In particular, we consider a  gauge such that the $(n-1)$-form potential vanishes at the horizon, which has the benefit of simplifying the boundary conditions there. Taking these considerations into account, we have the following ansatz for $A_{(n-1)}$: 
\beq
\begin{split}
\label{gaugefields}
(p,q) &= (5,5): \hs{0.5} A_{(4)} = L^4y_0^4\frac{y^2(2 - y^2)(2 - 2 y^2 + y^4)}{(1-y^2)^4}Q_7\dd t\wedge\dd \sigma_{(3)} - W\dd\sigma_{(4)}, \\
(p,q) &= (4,7): \hs{0.5} A_{(3)} = -L^3 y_0^3\frac{y^2 (3-3y^2+y^4)}{(1-y^2)^3}Q_7\dd t\wedge \dd\sigma_{(2)},
\end{split}\enq
where $\dd \sigma_{(r)}$ is the volume $r$-form of a round unit $r$-sphere, and $Q_7=Q_7(y,a)$ is an unknown function that encodes the deformations away from Schwarzschild-AdS$_p\times S^q$. For $Q_7(y,a) = 1$ (and a particular function $W$ for $(p,q) = (5,5)$), these expressions yield the field strengths given in \eqref{gaugeSchw} in the $(y,a)$ coordinates. Having fixed the gauge \textsl{a priori}, the equation of motion for the potential is elliptic; this equation together with the Einstein-DeTurck equations for the metric \eqref{EdT} and for suitable boundary conditions (see Section \ref{sssec:boundarylumpy1}) form  a well-posed boundary value problem.

There is a small caveat in the case $(p,q) = (5,5)$. Notice that we have not specified the particular form of the function $W(y,a)$ appearing in the ansatz. The reason is that the self-duality condition $F_{(5)} = \star F_{(5)}$ allows us to eliminate the first derivatives of $W$ in terms of the first derivatives of $Q_7$.  Then, the equation $\dd F_{(5)}=0$ can be written solely in terms of $Q_7$ and its first and second derivatives.   This is not surprising, since the self-duality condition by itself is non-dynamical. 

%~~~~~~~~~~~~~~~~~~~~~~~~~~~~~~~~~~~~~~~~~~~~~~~
\subsection{Boundary conditions}
\label{sssec:boundarylumpy1}
%~~~~~~~~~~~~~~~~~~~~~~~~~~~~~~~~~~~~~~~~~~~~~~

The boundary value problem has to be supplemented with suitable boundary conditions to form a well-posed problem. These are obtained by requiring regularity at the horizon and at the poles (or reflection symmetry at the equator for even values of $\ell$), and by requiring that the solution is asymptotically AdS$_p\times S^q$. In full, the boundary conditions that we impose are: \begin{itemize}
\item{{\bf\textsf{Horizon}} at $y = 0$: regularity at the horizon implies that all functions $Q$ must be even in $y$ and therefore we impose a Neumann boundary condition on all of them, except $Q_3$ which is Dirichlet. The condition $Q_1(0,a) = Q_2(0,a)$ ensures that the geometry is free of conical singularities and fixes the surface gravity of the lumpy black holes to be the same as that of the reference metric.}
\item{{\bf\textsf{Asymptotic boundary}} at $y = 1$: AdS$_p\times S^q$ asymptotics imply Dirichlet boundary conditions $Q_i(1,a) = 1$, $\forall i \neq 3$, and $Q_3(1,a) = 0$.}
\item{{\bf\textsf{North and south poles of S$^\textbf{q}$ (for $\ell$ odd)}} at $a = \pm 1$ respectively: all $Q$'s must be even in  $(1\mp a)$ respectively, and thus we impose Neumann boundary conditions on all of them except the crossed term $Q_3$, which must satisfy a Dirichlet boundary condition, $Q_3=0$. In addition, to avoid conical singularities at the poles we impose: $Q_4(y,\pm1) = Q_6(y,\pm1)$.}
\item{{\bf\textsf{Equator and north pole of S$^\textbf{q}$ (for $\ell$ even)}} at $a = 0$ and $a = 1$ respectively: all $Q$'s must be even there and thus we impose Neumann boundary conditions on all of them except the crossed term $Q_3$ which must satisfy a Dirichlet boundary condition, $Q_3=0$. In this case, we only need to further impose the absence a conical singularity at the pole $(a=1)$ by requiring $Q_4(y,1) = Q_6(y,1)$.}
\end{itemize}

%~~~~~~~~~~~~~~~~~~~~~~~~~~~~~~~~~~~~~~~~~~~~~~~
\subsection{Further considerations}
\label{sssec:boundarylumpy2}
%~~~~~~~~~~~~~~~~~~~~~~~~~~~~~~~~~~~~~~~~~~~~~~
Now we are in position to construct the numerical solutions for the lumpy black holes. We start with the Schwarzschild-AdS$_{p}\times S^q$ black hole, corresponding to $Q_i(y,a) = 1$, $\forall i \neq 3$, and $Q_3(y,a) = 0$ in \eqref{ansatzlumpy} and \eqref{gaugefields}, and we add a bit of the $\ell$th zero mode, depending on the lumpy family that we aim to find, to construct a good enough initial guess of the Newton-Raphson loop. Whereas for $\ell$ odd, adding or subtracting the zero mode connects to the same lumpy family, this is not the case for $\ell$ is even. This is better understood in terms of the perturbations as written in \eqref{zeromodeV}: under a parity transformation, the perturbations transform as $h_{\munu}^\ell \riga (-1)^\ell h_{\munu}^\ell$. Therefore,  we see that for odd $\ell$, the positive sign in the perturbation is equivalent to the negative sign by transforming  $\theta \riga \pi - \theta$, which is nothing but a redefinition of what we call `north' and `south' poles of the $S^q$. However, this redundancy does not apply when $\ell$ is even, and  the choice of the sign yields physically different perturbations that connect to different families of lumpy black holes. In this paper we have focused on finding the first four families of lumpy black holes, i.e., $\ell = 1, 2^+, 2^-$ and $3$. The upper sign for  $\ell = 2$  refers to the family that branches off from the Schwarzschild-AdS$_{p}\times S^q$ black hole by adding ($+$) or subtracting ($-$) the zero mode for seeding the first solution.

Given the boundary conditions above, the temperature of our lumpy black holes is the same as the Schwarzschild-AdS$_p\times S^q$ black hole used as a reference metric and is controlled by the parameter $y_0$ in \eqref{ansatzlumpy}. We move along a given branch of lumpy black holes by varying $y_0$. In analogy with non-uniform black strings in KK theory \cite{Gubser:2001ac}, we define a `lumpiness' parameter $\lambda$ (see Eq.~\eqref{lumpiness}), that measures the size of deformations away from the Schwarzschild-AdS$_p\times S^q$ black hole. Slightly non-uniform (on the $S^q$) lumpy black holes have $\lambda \ll 1$, while $\lambda \to \infty$ as we approach the critical regime, where lumpy black holes merge with another family of black holes with a different horizon topology. 
For non-critical lumpy black holes ($\lambda  \lesssim 1$), one single patch of Chebyshev grid points is enough to obtain accurate numerical results. However, critical solutions need a more careful treatment. In order to access the critical regime we employed three techniques that have been used successfully in the past \cite{Cardona:2018shd,Kalisch:2017bin} . For critical solutions but not the most critical ones, we supplement the single domain with mesh-refinement  in the region where it is necessary. As we approach the merger point, some functions develop very steep gradients near one (or both) pole(s) and mesh-refinement alone is insufficient to accurately construct such solutions; to resolve them, we simply redefine the unknowns as $Q^{\text{new}} = 1/Q^\text{old}$, for some functions (see below), and we split the domain into several (smaller) patches and apply mesh-refinement to increase the resolution where it is needed. The boundary conditions for the new unknowns are the same as the ones listed before, but the new functions remain bounded as we approach the critical regime. 

In general, we have tracked the value of $\xi^2$ to determine when the resolution needs to be increased. We have always required $\xi^2 < 10^{-10}$, despite the most critical solutions we reached do not satisfy this bound. Still, the numerical error is very localized and under control. In addition, we have also monitored some physical observables (see Section \ref{sec:results}) along a given family; when numerical errors are unacceptably large they are reflected in the physical quantities. With these considerations, we have managed to confidently construct families  of lumpy black holes up to following maximum values of 
$\lambda$: 
\beq
\label{maxlambda}
\begin{split}
(p,q) &= (5,5): \hs{0.5} \lambda_{\text{max}}^{\ell = 1} = 25, \hs{0.955} \lambda_{\text{max}}^{\ell = 2^+} = 8.1, \hs{0.83} \lambda_{\text{max}}^{\ell = 2^-} = 6.3, \hs{0.95} \lambda_{\text{max}}^{\ell = 3} = 26,  \\
(p,q) &= (4,7): \hs{0.5} \lambda_{\text{max}}^{\ell = 1} = 104, \hs{0.75} \lambda_{\text{max}}^{\ell = 2^+} = 109, \hs{0.75} \lambda_{\text{max}}^{\ell = 2^-} = 6.0, \hs{0.75} \lambda_{\text{max}}^{\ell = 3} = 100.
\end{split}
\enq

To construct solutions in the critical regime for the various families, we generically redefine the unknowns $Q_{1,2,7}$ as described above. For the families $\ell = 1,2^+$ and $3$, the singular behavior appears near a  pole and near the horizon. To resolve it, we increase the resolution in that  corner  by splitting the integration domain into several patches, with higher resolution  and applying mesh-refinement  in the regions where it is needed. We have considered up to 4 patches for the most critical solutions. See Appendix \ref{app:conve} for more details and some examples of the types of domains that we have used to construct critical solutions. 

 It turns out to be much harder to construct critical solutions for the $\ell = 2^-$ branch. The reason is that the singular behavior of the functions  is not localized at a corner but somewhere  near the horizon and in between the equator and the poles. Additionally, for the two cases that we have considered, namely $(p,q)=(5,5),\,(4,7)$, there is a turning point along the branch in the non-critical regime. Indeed, we find that starting at $y_0=y_0^{\ell=2}$ (see equation \eqref{GLvalues}) and decreasing $y_0$, at some point we can no longer find solutions. Examining the behavior of the physical quantities near this point, it is clear that we are not at the end of the branch; in fact, we observe that the derivative of any thermodynamical variable as a function of the temperature becomes infinite. It turns out that  this point corresponds to a local maximum of the temperature along the branch. To go past it and  to larger values of $\lambda$, we use the trick explained in Ref.~\cite{Dias:revst}.

%~~~~~~~~~~~~~~~~~~~~~~~~~~~~~~~~~~~~~~~~~~~~~~~
\section{Results}
\label{sec:results}
%~~~~~~~~~~~~~~~~~~~~~~~~~~~~~~~~~~~~~~~~~~~~~~

%~~~~~~~~~~~~~~~~~~~~~~~~~~~~~~~~~~~~~~~~~~~~~~~
\subsection{Thermodynamics}
\label{ssec:thermo}
%~~~~~~~~~~~~~~~~~~~~~~~~~~~~~~~~~~~~~~~~~~~~~~
In this section we display and discuss the phase diagrams of asymptotically global AdS$_p\times S^q$ spacetimes, including the lumpy black holes solutions that we have numerically constructed. To make the microcanonical and canonical phase diagrams easier to visualize, we plot the dimensionless differences of the entropy or the free energy with respect to the small Schwarzschild-AdS$_p\times S^q$ black hole with the same energy or temperature, depending on the ensemble.

In the $D=10$ case we can use the holographic relation,
\beq
\label{G10G5N}
N^2 = \frac{\pi^4L^8}{2G_{10}},
\enq
to express the physical quantities in terms of gauge theory units. In this way, we shall display 
$S/N^2$ versus $EL/N^2$ in the microcanonical ensemble and $FL/N^2$ versus $TL$ for the canonical ensemble. In $D = 11$, the Newton's constant is a fundamental quantity and it is only related to Planck's length. We can safely take units where $G_{11} = 1$, and then $S/L^9$, $E/L^8$ and $F/L^8$ are dimensionless.

Before proceeding to display the phase diagrams, we collect the expressions for the various physical quantities of interest. Given our ansatz for the metric \eqref{ansatzlumpy} and the boundary conditions that we impose, the temperature of lumpy black holes is the same as the reference Schwarzschild-AdS$_p\times S^q$ black hole and it is given by \eqref{EEE}. The entropy of a black hole is proportional to its horizon area, $S = A_H/(4G_D)$. For generic values of $p$ and $q$, the area of the horizon for the lumpy black holes is given by 
\beq
A_H = 2(Ly_0)^{p-2}R^{q}\Omega_{(p-2)}\Omega_{(q-1)}\int_{-1}^{+1}\dd a\,\frac{(1-a^2)^{q-1}}{\sqrt{2-a^2}}\sqrt{Q_5^{p-2}Q_4Q_6^{q-1}}\Big|_{y = 0},
\enq
for odd $\ell$. For even $\ell$ the integral ranges from 0 to 1 and we have to multiply by a factor of 2 due to the reflection symmetry about the equator of the internal $S^q$. As a consistency check, note that setting $Q_4 = Q_5 = Q_6 = 1$, yields the horizon area of the Schwarzschild-AdS$_p\times S^q$ solution given in \eqref{EEE}.

The most straightforward way to compute the energy of the solutions is by integrating the 1st law of thermodynamics. We can easily find the curve $S(T)$, since it only involves the computation of horizon area, and the families of lumpy black holes are parametrized by $T$. Then the energy is given by: 
\beq
\label{1stlaw}
E(T) = E_{\text{GL}} + \int_{T_{\text{GL}}}^T\deriv[S(T')]{T'}\dd T'\,,
\enq
where $E_\text{GL}$ and $T_{\text{GL}}$ are the energy and temperature of the Schw-AdS$_p\times S^q$ black hole at the $\ell$th GL threshold point. This procedure, however, fails if there is a turning point along a given family, since at this point  the derivative $\dd S(T')/\dd T'$ blows up. This is the case for the $\ell = 2^-$ branches. In this case, we  can still integrate over the rest of the branch, where the first derivative is finite, and leave the contribution coming from the solutions around the turning point region as an undetermined constant. Then we can try to `guess' such a constant using the phase diagrams: in the microcanonical ensemble the phase diagram must be smooth, whereas in the canonical ensemble the turning point must be a cusp. This trick works well as long as we have enough solutions sufficiently close to the turning from both sides. In both $D=10$ and $D=11$, the phase diagrams for the $\ell = 2^-$ families have been completed using this trick.

An alternative way to compute the energy $E$ of the solutions is by performing an asymptotic expansion of the fields near the AdS boundary and apply KK holography \cite{Skenderis:2006kkh} and holographic renormalization  \cite{deHaro:2000hrec}. 
In this paper we only have worked out this alternative way of finding $E$ for $D = 10$ case. The calculation is long and tedious and the details can be found in Appendices \ref{app:KKhol} and \ref{applyHolo}. The final  expression is given by: \beq\label{Eholo10D}
\frac{E\,L}{N^2} = \frac{3}{16}(1+2y_0^2)^2 + \frac{y_0^4}{512}\(\partial_y^4Q_5 - \partial_y^4Q_1\)\Big|_{y = 1}.
\enq
Note that for $Q_4 = Q_5=1$, this expression reduces to the energy  in \eqref{EEE} for $(p,q) = (5,5)$, after using (\ref{G10G5N}) and identifying the well-known dimensionless vacuum energy of global AdS$_5$ \cite{Balasubramanian:1999re}. The drawback of using \eqref{Eholo10D} to compute $E$ is that it implies computing four derivatives of the numerical solutions,  which inevitably leads to a loss of accuracy.  

%~~~~~~~~~~~~~~~~~~~~~~~~~~~~~~~~~~~~~~~~~~~~~~~
\subsubsection*{(p,q) = (5,5)}
\label{sssec:thermo55}
\begin{figure}[t!]
\begin{center}
\begin{minipage}{\textwidth}
\begin{center}
\input{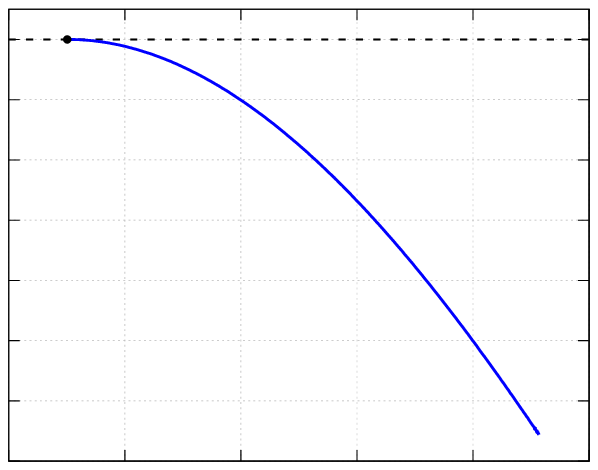}
\input{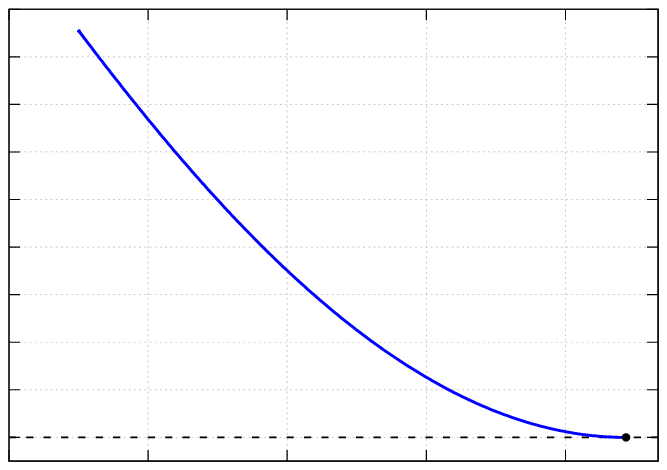}
\end{center}
\end{minipage}
\captionsetup{width=0.9\textwidth}
\captionof{figure}{\textsl{Phase diagram in the microcanonical ensemble (left) and the canonical ensemble (right), for $\ell = 1$ lumpy black holes. The GL threshold point is indicated with a black solid disc. The zero is taken to be the Schwarzschild-AdS$_5\times S^5$ black hole phase. Lumpy black holes are always subdominant in both ensembles. The phase diagrams of the other families of lumpy black holes that we have constructed are qualitatively similar, except for the $\ell = 2^-$ family.}}
\label{fig:10-1canomicro}
\end{center}

\begin{center}
\begin{minipage}{\textwidth}
\begin{center}
\input{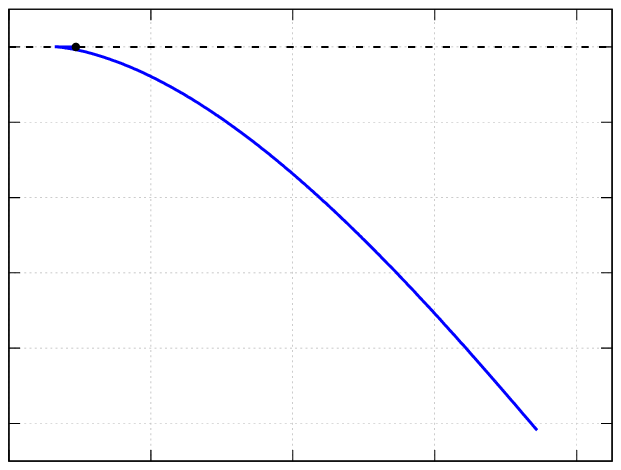}
\input{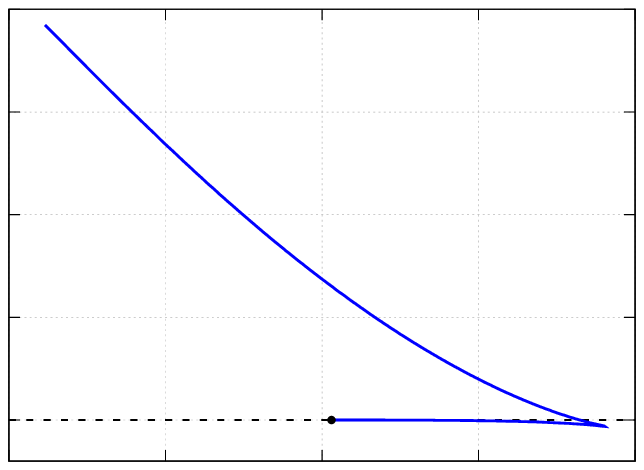}
\end{center}
\end{minipage}
\captionsetup{width=0.9\textwidth}
\captionof{figure}{\textsl{Phase diagram in the microcanonical ensemble (left) and canonical ensemble (right), of $\ell = 2^-$ lumpy black holes. The GL threshold point is indicated with a black solid disc. The zero is taken to be the Schwarzschild-AdS$_5\times S^5$ black hole phase. The $\ell = 2^-$ lumpy black holes dominate the microcanonical ensemble near the GL threshold. In the canonical ensemble, the $\ell = 2^-$ lumpy black holes are also dominant with respect to the small Schwarzschild-AdS$_5\times S^5$ black hole for temperatures near the GL point, but thermal AdS$_5\times S^5$ is always the preferred phase in this range of temperatures.}}
\label{fig:10-2canomicro}
\end{center} 
\end{figure}

%~~~~~~~~~~~~~~~~~~~~~~~~~~~~~~~~~~~~~~~~~~~~~~
In Fig.~\ref{fig:10-1canomicro} we display the phase diagrams for both  the microcanonical and canonical ensembles for the $\ell = 1$ lumpy family. The branch emanates from the GL threshold point, and it moves towards  higher energies or lower temperatures. As this figure shows, the $\ell = 1$ lumpy solutions are always subdominant in both the microcanonical and canonical ensembles. Furthermore, the phase diagrams do not present any turning points for the range of temperatures that we explored.  As we shall see in Section \ref{ssec:crit}, the local cone model for the merger predicts that in $D=10$ there should not be any turning points sufficiently close to the merger point, i.e., in the limit $\lambda\to\infty$. For this particular family it turns out that there are no turning points at all. 

The phase diagram for the $\ell=1$ lumpy AdS$_5\times S^5$ black holes was previously discussed in \cite{Dias:2015pda}. We now compare their results with ours. In the region near the GL threshold (i.e., $\lambda \lesssim1$), our results and theirs are in perfect agreement. However, we find some disagreement for $\lambda \sim \mathcal{O}(1)$. According to Fig.~9 in their paper, the solution with the lowest temperature (i.e.,~the most critical) that they managed to find has  $T\,L= 0.49444$ and energy $EL/N^2 \sim 0.388$; using the data in their figure, we see that $\text{Min}[R_{S^3}] \sim 0.09$ and $\text{Max}[R_{S^3}] \sim 0.56$, which implies $\lambda \sim 2.6$. On the other hand, the most critical solution that we have found for this branch has $\lambda \simeq 25$, which corresponds to a temperature $T\,L= 0.49575$ and energy $EL/N^2 = 0.38540$. Clearly our results and theirs are not compatible. The extent of the $\ell=1$ family in the phase diagram that we find is significantly shorter than previously reported \cite{Dias:2015pda}. The most reasonable explanation for this discrepancy is that the resolution used in \cite{Dias:2015pda} is insufficient to accurately construct the lumpy black holes with large $\lambda$. Indeed, we had to use various tricks (see Section \ref{sssec:boundarylumpy2}) to be able to numerically construct critical lumpy black holes, whilst  \cite{Dias:2015pda} used a single patch with  moderate resolution. According to our estimates, the grid setup used in  \cite{Dias:2015pda} rapidly becomes insufficient to accurately resolve the lumpy black holes with $\lambda\sim\mathcal{O}(1)$. Therefore, it seems reasonable to  conclude that part of the phase diagram reported in \cite{Dias:2015pda} is a numerical artefact and hence unphysical.

\begin{figure}[t!]
\begin{center}
\begin{minipage}{\textwidth}
\begin{center}
\input{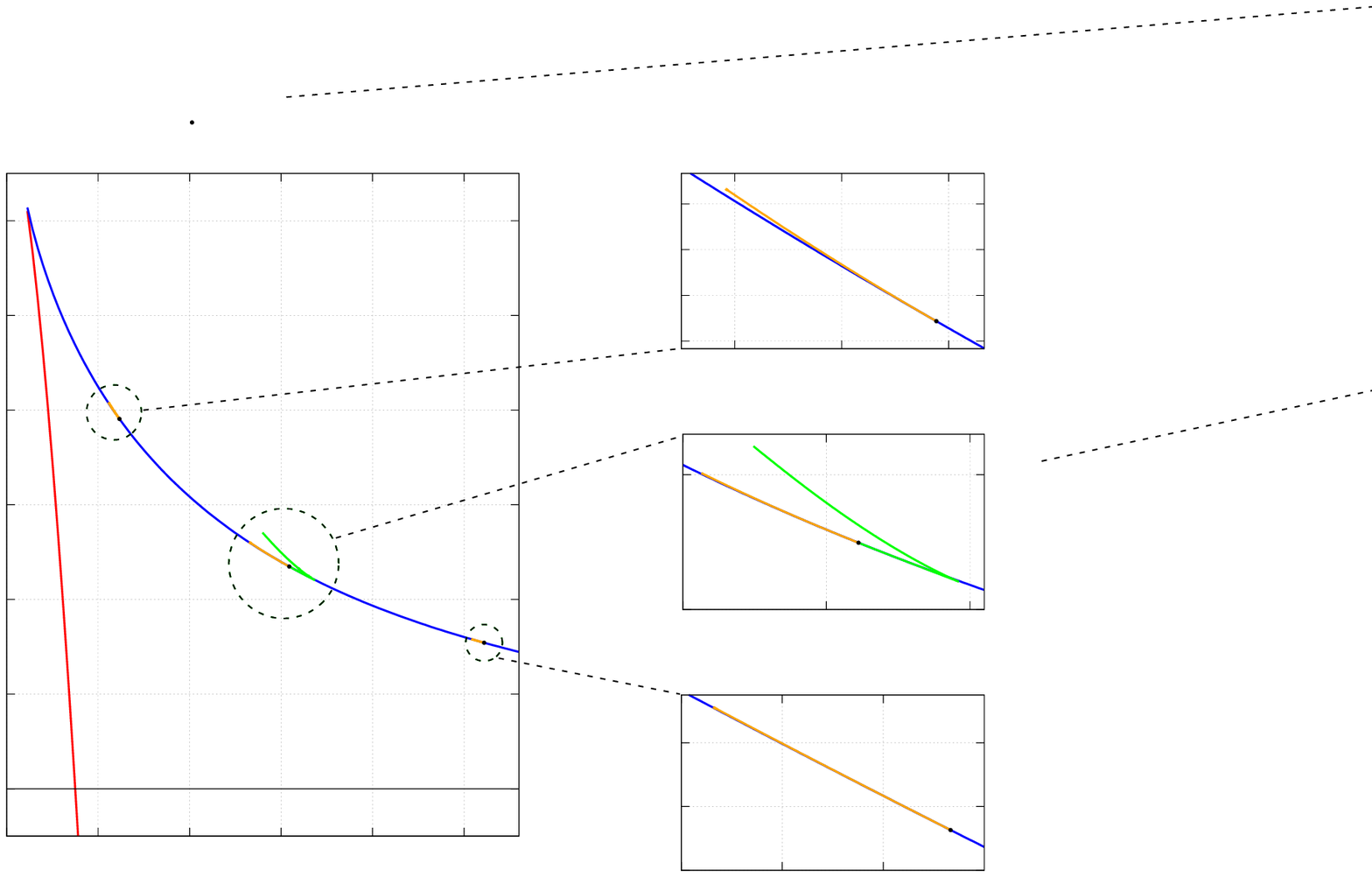}
\end{center}
\end{minipage}
\captionsetup{width=0.9\textwidth}
\captionof{figure}{\textsl{Phase diagram in the canonical ensemble containing all asymptotically AdS$_5\times S^5$ spacetimes that we consider in this paper. The GL threshold point for the various modes is indicated with a solid black disk. The zero of the free energy is taken to be the background thermal AdS$_5\times S^5$, which has free energy $F_\text{AdS} =3N^2/(16L)$. All lumpy black holes are subdominant, even though the $\ell=2^-$ family has lower free energy than the small Schwarzschild-AdS$_5\times S^5$ black hole for some temperatures.}}
\label{fig:10-fullcano}
\end{center}
\end{figure}

The phase diagrams for the branches $\ell = 2^+,3$  are qualitatively the same as those for the $\ell = 1$ family and we do not discuss them any further. We simply stress that these other families of lumpy black holes are also always subdominant in both the microcanonical and canonical ensembles. In order to `locate'  in the phase diagram the various families of lumpy black holes that we have studied, in Fig.~\ref{fig:10-fullcano} we display the phase diagram in the canonical ensemble with all solutions that we have found. In this case we compare the free energy with respect to that of global AdS$_5\times S^5$. As this figure illustrates, lumpy black holes occupy a very small portion of the entire phase diagram. 

Before we move on to discuss the phase diagrams for the $\ell=2^-$ branch, we briefly comment on the previous results by \cite{Dias:2015pda} on the $\ell = 2^+$ and $\ell = 2^-$ branches. In particular, their most critical solution has temperature $T\,L = 0.57278$. We infer from their Fig.~11 (right panel) that this solution has $\lambda \sim 3.1$. On the other hand, the left panel of their  Fig.~11 indicates that their most critical solution for the $\ell = 2^-$ branch has temperature  $T\,L = 0.60849$ and $\lambda \sim 0.5$. In both cases, this agrees with our results and up to these solutions, our phase diagrams and theirs match very well. With our more accurate numerics, we have been able extend both the $\ell=2^+$ and $\ell=2^-$ families to larger values of $\lambda$, closer to the critical point.

It turns out that for the $\ell = 2^-$ branch, the solution with $TL = 0.60849$ is very close to a turning point, and more specifically to a local maximum of the temperature. This solution has  $\lambda \sim 0.54$, and hence it is still very far (in solution space) from the critical regime. We have managed to continue beyond this turning point, and the full phase diagram for the $\ell=2^-$ family is displayed in Fig.~\ref{fig:10-2canomicro}.  Note that whilst the $\ell = 2^+$ branch never dominates any ensemble,  the $\ell = 2^-$ lumpy black holes can be favored with respect to the small Schwarzschild-AdS$_5\times S^5$ black hole.  In the microcanonical ensemble, the $\ell = 2^-$ phase has larger entropy than the Schwarzschild-AdS$_5\times S^5$ black hole (for the same energy) in the region near the GL point. Likewise, in the canonical ensemble the $\ell = 2^-$ lumpy black holes dominate over the small Schwarzschild-AdS$_5\times S^5$ black hole near the GL threshold point and until slightly beyond the turning point. However, in this range of temperatures, thermal AdS is always the preferred phase in the canonical ensemble. It is interesting to note that for this particular branch of lumpy black holes, the microcanonical and the canonical ensembles seem to give slightly different results. This minor observation adds to the puzzles raised in \cite{Yaffe:2017axl} related to the apparent differences between the two ensembles in the thermodynamic limit.
 
%~~~~~~~~~~~~~~~~~~~~~~~~~~~~~~~~~~~~~~~~~~~~~~~
\subsubsection*{(p,q) = (4,7)}
%~~~~~~~~~~~~~~~~~~~~~~~~~~~~~~~~~~~~~~~~~~~~~~

\begin{figure}[t!]
\begin{center}
\begin{minipage}{\textwidth}
\begin{center}
\input{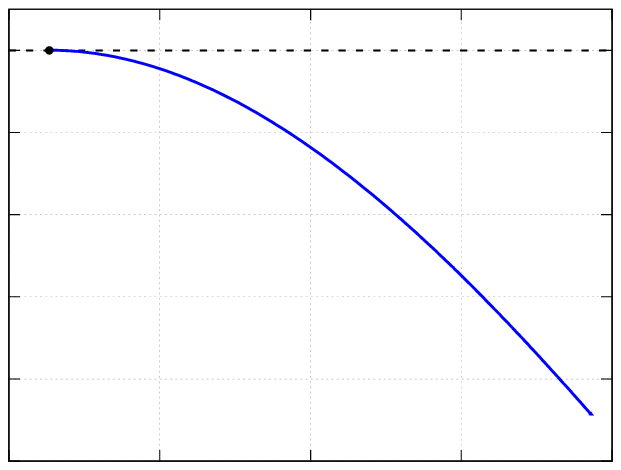}
\input{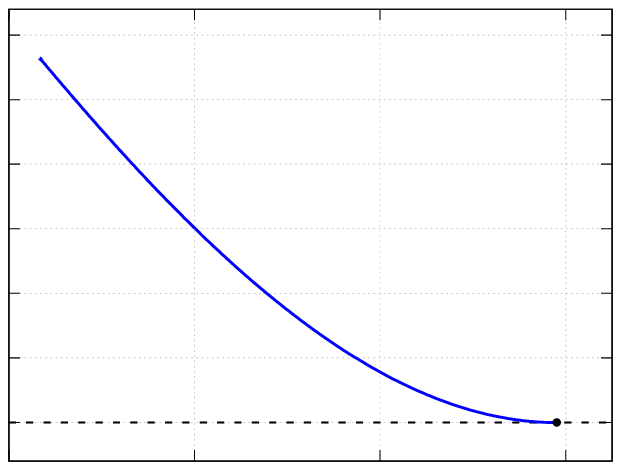}
\end{center}
\end{minipage}
\captionsetup{width=0.9\textwidth}
\captionof{figure}{\textsl{Phase diagram in the microcanonical ensemble (left) and canonical ensemble (right), of $\ell = 1$ lumpy black holes. The GL threshold point is indicated with a black solid disc. The zero is taken to be the Schwarzschild-AdS$_4\times S^7$ black hole phase. The lumpy black holes are subdominant in both ensembles. The phase diagrams of the other families of lumpy black holes that we have constructed are qualitatively similar, except for the $\ell = 2^-$ family.}}
\label{fig:11-1canomicro}
\end{center}

\begin{center}
\begin{minipage}{\textwidth}
\begin{center}
\input{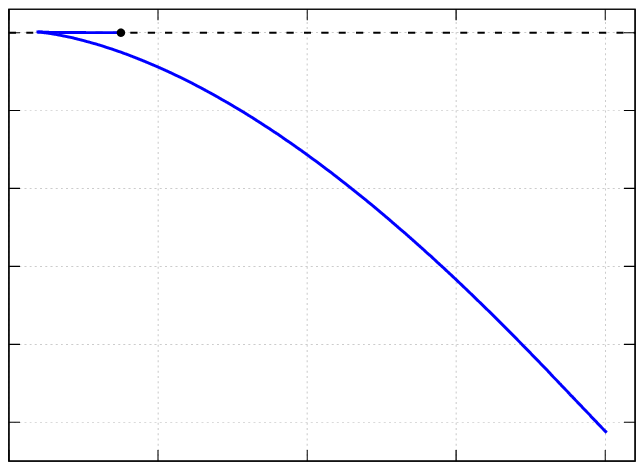}
\input{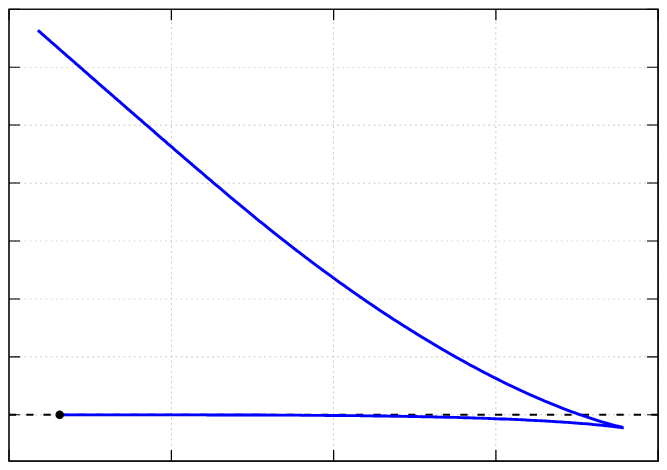}
\end{center}
\end{minipage}
\captionsetup{width=0.9\textwidth}
\captionof{figure}{\textsl{Phase diagram in the microcanonical ensemble (left) and canonical ensemble (right), of $\ell = 2^-$ lumpy black holes. The GL threshold point is indicated with a black solid disk. The zero is taken to be the Schwarzschild-AdS$_4\times S^7$ black hole phase. The $\ell = 2^-$ dominate over the small Schwarzschild-AdS$_4\times S^7$ black hole near the GL threshold point in both ensembles. In the canonical ensemble, thermal AdS$_4\times S^7$ is the dominant phase in this range of temperatures. }}
\label{fig:11-2canomicro}
\end{center}
\end{figure}

\noindent The phase diagrams in the microcanonical and canonical ensembles for $\ell = 1$ lumpy black holes in AdS$_4\times S^7$ are depicted in Fig.~\ref{fig:11-1canomicro}. According to this figure, the $\ell = 1$ lumpy phase exists beyond the corresponding GL threshold point towards larger energies or lower temperatures, depending on the ensemble under consideration. Just as in the $(p,q)=(5,5)$ case, this phase is always subdominant with respect to the small Schwarzschild-AdS$_4\times S^7$ black hole (and of course with respect to thermal AdS$_4\times S^7$, which is the dominant phase in the canonical ensemble in this range of temperatures). It is interesting to note that even though we have managed to follow this branch up until $\lambda=104$, and hence very close to the merger point, we did not find any turning points along the family. In the critical regime (i.e., large values of $\lambda$) the local cone model predicts the absence of turning points and our numerical construction confirms this model beautifully (see Section \ref{ssec:crit} for more details). The phase diagrams for the $\ell = 2^+,\,3$  lumpy AdS$_4\times S^7$ black hole families are qualitatively similar to those of the $\ell = 1$ black holes and we do not  present them here.

On the other hand,  the phase diagrams for the $\ell = 2^-$ family exhibit some notable differences. In Fig.~\ref{fig:11-2canomicro} we display the microcanonical and canonical phase diagrams for the $\ell=2^{-}$ lumpy AdS$_4\times S^7$ black holes. Using the trick explained in the discussion after equation \eqref{1stlaw}, we can find the complete phase diagram beyond the turning point and up to the most critical solution that we have been able to construct, even if we are integrating the 1st law. Fig.~\ref{fig:11-2canomicro}  includes the whole $\ell = 2^-$ family, up to $\lambda=6.00374$ corresponding to $TL = 0.37907$.

\begin{figure}[t!]
\begin{center}
\begin{minipage}{\textwidth}
\begin{center}
\input{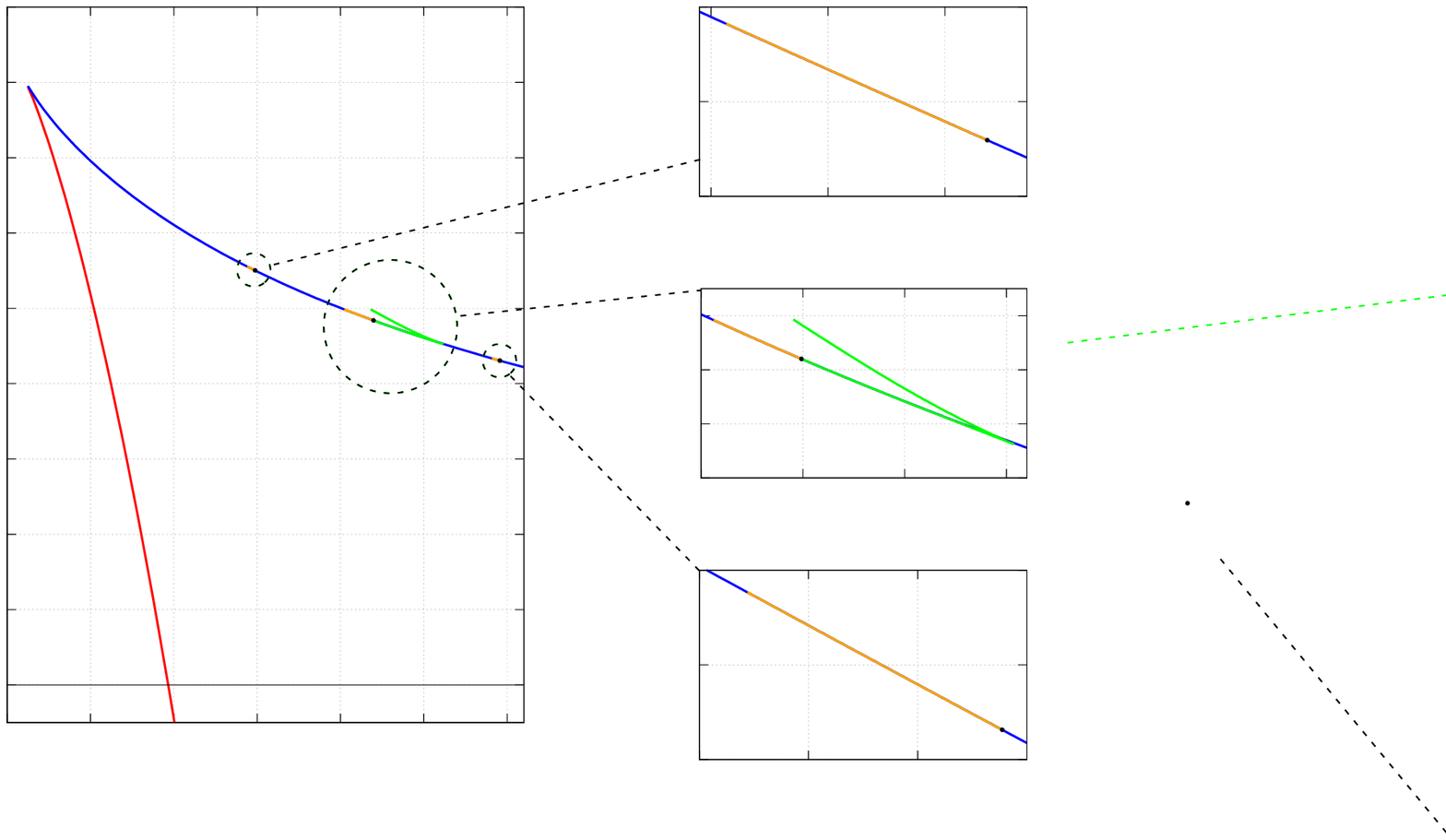}
\end{center}
\end{minipage}
\captionsetup{width=0.9\textwidth}
\captionof{figure}{\textsl{Phase diagram in the canonical ensemble of all lumpy black holes we have found in $D = 11$. The GL threshold points for the various modes are indicated with black solid disks. The zero of the free energy is taken to be the background AdS$_4\times S^7$. As this figure shows, all lumpy black holes are subdominant with respect to thermal AdS$_4\times S^7$ for the range of temperatures that they exist. }}
\label{fig:11-fullcano}
\end{center}
\end{figure}

In the microcanonical ensemble (left panel in Fig. \ref{fig:11-2canomicro}), the $\ell=2^-$ black holes dominate for energies near the GL point. This suggests that this family of black holes could be dynamically stable in some region of parameter space, and a certain (perhaps fine-tuned) class of perturbations of the Schwarzschild-AdS$_4\times S^7$ black holes could evolve into them at the non-linear level. There is a minimum energy along this branch at $E_\text{min}/L^8 = 519.30461$; presumably, at this point the stability properties of these black holes change. The branch continues to higher energies again while the entropy decreases until it reaches $E^*/L^8 = 521.80046$. At this energy, the lumpy black holes and the Schwarzschild-AdS$_4\times S^7$ black holes have the same entropy. Continuing along the $\ell=2^-$ branch, for energies higher than $E^*$, the Schwarzschild-AdS$_4\times S^7$  black hole is the phase with the highest entropy. The last solution we have found has energy $710.39314\times L^8$, corresponding to $\lambda = 6$, but the phase diagram extends way beyond this value.

In the canonical ensemble, right panel in Fig. \ref{fig:11-2canomicro}, the $\ell=2^-$ black holes dominate over the small  Schwarzschild-AdS$_4\times S^7$  near the GL threshold. There is a turning point along this branch at $T_{\text{turn}}L = 0.400824$ ($\lambda_{\text{turn}} = 1.37424$) corresponding to a local maximum of the temperature. The difference of free energies increases at lower temperatures beyond the turning point and it becomes zero at $TL = 0.399124$. At even lower temperatures,  the small Schwarzschild-AdS$_4\times S^7$ dominates over the lumpy black hole phase. However, in the range of temperatures that  the $\ell=2^-$  black holes exist, thermal AdS$_4\times S^7$ is always the dominant phase in the canonical ensemble.

The phase diagram in the canonical ensemble showing all the black hole phases that we have considered in this paper is displayed in Fig. \ref{fig:11-fullcano}. As this figure shows, all families of lumpy black holes that we have studied are subdominant with respect to thermal AdS$_4\times S^7$ in the range of temperatures that they exist. However, the $\ell=2^-$ branch locally dominates  over the small Schwarzschild-AdS$_4\times S^7$ black holes for a small range of temperatures. It is clear from Fig. \ref{fig:11-fullcano} that the families of lumpy black holes occupy a very small portion of the phase diagram. 

%~~~~~~~~~~~~~~~~~~~~~~~~~~~~~~~~~~~~~~~~~~~~~~~
\subsection{Kaluza-Klein holography}
\label{ssec:kkholo}
%~~~~~~~~~~~~~~~~~~~~~~~~~~~~~~~~~~~~~~~~~~~~~~
An important class of observables that characterize the lumpy black holes are the vev's of the dual scalar operators of different conformal dimensions. These essentially parametrize the deformations of the internal space and they can be obtained using KK holography \cite{Skenderis:2006kkh}. In this subsection we will focus on the $(p,q)=(5,5)$, i.e.,~type IIB supergravity in AdS$_5\times S^5$,  since in this case the details of this procedure have been fully worked out in \cite{Skenderis:2006kkh}. We provide the details of our calculation in the Appendices \ref{app:KKhol} and \ref{applyHolo} to correct some typos in the literature and because we find some disagreements with the calculation presented in \cite{Dias:2015pda}. The generalization to the $(p,q)=(4,7)$ is beyond the scope of this paper, even though there are some partial results in the literature \cite{Jang:2016exh,Jang:2018nle}.

\begin{figure}[t!]
\begin{center}
\begin{minipage}{\textwidth}
\begin{center}
\hs{-0.4}\input{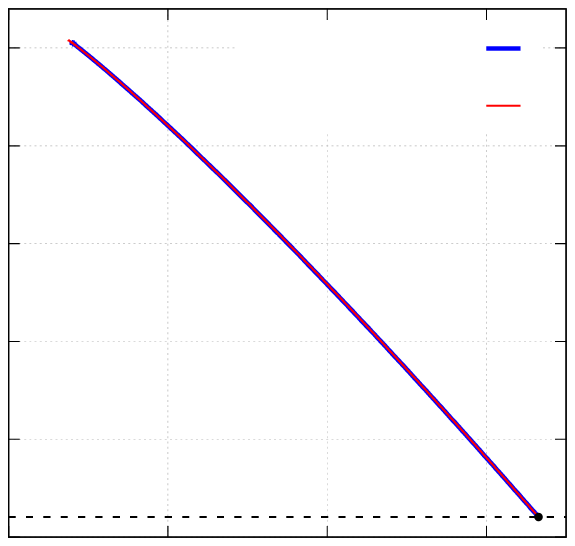} 
\input{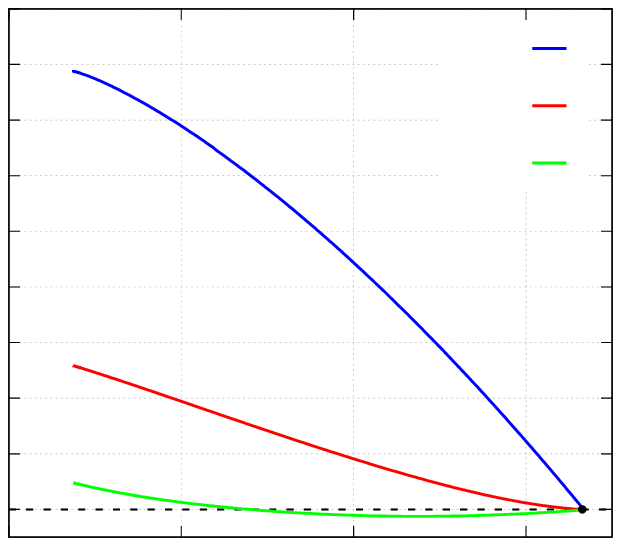}
\end{center}
\end{minipage}
\captionsetup{width=0.9\textwidth}
\captionof{figure}{\textsl{Left:  energy comparison for the $\ell = 1$ lumpy black hole family when computed by integrating the 1st law of thermodynamics (red line), and when computed using holography  \eqref{Eholo10D} (blue line). Right: Dimensionless vev of the dual lower-dimensional scalar field operators $\mc{S}^{2,3,4}$ as a function of the dimensionless temperature.}}
\label{fig:10-holo1}
\end{center}

\begin{center}
\begin{minipage}{\textwidth}
\begin{center}
\input{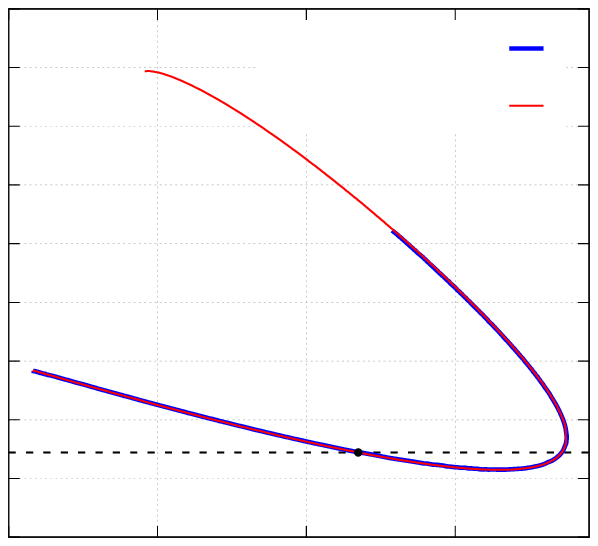}
\input{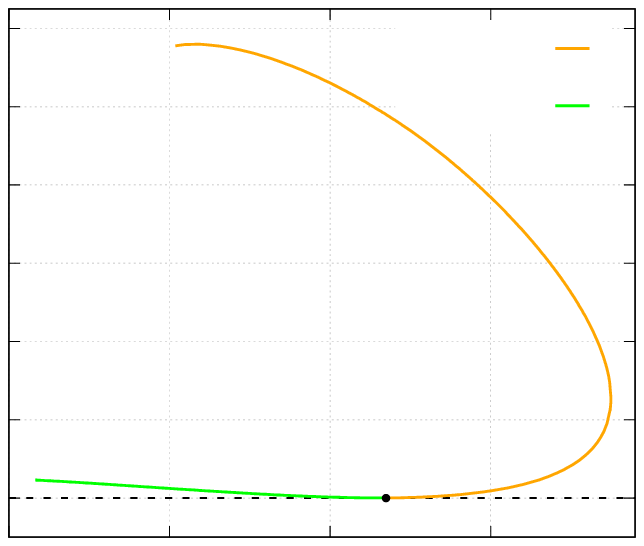}
\end{center}
\end{minipage}
\captionsetup{width=0.9\textwidth}
\captionof{figure}{\textsl{Left: energy comparison for the $\ell = 2$ lumpy black hole family when computed by integrating the 1st law of thermodynamics (red line) and when computed using holography \eqref{Eholo10D} (blue line). Right: Dimensionless vev of the dual lower-dimensional scalar operator $\mc{S}^2$ as a function of the dimensionless temperature.}}
\label{fig:10-holo2}
\end{center}
\end{figure}

The lumpy black holes in AdS$_5\times S^5$ have horizons that are non-uniformly smeared over the $S^5$, breaking the SO$(6)$ symmetry of the latter down to SO$(5)$. Holographically, this corresponds to the spontaneous breaking of the R-symmetry in the dual $\mathcal N=4$ SYM theory and is reflected in the condensation of an infinite tower of scalar operators. The vev's of these operators encode the deformations of the internal $S^5$ and they can be computed, using  KK holography, from the asymptotic expansion of the unknown functions $Q_1,\dots, Q_7$ near the boundary of AdS$_5$ (see Appendix \ref{applyHolo}). We find that the operators with the lowest conformal dimension that develop a non-trivial vev are $\mc{S}^2$, $\mc{S}^3$ and $\mc{S}^4$. From (\ref{scalarVEVsLumpy}), the vev's of these operators are given by: 
\beq
\label{eq:vevs}
\begin{split}
\la \mc{O}_{\mc{S}^2}\ra &= \frac{N^2}{2\pi^2}\frac{\sqrt{2}}{15}y_0^2\beta_2, \hs{0.75} \la \mc{O}_{\mc{S}^3}\ra = \frac{N^2}{2\pi^2}\frac{1}{\sqrt{3}}y_0^3\gamma_3, \hs{0.75} \la \mc{O}_{\mc{S}^4}\ra = \frac{N^2}{2\pi^2}\frac{1}{\sqrt{3}}y_0^4\Big(\frac{19\sqrt{7}}{1500}\beta_2^2-\delta_4\Big),
\end{split}
\enq
where $\beta_2$, $\gamma_3$ and $\delta_4$ are constants that can be obtained from our numerical solutions. On the other hand, the other family of scalar operators of low dimension that could have contributed, namely $\{\mc{T}^0,\mc{T}^1,\mc{T}^2,\mc{T}^3,\mc{T}^4\}$, all have vanishing vev's.\footnote{From the expressions in equation  (A.42) of  \cite{Dias:2015pda}, it would seem that $\langle \mc{O}_{\mc{T}^4}\rangle\neq 0$. We believe that this is incorrect. In fact, we have checked from the near boundary expansions in equation (A.10)  of \cite{Dias:2015pda} that the equations of motion for some of the gauge invariant scalar fields at linear and quadratic order are not satisfied. Therefore, it is possible  that some of the expressions for the vev's given in \cite{Dias:2015pda} are not correct.}

There are various ways to extract the coefficients $\beta_2$, $\gamma_3$ and $\delta_4$ (and $\delta_0$) appearing in \eqref{eq:vevs} from our numerical solutions. One possibility consists of taking derivatives of the $Q$'s at the boundary and then project onto the harmonic basis. For instance, from \eqref{Q1expanded} the constant $\beta_2$ is given by: \beq\label{beta22}
\beta_2 \sim \int_{-1}^{+1} \partial_y^{2}Q_{1,2,4,5,6,7}\big|_{y=1}Y_2(x)\dd x, \hs{0.75} \text{(for $\ell$ odd)},
\enq
where $Y_2(x)$ is the $\ell = 2$ SO$(5)$ scalar harmonic (see Appendix \ref{harm}) and the precise coefficient in front of the integral depends on the function $Q$ that one uses to do the extraction. Another possibility to find the undetermined constants is to fit our numerical data according to the asymptotic expansion (\ref{Q1expanded}). This latter option has been shown to be more accurate and stable towards the end of the branches, specially for the coefficients that appear at 3rd and 4th order. The reason is that too close to the merger, the accuracy of the numerical solution deteriorates and we could not reliably compute the numerical derivatives needed to calculate the coefficients. In practice, we compute $\beta_2$ using both: (\ref{beta22}) for all $Q_{i}$'s, $i=1,2,4,5,6,7$ and performing data fits to check the consistency of the results and estimate the error in this quantity. The other coefficients, $\gamma_3$ and $\delta_4$ (and $\delta_0$), have generically been found fitting our numerical data.

The temperature dependence of $\la\mc{O}_i\ra \equiv \la\mc{O}_{\mc{S}^i}\ra$, $i=2,3,4$, along the $\ell=1$ lumpy black hole family is shown in Fig.~\ref{fig:10-holo1} (right). The vev's vanishes at the GL threshold point, which lies on the Schwarzschild-AdS$_5\times S^5$ solution, and attains a finite value at the merger. The behavior of $\la\mc{O}_1\ra$ is consistent with the plot shown in Ref.~\cite{Dias:2015pda}, but $\la\mc{O}_3\ra$ differs in sign and value. We find that the behavior of the vev's along the $\ell = 3$ branch is qualitatively the same as that for the $\ell = 1$ family. As we show in Section \ref{ssec:crit}, the approach of all three vev's to the merger is controlled by the local (singular) cone geometry in the bulk; in particular, it follows the same power law (with a logarithmic correction) and the same critical exponents as the other physical observables.

For the families with even $\ell$ we have only been able to accurately  compute the vev $\la\mc{O}_2\ra$ and the result is shown in Fig.~\ref{fig:10-holo2} (right). Again, it vanishes at the GL threshold point and becomes positive along the extend of our branch. Unfortunately, in this case we could not find enough critical solutions to recover the critical exponents predicted by the local cone model.

We have also included an energy comparison when it is computed via integration of the 1st law and using the holographic expression (\ref{Eholo10D}). In Fig.~\ref{fig:10-holo1} (left) and \ref{fig:10-holo2} (left) we present the energy as a function of the temperature for $\ell = 1, 2^+$ and $2^-$ lumpy black holes. For the $\ell = 1$ and $2^+$ families we check the consistency of the energy values almost for the whole branch. For the $\ell = 2^-$ family we could only do it partially, but far enough to cover the turning point, which is the main interest in this case. Clearly, the turning point is well-resolved by the energy computed using the 1st law.

%~~~~~~~~~~~~~~~~~~~~~~~~~~~~~~~~~~~~~~~~~~~~~~~
\subsection{Geometry}
\label{ssec:geo}
%~~~~~~~~~~~~~~~~~~~~~~~~~~~~~~~~~~~~~~~~~~~~~~
In this subsection we study in detail the geometry of the horizon of the lumpy AdS$_p\times S^q$ black holes. We construct embedding diagrams of the horizon geometry into Euclidean space to visualize how the shapes of the various spheres change as one moves along a given family of solutions. This allows us to get a better understanding of the possible topology changes at the critical points, where the lumpy black holes merge with new families of black holes. 

The induced metric on the spatial cross-sections of the horizon of the lumpy AdS$_p\times S^q$ black holes is:
 \beq\label{indlump}
\dd s^2\Big|_H = L^2y_0^2Q_5(0,a)\dd\Omega_{(p-2)}^2  + R^2\(\frac{4Q_4(0,a)}{2-a^2}\dd a^2 + (1-a^2)^2Q_6(0,a)\dd\Omega_{(q-1)}^2\).
\enq
At the horizon, the radii of the $S^{p-2}$ within the Anti-de Sitter factor and the radii of the $S^{q-1}$ within the internal $S^q$ are given by: 
\beq
R_{p-2}(a) = Ly_0\sqrt{Q_5(0,a)}, \hs{0.75} R_{q-1}(a) = R(1-a^2)\sqrt{Q_6(0,a)},
\label{eq:radii}
\enq
respectively. These are gauge invariant quantities within the symmetry class of spacetimes that we consider. The behavior of these two quantities  (and in particular their zeroes) as functions of the polar angle $a$ dictate the topology of the horizon. Indeed, the geometry \eqref{indlump} can be understood as a fibration of an $S^{p-2}$ and an $S^{q-1}$ over an interval parametrized by the coordinate $a$ with $a\in[-1,1]$. For all lumpy black holes, we see from \eqref{eq:radii} that $R_{q-1}(a)$ smoothly shrinks to zero size at both ends of the interval  whilst $R_{p-2}$ is always finite there. This results in the familiar horizon topology $S^{p-2}\times S^q$, as the Schwarzschild-AdS$_p\times S^q$ black hole.  In the latter case,  both radii  $R_{p-2}(a)$ and $R_{q-1}(a)$ are constant everywhere. As we shall see momentarily, for the lumpy black holes  these radii  develop a non-trivial profile as we move along a given family, away from the GL threshold. In particular, $R_{p-2}(a)$ approaches zero at a point (or several points, depending on the harmonic $\ell$ labelling the family) on the horizon, becoming precisely zero at the merger. At the zero(s) of $R_{p-2}(a)$, the horizon becomes singular and the branch of lumpy black holes presumably merges with a new branch of topologically distinct black holes in a conifold-type-of transition. In analogy with the non-uniform black strings \cite{Gubser:2001ac}, we may define the `lumpiness' parameter in terms of $R_{p-2}(a)$ as a measure of the (inverse) `distance' to the merger: 
\beq
\label{lumpiness}
\lambda = \frac{1}{2}\(\frac{\text{max}[R_{p-2}(a)]}{\text{min}[R_{p-2}(a)]}-1\).
\enq

For solutions close to the threshold of the GL $\ell$th zero mode, $R_{p-2}(a)$ is approximately constant on the horizon and consequently $\lambda \sim 0$, with $\lambda=0$ precisely at the GL threshold point. Conversely, as we move along the family of lumpy black holes and towards the merger point, $\text{max}[R_{p-2}(a)]$  remains finite whilst $\text{min}[R_{p-2}(a)]\to 0$  and hence $\lambda\rightarrow \infty$, just as in the standard KK setting. The vanishing of $R_{p-2}$ at a certain point $a=a_\ast$ at the merger happens in a singular way. We should expect that at the other side of the transition, the vanishing of $R_{p-2}$ occurs in a smooth manner, as it should if there exists a smooth family of black holes with a different horizon topology. We point out that even though the radius  $R_{q-1}(a)$ of the horizon $S^{q-1}$ vanishes smoothly at the endpoints of the interval parametrized by $a$ for all lumpy black holes, for some families (but not all) the horizon $S^{q-1}$ can also become singular at the merger. However, for all cases that we have considered, the horizon $S^{p-2}$ shrinks to zero size in a singular manner somewhere on the horizon precisely at the merger, and hence the lumpiness parameter defined in \eqref{lumpiness} is a good measure of the degree of deformation of the horizon geometry.

\begin{figure}[t]
\begin{center}
\begin{minipage}{\textwidth}
\begin{center}
\input{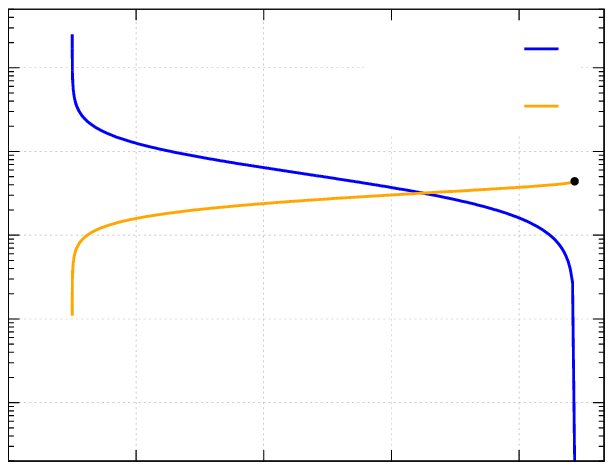}
\input{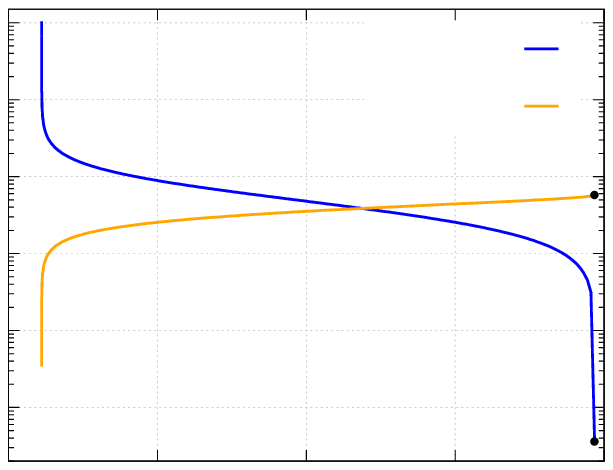}
\end{center}
\end{minipage}
\captionsetup{width=0.9\textwidth}
\captionof{figure}{\textsl{Lumpiness parameter $\lambda$ and minimum radius of the $S^{p-2}$, $\text{min}[R_{p-2}(a)]$, for the $\ell=1$ family of lumpy black holes as a function of the dimensionless temperature for the $(p,q) = (5,5)$ (left) and $(p,q) = (4,7)$ (right) cases. These plots show that $\lambda$ diverges as we approach the merger point. Lumpy black hole families labelled by other harmonics $\ell$ exhibit a qualitatively similar behavior.}}
\label{fig:lambda5}
\end{center}

\begin{center}
\begin{tabular}{|r|r|r|r|r|}
\cline{2-5}
\multicolumn{1}{r|}{} & $\ell$ & $1$ & $2^+$ & $3$ \\ \hline\hline
\multirow{4}{*}{$L_{\text{hor}}/L$} 
& \multirow{2}{*}{$D = 10$} & \multirow{2}{*}{$3.26912$} & \multirow{2}{*}{$-$} & \multirow{2}{*}{$3.21607$} \\
& & & & \\ \cline{2-5}
& \multirow{2}{*}{$D = 11$} & \multirow{2}{*}{$6.34959$} & \multirow{2}{*}{$6.39356$} & \multirow{2}{*}{$6.33214$} \\
& & & & \\ \hline
\end{tabular}
\captionsetup{width=0.9\textwidth}
\captionof{table}{\textsl{Dimensionless horizon's length values of the most critical solutions that we have found for the families $\ell = 1,3$ in $D = 10$ and $\ell = 1,2^+,3$ in $D = 11$. We did not construct critical enough solutions of the $\ell=2^+$ family in $D=10$.}}
\label{Lhor}
\end{center}
\end{figure}

In Fig.~\ref{fig:lambda5} we plot $\lambda$ and the minimum of $R_{p-2}$ on the horizon, as functions of the temperature. This figure shows that $\lambda$ defined in \eqref{lumpiness} behaves in the same way as the analogous quantity for the non-uniform black strings: $\lambda$ increases monotonically along the family and diverges at the merger point. The behavior of $\lambda$ for the other families of lumpy black holes labelled by different harmonics $\ell$ is qualitatively similar  and we will not display them here. Note that for the $\ell = 2^-$ family, $\lambda$ (and $\text{min}[R_{p-2}(a)]$) are not uni-valued functions of the temperature because there is a turning  point at some temperature. However, the corresponding $\lambda$ still  increases monotonically along the family and it diverges in a qualitatively similar manner as in the other families as we approach the merger point.

One may also define the horizon length along the $S^q$: 
\beq
L_{\text{hor}} = 2R\int_{-1}^{+1}\dd a\sqrt{\frac{Q_4(0,a)}{2-a^2}}.
\enq
Again, for $\ell = 2$ solutions the integral is from 0 to 1 and there is an extra factor of 2. The Schwarzschild-AdS$_p\times S^q$ black hole has $L_{\text{hor}} = \pi R$, which is nothing but the geodesic distance between the poles of the $S^q$ of radius $R$. The behavior of this physical quantity along the family of lumpy black holes is quite generic. For a given $\ell$, the curve starts at the GL point where  $L_\text{hor}/L=\pi$ or $L_\text{hor}/L = 2\pi$ in $D = 10$ or $11$ respectively, and it is a monotonically increasing function until it attains a finite value at the merger point. In Table~\ref{Lhor} we list the different values at the merger for the families for which we have constructed critical enough solutions. As this table shows, the value of $L_\text{hor}/L$ at the merger point is unique to each family. Critical localized-type black holes that merge with the various family of lumpy solutions studied here should give the same values of $L_\text{hor}/L$.

Ref.\ \cite{Dias:2015pda} considered the possible topologies of the localized black hole solutions that would merge with a given family $\ell$ of lumpy black holes in $D = 10$. This reference based their analysis on the behavior of the induced Ricci scalar at the poles of the $S^q$, and concluded that the divergence of the Ricci scalar there indicated localization. In this paper, we follow a different route and instead we consider embedding diagrams into Euclidean space as they are more intuitive and the topology changes become apparent. 

One may consider two possible embeddings: at fixed coordinates of the $S^{p-2}$ or fixed coordinates of the $S^{q-1}$. For fixed $S^{q-1}$  coordinates, the horizon geometry may be embedded  into $\E^{p}$, whereas for fixed $S^{p-2}$ coordinates the horizon geometry may be visualized as an embedded surface in $\E^{q+1}$. Generically, the metric of $\E^{n}$ reads 
\beq
\dd s^2_{\E^{n}} = \dd X^2 + \dd Y^2 + Y^2\dd\Omega_{(n-2)}^2.
\label{eq:En}
\enq
The embedded surface has the form: $X = X(a)$ and $Y = R_{p-2}(a)$ or $Y=R_{q-1}(a)$ for fixed $S^{q-1}$ or $S^{p-2}$ coordinates respectively. Then, matching \eqref{eq:En}  with \eqref{indlump}, we obtain $X(a)$: \beq\begin{split}
\text{Fixed $S^{q-1}$: } &\hs{0.25} X(a) = R\int_{-1}^{+1}\dd a\sqrt{\frac{4Q_4(0,a)}{2 - a^2} - y_0^2\frac{(L/R)^2}{4Q_5(0,a)}\bigg(\deriv[Q_5(0,a)]{a}\bigg)^2}, \\
\text{Fixed $S^{p-2}$: } &\hs{0.25} X(a) = R\int_{-1}^{+1}\dd a\sqrt{\frac{4Q_4(0,a)}{2 - a^2} - 4Q_6(0,a)\(a -\frac{1-a^2}{4}\deriv[\ln Q_6(0,a)]{a}\)^2}.
\end{split}\enq
Recall that for $\ell = 2$ lumpy black holes the range of the integral is from 0 to 1, and to represent the full embedding we just add the mirror image with respect to the equator  ($a=0$). In our plots we present $X/L$ versus $Y/L$ for the three families $\ell = 1,2^+$ and $2^-$. For each of these families, we show together the embedding plots for fixed $S^{p-2}$ and fixed $S^{q-1}$ coordinates respectively, to make the topology of the horizon apparent. The qualitative behavior of the $\ell=3$ family is essentially the same as that of $\ell = 1$; we comment on it below but we do not display any embedding plots for this case.

\begin{figure}[!t]
\begin{center}
\begin{center}
{\bf \textsf{$\ell$ = 1}}
\end{center}
\begin{minipage}{\textwidth}
\begin{center}
\hs{-1.25}\input{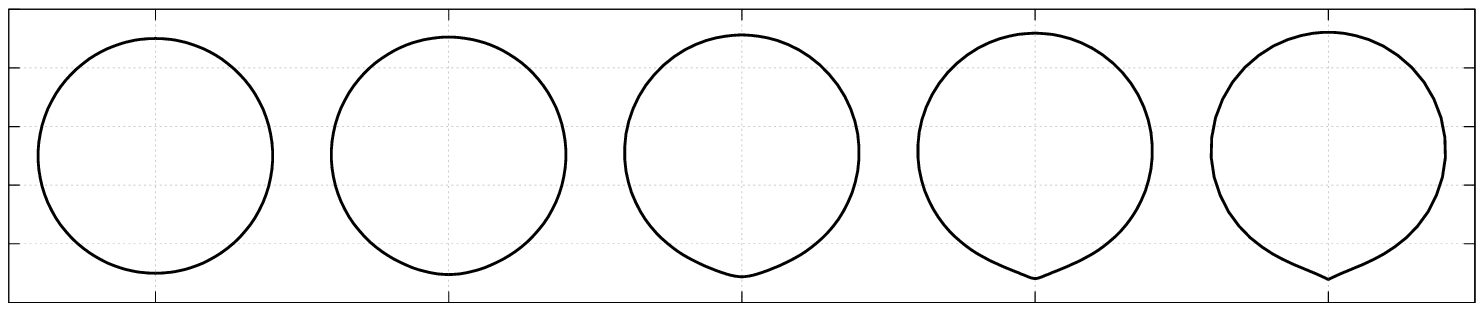}

\hs{-1}\input{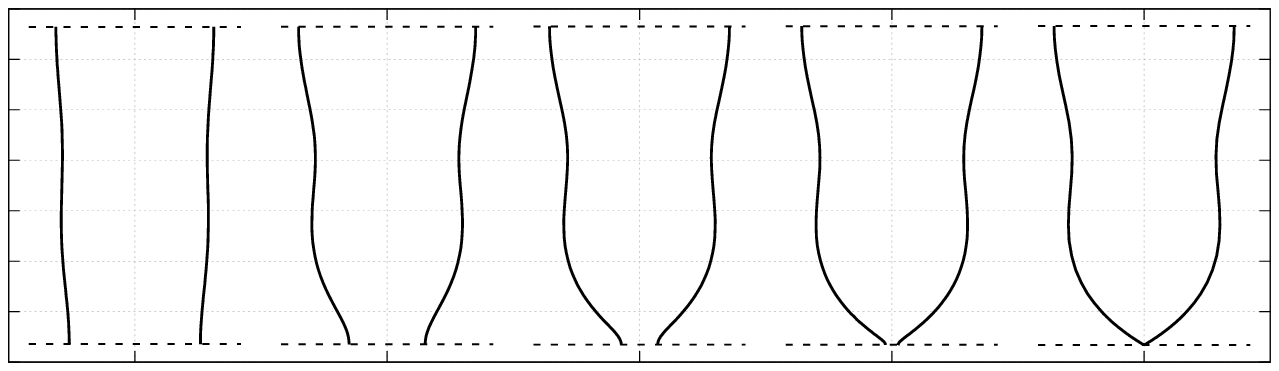}
\end{center}
\end{minipage}
\captionsetup{width=0.9\textwidth}
\captionof{figure}{\textsl{Embedding plots for fixed $S^2/S^6$ coordinates of lumpy black holes in AdS$_4 \times S^7$ with $\ell = 1$. This family merges with a localized black hole with $S^9$ horizon topology. The corresponding plots for $\ell = 3$ are similar to those for $\ell = 1$, but the pinch-off appears at the opposite pole.}}
\label{fig:1AdS4}
\end{center}
\end{figure}

\begin{figure}[!t]
\begin{center}
\begin{center}
{\bf \textsf{$\ell$ = 2$^\text{+}$}}
\end{center}
\begin{minipage}{\textwidth}
\begin{center}
\hs{-1.25}\input{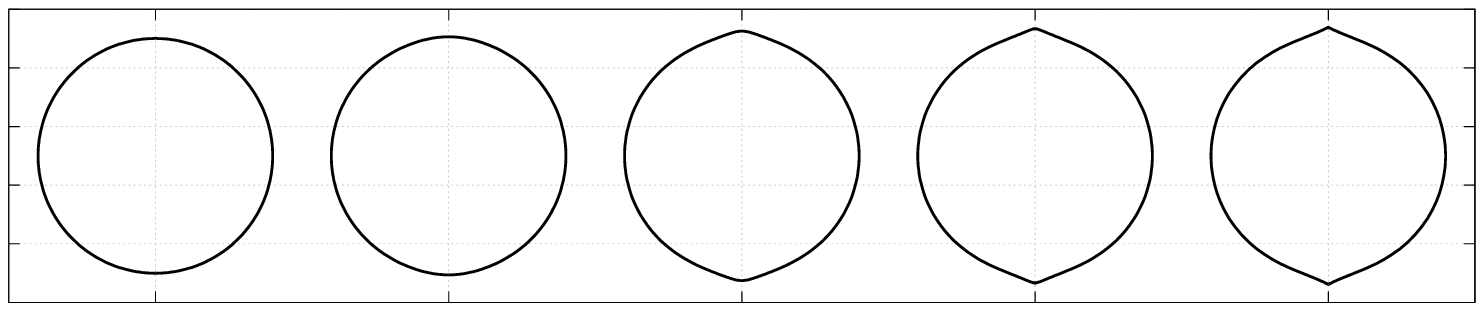}

\hs{-1}\input{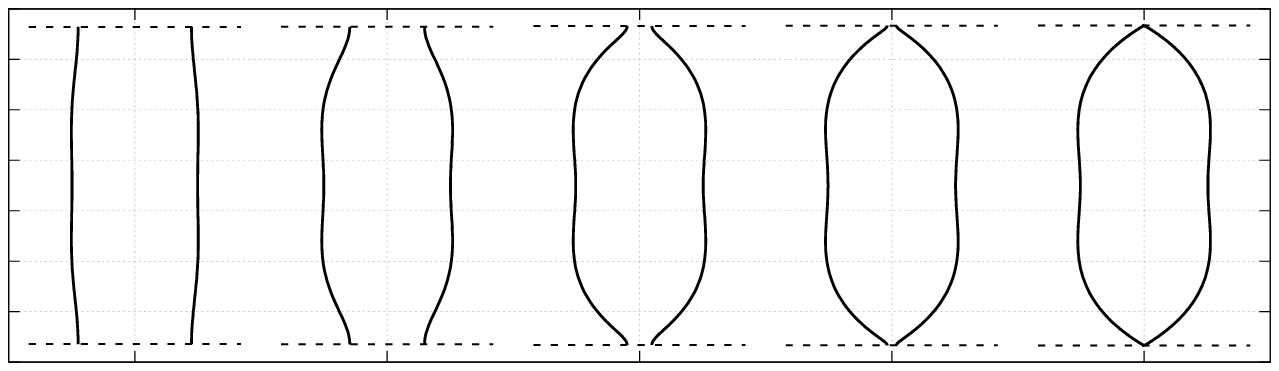}
\end{center}
\end{minipage}
\captionsetup{width=0.9\textwidth}
\captionof{figure}{\textsl{Embedding plots for fixed $S^2/S^6$ coordinates of lumpy black holes in AdS$_4 \times S^7$ with $\ell = 2^+$. This family merges with a black belt with horizon topology $S^3\times S^6$.}}
\label{fig:2pAdS4}
\end{center}
\end{figure}

\begin{figure}[!t]
\begin{center}
\begin{center}
{\bf \textsf{$\ell$ = 2$^\mathbf{-}$}}
\end{center}
\begin{minipage}{\textwidth}
\begin{center}
\hs{-1.25}\input{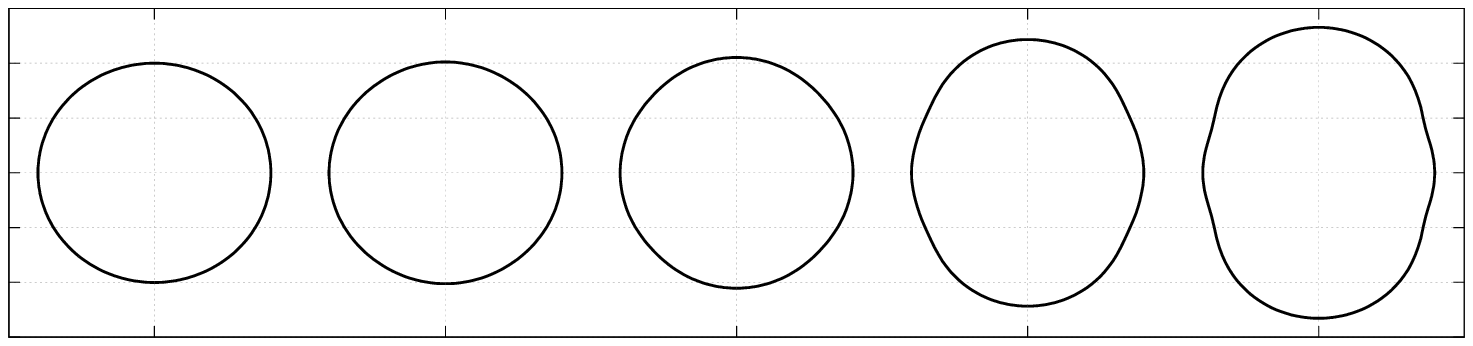}

\hs{-1}\input{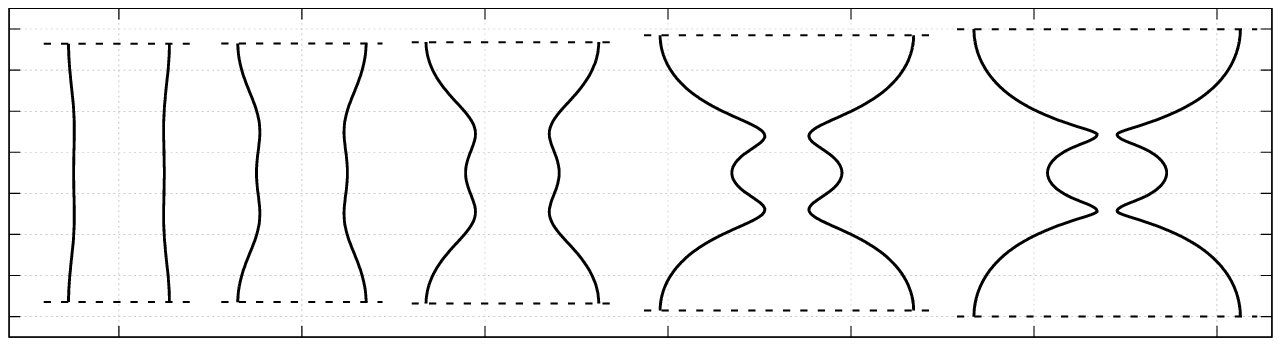}
\end{center}
\end{minipage}
\captionsetup{width=0.9\textwidth}
\captionof{figure}{\textsl{Embedding plots for fixed $S^2/S^6$ coordinates of lumpy black holes in AdS$_4 \times S^7$ with $\ell = 2^-$. This family merges with a black hole whose horizon topology is the connected sum $S^9\#(S^3\times S^6)\# S^9$.}}
\label{fig:2mAdS4}
\end{center}
\end{figure}

In Figs.~\ref{fig:1AdS4},  \ref{fig:2pAdS4} and \ref{fig:2mAdS4} we show the embeddings for different families of the AdS$_4\times S^7$ lumpy black holes. The analogous embeddings for the AdS$_5\times S^5$ lumpy black holes are qualitatively similar and we display them in Appendix \ref{appEmb10d}. The only exception is the $\ell=2^-$ family, which we comment on below. 

In Fig.~\ref{fig:1AdS4} we show representative embeddings for the $\ell=1$ AdS$_4\times S^7$ lumpy black holes for fixed $S^2$ coordinates (top) and fixed $S^6$ coordinates (bottom). This figure shows that whilst the $S^6$ smoothly shrinks to zero size at both poles of the $S^7$, as $\lambda$ increases it develops a conical singularity at the south pole. Similarly, the horizon $S^2$ has finite size everywhere on the $S^7$ for all lumpy black holes, but as $\lambda$ increases it also develops a conical singularity at the south pole, suggesting that it pinches-off there. At the other side of the transition, both of these singularities will be resolved, resulting on a $S^6$ that smoothly shrinks to zero size at the north pole of the $S^7$, whilst it has finite size on the south pole; conversely, the $S^2$ will continue to have finite size at the north pole but it shrinks smoothly at the south pole. The induced geometry at the other side of the merger can hence be modeled by the line element, 
\beq
\sim f_1(a)\dd a^2 + \cos^2 a f_2(a)\dd\Omega_{(p-2)}^2 + \sin^2a f_3(a)\dd\Omega_{(q-1)}^2,
\enq
with the poles located at $a =0$ and $a = \pi/2$, and non-trivial smooth and finite functions $f_i(a)$, $i = 1,2,3$, such that $f_3(0)=f_1(0)$ and $f_2(\pi/2)=f_1(\pi/2)$. Indeed, this geometry has the topology of $S^{D-2}$ and hence it corresponds to a single localized black hole, as expected.

In Fig.~\ref{fig:2pAdS4} we consider the embeddings for different values of $\lambda$ for the $\ell=2^+$ family of AdS$_4\times S^7$ lumpy black holes. From the top panel in this figure, (i.e.,~fixed $S^2$ coordinates,) we see that the $S^6$ will develop conical singularities at the merger point at both the north and the south pole of the $S^7$. Likewise, the bottom panel (i.e.,~fixed $S^6$ coordinates) shows that the $S^2$ will also develop conical singularities at both poles for $\lambda\to\infty$. Then, we expect that the resolved geometry at the other side of the merger will have an $S^6$ that is finite in the whole $a$-interval, whilst the $S^2$ will smoothly shrink to zero size at both endpoints, giving an $S^3$. Therefore, we conclude that the topology of the horizon of the family of black holes sitting at the other side of the merger is $S^3\times S^6$, corresponding to a black belt.  

The sequence of embeddings for the $\ell=2^-$ family are shown in Fig.~\ref{fig:2mAdS4}. For fixed $S^2$ coordinates (top panel), we see that see that the $S^6$ becomes more deformed, but it continues to shrink to zero size smoothly at the endpoints of the $a$-interval, even in the $\lambda\to\infty$ limit. On the other hand, for fixed $S^6$ coordinates (bottom panel), we see that for $\lambda\to\infty$ the $S^2$ will pinch off at two different points, $a_\ast^{N}$ and $a_\ast^{S}$, near the north and the south poles of the $S^7$ respectively. Therefore, in the resolved geometry at the other side of the merger, we expect that the $a$-interval will be divided in three subintervals, $-1< a^S_\ast < a^N_\ast<1$, such that: For $-1\leq a \leq a^S_\ast$, the $S^6$ shrinks to zero size smoothly at $a=-1$, whilst it remains finite at $a=a_\ast^S$; on the other hand, the $S^2$ is finite at $a=-1$ and smoothly shrinks to zero at  $a=a_\ast^S$. Therefore, the horizon topology in this region is $S^9$. Likewise, a similar reasoning shows that the horizon topology in the $a_\ast^N\leq a \leq 1$ interval is also that of an $S^9$. On the other hand, in the middle interval, $a_\ast^S\leq a \leq a_\ast^N$, we have that the $S^6$ is finite everywhere, whilst the $S^2$ smoothly shrinks at the endpoints, giving a horizon topology of $S^3\times S^6$. Putting everything together, we conclude that the $\ell=2^-$ family of lumpy black holes merges with a family of black holes whose horizon topology is a connected sum $S^9\#(S^3\times S^6)\# S^9$, that is, two localized black holes with a black belt joining them.  Interestingly, the $\ell=2^-$ family in the AdS$_5\times S^5$ seems to merge with a different family of black holes. Indeed, as \cite{Dias:2015pda} pointed out and our embedding diagrams confirm (see Fig.~\ref{fig:12AdS5}), in this case the lumpy black holes merge with a double localized black hole, with horizon topology  $S^8\# S^8$. It would seem that it should also be possible to connect the two localized black holes with a black belt, giving a topology $S^8\# (S^4\times S^4)\# S^8$. However, if such a family of  black holes exists, it does not merge with the $\ell=2^-$ lumpy AdS$_5\times S^5$ black holes.

The $\ell = 3$ solutions do not differ much with respect to the $\ell = 1$ ones in terms of the topology change across the transition. Although the profile of the embeddings presents more oscillations (since it belongs to a higher harmonic family), our data seems to indicate that the branch merges with another localized black hole with $S^9$ horizon topology, but potentially having lumpiness along its horizon and hence distinct from the localized black hole family that merges with the $\ell = 1$ lumpy branch. Presumably this new family of deformed localized black holes would merge with new families of multiple localized black holes and black belts. This is quite analogous to the situation in higher dimensional asymptotically flat singly spinning black holes, where different families of topologically spherical bumpy black holes, corresponding to different harmonics,  merge with different families of black rings and bumpy black rings \cite{Figueras:bumpy}.

Summarizing, we have the following topology-change phase transitions for the different families of lumpy black holes considered in this paper: 
\beq\begin{split}
\ell = 1&: \hs{0.75} S^{p-2}\times S^{q} \hs{0.5} \riga \hs{0.5} S^{p+q-2}\,, \\
\ell = 2^+&: \hs{0.75} S^{p-2}\times S^{q} \hs{0.5} \riga \hs{0.5} S^{p-1}\times S^{q-1}\,, \\
\ell = 2^-&: \hs{0.75} S^{p-2}\times S^{q} \hs{0.5} \riga \hs{0.5} S^{p+q-2}\#(S^{p-1}\times S^{q-1})\#S^{p+q-2} \hs{0.5}  \text{or} \hs{0.5}  S^{p+q-2}\#S^{p+q-2}\,,\\
\ell = 3&: \hs{0.75} S^{p-2}\times S^{q} \hs{0.5} \riga \hs{0.5} S^{p+q-2}\,.
\end{split}\enq
Given the richness of the various topology changes that we have uncovered, it is not inconceivable that other topology changes are possible  for  values of $(p,q)$ not considered in this paper. 

We see that except for the $\ell=2^-$ family, the topology-changing transitions depend in a trivial way on the number of spacetime dimensions. The details of the underlying theory in each of the cases that we have considered, either type IIB supergravity or 11$D$ supergravity, do seem not play a significant role. However, the particular way in which physical quantities approach the critical point does depend on the spacetime dimension, and is predicted by the local cone model governing the change of topology. 

%~~~~~~~~~~~~~~~~~~~~~~~~~~~~~~~~~~~~~~~~~~~~~~~
\subsection{Critical behavior}
\label{ssec:crit}
%~~~~~~~~~~~~~~~~~~~~~~~~~~~~~~~~~~~~~~~~~~~~~~
Topology-changing transitions in the space of equilibrium solutions have been previously studied in higher-dimensional general relativity. The first and most studied system is the transition between non-uniform black strings and localized black holes in standard KK theory. Kol  \cite{Kol:top} proposed that, in this case, the transition  is governed by a Ricci-flat double cone over $S^2\times S^{D-3}$. The $S^{D-3}$ appears explicitly in both the non-uniform black strings and in the localized black holes; the $S^2$ on the other hand comes from the fact that the Euclidean time is fibered over an interval whose endpoints are on the horizon. The $S^{D-3}$ is non-contractible in the black string phase, while it is contractible in the localized black hole phase; the opposite happens with the $S^2$, which is contractible in the black string phase and non-contractible in the black hole phase. There is now detailed numerical evidence that this model is correct in $D = 5,6$ \cite{Kalisch:2016fkm,Kalisch:2017bin} and in $D = 10$ \cite{Cardona:2018shd,Ammon:2018sin} (see \cite{Emparan:top2} for an analytical approach in the  large-$D$ limit of general relativity). Beyond KK theory, Ref.\ \cite{Emparan:top,Figueras:bumpy} considered mergers of singly spinning black hole solutions in asymptotically flat space and described local Lorentzian double cone models for the critical geometries controlling them. These papers considered mergers between rotating black holes, black rings, di-rings and black saturns in $D \geq 6$, providing strong evidence that such double-cone models do indeed describe the topology-changing phase transitions for asymptotically flat stationary black holes as well.

The topology-changing transitions that have been considered so far involve black hole spacetimes in pure Einstein gravity, and hence it is natural to consider local Ricci-flat cones to model the local singularity at the merger point. However, in this paper we consider Einstein gravity in asymptotically AdS$_p\times S^q$ coupled  to a certain form field strength so it is not obvious that  local Ricci-flat cones should still control the topology-change transitions in this case. In this section we will show that even though the topology change is still controlled by a conical geometry, the latter is no longer Ricci flat due to the presence of fluxes in the internal $S^q$. 

We will focus on the local merger between the $\ell=1$ lumpy black holes with horizon topology $S^{p-2}\times S^{q}$ and a localized black hole with horizon topology $S^{p+q-2}$, but the discussion also applies to the other $\ell$'s. Based on previous studies, the double-cone inherits a huge amount of spherical symmetry from the full solution: the $S^{p-2}$ and $S^{q-1}$  that appear explicitly in the metric, plus an additional $S^r$  that comes from the fibration, in the Wick rotated geometry, of the Euclidean time circle over some base manifold that depends on the particular merger. In the canonical non-uniform black string/localized black hole merger, the base is an interval with the two endpoints on the horizon; this, together with the fibered Euclidean time circle, gives $r = 2$;  in the black ring/black hole merger the base is a 2-disk giving $r = 3$. In the case under consideration the base is also an interval. This consists of any path that connects two points on the horizon along the polar direction and thus $r=2$, just as in non-uniform black string/localized black hole transition. Similarly, this $S^2$ is contractible in the lumpy black hole phase, but it is not contractible in the localized phase.

By symmetry, the merger should be described by a cone over a triple direct product of spheres, $S^2\times S^{p-2}\times S^{q-1}$:
\beq
\label{triple}
\dd s^2 = \dd\rho^2 + \frac{\rho^2}{D-2}\(\dd\Omega_{(2)}^2 + (p-3)\dd\Omega_{(p-2)}^2 +\alpha\,\dd\Omega_{(q-1)}^2\).
\enq
With the presence of a non-trivial fluxes, it is not obvious that the cone that controls the merger should be Ricci flat and, as we shall see shortly, our data indicates that it is not. We will leave to future work the first principles understanding of the types of cones that control the topology changing transitions in the presence of fluxes. The numerical factors in \eqref{triple} have been chosen so that the embeddings agree with our numerical solutions in the critical regime. The factors in front of the $S^2$ and the $S^{p-2}$ in \eqref{triple} ensure that the components  of the Ricci tensor of the cone along these directions vanish; on the other hand, our data suggests that \beq\label{alphflux}
\alpha=q-2+\frac{D}{4},
\enq
which implies that the components of the Ricci tensor along the $S^q$ do not vanish. 

\begin{figure}[t!]
\centering
\input{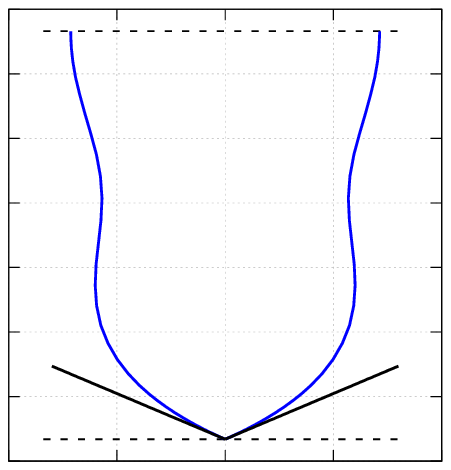}
\input{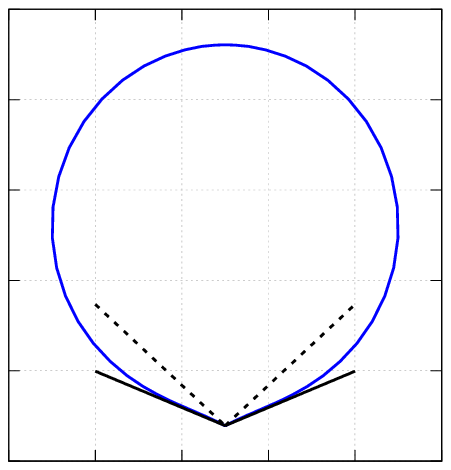}
\input{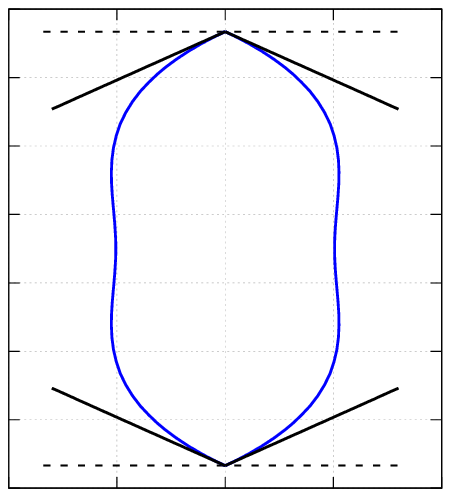}
\input{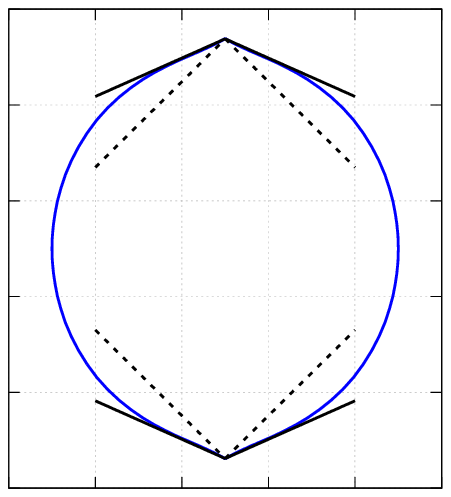}
\captionsetup{width=0.9\textwidth}
\captionof{figure}{\textsl{Embedding diagrams of the horizon for the $\ell = 1$ (top) and $\ell = 2^+$ (bottom) lumpy black holes (blue line) in AdS$_4\times S^7$ for the largest lumpiness parameter  that we have reached. The plots on the left correspond to  fixed $S^6$ coordinates while the plots on the right correspond  to fixed $S^2$ coordinates. The black lines correspond to the embeddings of the local cone that mediates the topology change transition. For fixed $S^6$ coordinates, the agreement between the cone model and the local horizon geometry is very good. For fixed $S^2$ coordinates, demanding that the cone model agrees with the local geometry suggests $\alpha = q-2+\frac{D}{4}=\frac{31}{4}$. In these two plots the embeddings dictated by the Ricci-flat cone are indicated with dashed lines. Clearly, the latter do not agree with our data.}}
\label{fig:embcone}
\end{figure}

To test the validity of the local model \eqref{triple}, we compare the embeddings of the cone:
\beq\begin{split}
\begin{aligned}
\text{Fixed $\Omega_{(q-1)}$: }& \quad X(\rho) = \rho\sqrt{\frac{q+1}{D-2}}, &\quad Y(\rho) = \rho\sqrt{\frac{p-3}{D-2}}, \\
\text{Fixed $\Omega_{(p-2)}$: }& \quad X(\rho) = \rho\sqrt{\frac{D-2-\alpha}{D-2}}, &\quad Y(\rho) = \rho\sqrt{\frac{\alpha}{D-2}},
\end{aligned}
\end{split}\enq
with the embeddings of the horizon geometry of the critical lumpy black holes described in Section \ref{ssec:geo}.  The results are shown in Fig. \ref{fig:embcone} for the AdS$_4\times S^7$ case and in Fig. \ref{fig:embcone2} for the AdS$_5\times S^5$ one. For fixed $S^6$ coordinates, Fig. \ref{fig:embcone} shows that there is a very good agreement between the local cone model and the actual geometry of the horizon. On the other hand, for fixed $S^2$ coordinates along the AdS factor, we can demand that the local geometry of the horizon near the pinch off region is well described by \eqref{triple} and thereby fix the constant $\alpha$. By doing this, we find that (\ref{alphflux}) is preferred. The dashed line in this figure shows the embedding of a Ricci flat cone. Clearly, our numerical data is clearly incompatible with a Ricci flat cone. 

The local cone model not only captures the shape of the horizon near the pinch off, but it also predicts the behavior of the physical quantities near the merger \cite{Kol:2005vy}. The latter follows from considering perturbations of \eqref{triple}  that preserve the various spheres while they change their relative sizes: 
\beq
\label{perturbtriple}
\dd s^2 = \dd\rho^2 + \frac{\rho^2}{D-2}\(e^{\epsilon(\rho)}\dd\Omega_{(2)}^2 + (p-3)e^{-\xi\epsilon(\rho)}\dd\Omega_{(p-2)}^2 + \alpha\,e^{-\eta\epsilon(\rho)}\dd\Omega_{(q-1)}^2\),
\enq
where $\xi$ and $\eta$ are free parameters. 
Imposing that the $\rho\rho$, $S^2$ and $S^{p-2}$ components of the Ricci tensor of the perturbed cone \eqref{perturbtriple} vanish  implies that 
\beq
\xi = \frac{2}{p-2}, \hs{0.75} \eta = 0,
\enq
or
\beq
\xi = 0, \hs{0.75} \eta = \frac{2}{q-1}.
\enq
Either choice leads to the same linear equation of motion for $\epsilon(\rho)$ and, unsurprisingly, it coincides with the equation of motion for the perturbation mode of the double-cone of \cite{Kol:top}. Note that in our case, the first solution is the relevant one since it implies that the metric on the internal $S^q$ is not modified at linear order; this is consistent with the vanishing of the components of the Ricci tensor along the  $S^2$ and the $S^{p-2}$ directions.  On the other hand,  the components along the $S^q$ are affected by the leftover fluxes on this internal sphere, and hence they should be sourced at second order in perturbation theory. The general solution is given by: 
\beq
\epsilon(\rho) =  c_+\rho^{s_+} + c_-\rho^{s_-},
\label{eqn:conelin1}
\enq
with critical exponents \beq
s_\pm = \frac{D-2}{2}\bigg(-1\pm \textrm{i}\sqrt{\frac{8}{D-2}-1}\bigg).
\enq
For $D<10$, the solutions have a non-zero imaginary part,  which implies that they will inspiral infinitely many times as they approach the origin $\rho=0$; $D=10$ is a marginal dimension for this model, where the imaginary part vanishes and the solution has a quartic power-law with an additional logarithmic term. For $D  > 10$, the critical exponents become purely real and the approach to the origin is govern by two powers. In $D=11$, we have  $s_+ = 6$ and $s_- = 3$.

The prediction of  \cite{Kol:top} is that the deformations of the cone metric \eqref{eqn:conelin1} should be reflected in the behavior of the physical quantities of the black holes sufficiently close to criticality, with the zero mode $\epsilon(\rho)$ measuring the deviation from the cone. According to this prediction, any physical quantity $Q$ near the critical solution should behave as
\beq\begin{split}
\label{predictcone}
(p,q) = (5,5)&: \hs{0.75} \delta Q =  \rho^4_0(C_1 + C_2\log \rho_0), \\
(p,q) = (4,7)&: \hs{0.75} \delta Q =  C_+ \rho_0^6 + C_- \rho_0^3\,,
\end{split}\enq
where $\delta Q \equiv Q-Q_c$ and $\rho_0$ is any typical length scale that measures the deviation from the actual cone. In principle, sufficiently close to the merger, the topology change develops in the same manner independently of the boundary data that one considers.

\begin{figure}[t!]
\begin{center}
\begin{minipage}{\textwidth}
\begin{minipage}[h!]{0.5\textwidth}
\begin{center}
\end{center}
\end{minipage}
\begin{minipage}[h!]{0.5\textwidth}
\begin{center}
\end{center}
\end{minipage}
\begin{minipage}[h!]{0.5\textwidth}
\input{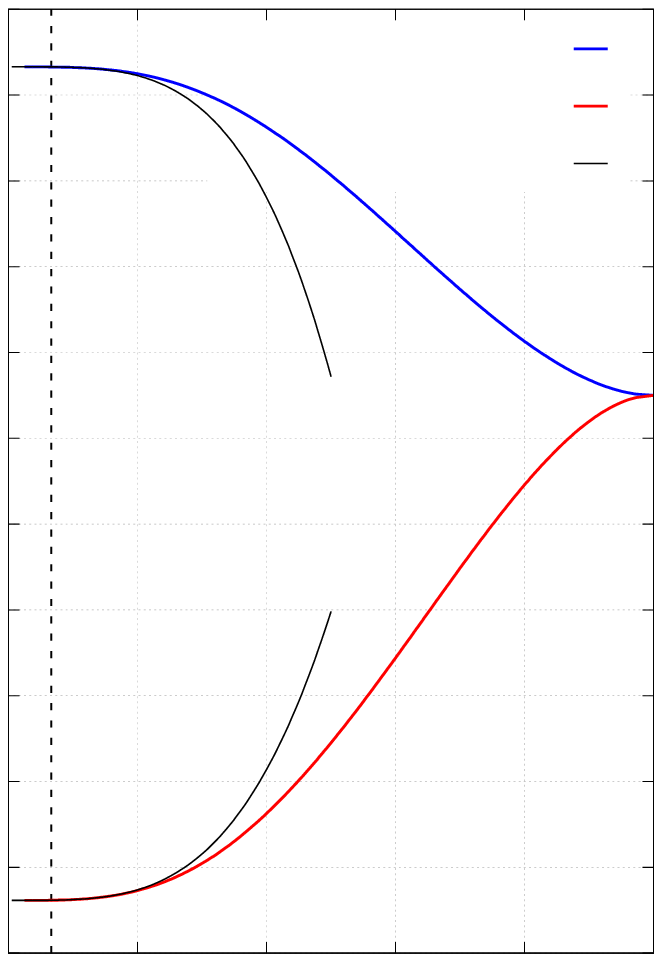}
\end{minipage}
\hfill
\begin{minipage}[h!]{0.5\textwidth}
\input{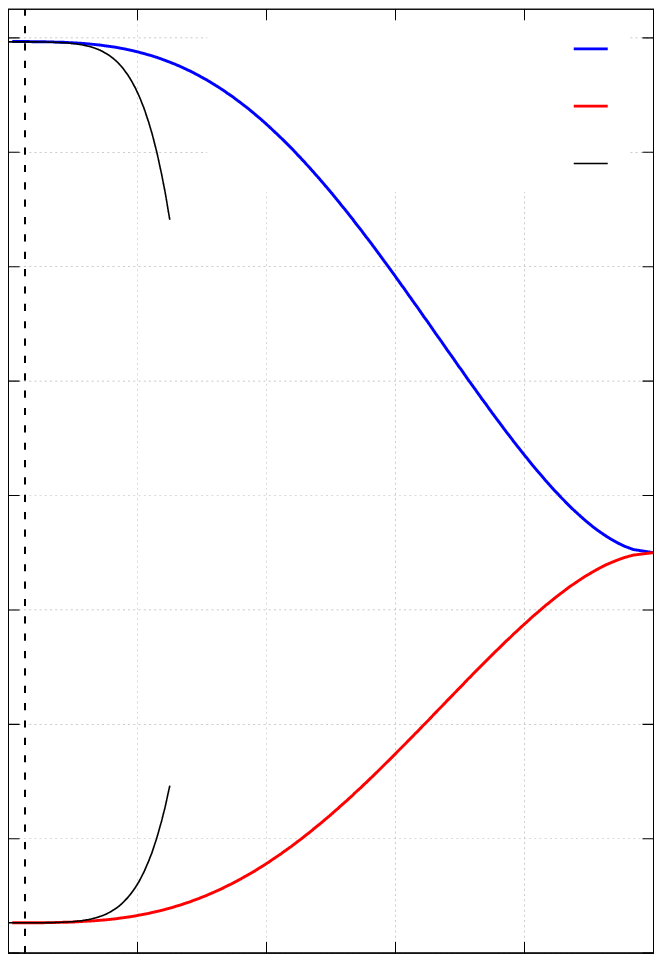}
\end{minipage}
\captionsetup{width=0.9\textwidth}
\captionof{figure}{\textsl{Normalized free energy (blue) and temperature (red) as a functions of $x_\text{lumpy}$ for $\ell = 1$ lumpy black holes in AdS$_p\times S^q$. Data points to the left of the dashed vertical line are the ones used for the fit (in black). Other thermodynamic quantities such as the entropy or the energy exhibit a similar behavior.}}
\label{critdataplots}
\end{minipage}
\end{center}
\end{figure}

\begin{table}[t!]
\begin{center}
\begin{tabular}{|r|r|r|r|r|r|}
\cline{2-6}
\multicolumn{1}{r|}{} & $\ell$ & $Q_c$ & $A$ & $B$ & $C$ \\ \hline\hline
\multirow{2}{*}{$T/T_{\text{GL}}$} & $1$~ & {\bf 0.98823} & {\bf 4.00002} & 0.07251 & $-0.05082$ \\
						   & $3$~ & {\bf 0.98996} & {\bf 4.00017} & $-0.01765$ & $-0.08789$  \\ \hline\hline
\multirow{2}{*}{$E/E_{\text{GL}}$} & $1$~ & {\bf 1.06752} & {\bf 4.00005} & $-0.75232$ & $-0.00001$ \\
			                            & $3$~ & {\bf 1.00739} & {\bf 3.99931} & $-0.02129$ & $0.04625$ \\ \hline\hline
\multirow{2}{*}{$S/S_{\text{GL}}$} & $1$~ & {\bf 1.18230} & {\bf 4.00004} & $-2.04392$ & $-0.00002$ \\
					            & $3$~ & {\bf 1.04771} & {\bf 3.99951} & $-0.38000$ & 0.21411 \\ \hline\hline
\multirow{2}{*}{$\la\mc{O}_2\ra/N^2$} & $1$~ & $\mathbf{1.571\times10^{-4}}$ & {\bf 3.99997} & $-0.13485$ & $-0.05648$ \\ 
					            & $3$~ & ${\bf 9.464\times10^{-6}}$ & {\bf 3.99996} & 0.00029 & 0.00016 \\ \hline\hline
\multirow{2}{*}{$\la\mc{O}_3\ra/N^2$} & $1$~ & ${\bf 5.168\times10^{-5}}$ & {\bf 3.99949} & $-0.00052$ & $0.00018$ \\
		 			            & $3$~ & ${\bf 1.090\times10^{-6}}$ & {\bf 4.00060} & $-6\times10^{-6}$ & $9\times10^{-6}$ \\ \hline\hline
\multirow{2}{*}{$\la\mc{O}_4\ra/N^2$} & $1$~ & ${\bf 9.499\times10^{-6}}$ & {\bf 3.99994} & $-0.00039$ & $0.00015$ \\
		 			            & $3$~ & ${\bf 7.882\times10^{-6}}$ & {\bf 4.00028} & $-0.00001$ & $0.00003$ \\ \hline\hline
\end{tabular}
\captionsetup{width=0.9\textwidth}
\captionof{table}{\textsl{Critical exponents and other parameters obtained from the fit of the temperature, energy, entropy and scalar vev's of the different lumpy solutions in AdS$_5 \times S^5$.}}
\label{tablecrit10}
\end{center}
\end{table}

\begin{table}[t]
\begin{center}
\begin{tabular}{|r|r|r|r|r|r|}
\cline{2-6}
\multicolumn{1}{r|}{} & $\ell$ & $Q_c$ & $A$ & $B$ & $C$ \\ \hline\hline
\multirow{4}{*}{$T/T_{\text{GL}}$} & $1$~ & {\bf 0.99353} & {\bf 3.00005} & 0.01318 & 8.97006 \\
						   & $2^+$ & {\bf 0.97730} & {\bf 2.99940} & 0.02580 & 9.55439 \\		                              					            & $3$~ & {\bf 0.99443} & {\bf 3.00003} & 0.01290 & 8.68568 \\ \hline\hline
\multirow{4}{*}{$E/E_{\text{GL}}$} & $1$~ & {\bf 1.01374} & ${\bf 3.00000}$ & $-0.02690$ & $-16.87280$  \\
						   & $2^+$ & {\bf 1.03911} & {\bf 2.99982} & $-0.04605$ & $-35.34570$ \\		                              					            & $3$~ & {\bf 1.00826} & {\bf 2.99996} & $-0.01918$ & $-12.96730$ \\ \hline\hline
\multirow{4}{*}{$S/S_{\text{GL}}$} & $1$~ & {\bf 1.02389} & {\bf 3.00000} & $-0.04691$ & $-29.42860$ \\
						   & $2^+$ & {\bf 1.07090} & {\bf 2.99982} & $-0.08444$ & $-64.81000$ \\		                              					           & $3$~ & {\bf 1.01520} & {\bf 2.99996} & $-0.03538$ & $-23.91850$ \\ \hline
\end{tabular}
\captionsetup{width=0.9\textwidth}
\captionof{table}{\textsl{Critical exponents and other parameters obtained from the fits of the temperature, energy and entropy of the different lumpy solutions in AdS$_4 \times S^7$. Note that for a given $\ell$, the critical exponents for the energy and the entropy are exactly the same. This is not surprising, since they are not independent: recall that in $D = 11$ the energy is found by integrating the 1st law of black hole mechanics.}}
\label{tablecrit11}
\end{center}
\end{table}

To compute the critical exponents from our numerical data and compare them with the predictions of the cone model \eqref{predictcone}, we have fitted the temperature, energy, entropy and free energy of critical enough solutions according to \beq\begin{split}\label{first}
(p,q) = (5,5)&: \hs{0.75} f(x) = f_c + x^A(B+C\log x), \\
(p,q) = (4,7)&: \hs{0.75} f(x) = f_c + x^A(B+Cx^3),\,
\end{split}\enq
with
\beq
x_{\text{lumpy}} = \frac{\text{min}[R_{p-2}(a)]}{Ry_0^{\ell}}\,,
\enq
measuring the deviation from the cone. Notice that in $D=11$ we have already assumed that one of the exponents is $A+3$. This exponent corresponds precisely to the smallest contribution to $f(x)$. In this case one would ideally want to fit the near critical data to a function of the form $f(x) = f_c + A x^B + C x^D$, so that both exponents can be independently extracted and tested. However, in practice this turns out to be quite hard. The reason is that to carry out the fits, we only consider the solutions that are close to the merger and for those we have $x \sim 10^{-3}$. Therefore, the behavior of $f(x)$ near the critical point is completely dominated by the smallest exponent, 3 in this case, and the corrections introduced by the other exponent (6 in $D=11$) are too small to be reliably detected.

In Figure \ref{critdataplots} we present the fits for the free energy and the temperature (normalized with respect to the values at the marginal GL point) for the $\ell = 1$ lumpy family in $D = 10$ and $D = 11$. The other physical quantities behave in a qualitatively similar way and we do not present the fits here. We have critical enough solutions for the families $\ell = 1,3$ in $D = 10$ and $\ell = 1,2^+,3$ in $D = 11$. The  fit parameters that we have extracted from the various physical quantities are shown in Tables \ref{tablecrit10} and \ref{tablecrit11} respectively. As these tables show, the values of the exponents obtained from fitting different physical quantities are perfectly compatible with each other and match the predictions of the cone model. With our fitting procedure we also extract the  values of the different thermodynamic quantities  (normalized with respect to the values of the $\ell$th family at the marginal GL point) at the merger point. In the different cases studied in this paper, the agreement with the cone model is very good, with deviations of $0.1\%$ in the worst case. Working with a denser grid and/or including more data points, i.e.,~more solutions near the merger region, may improve the accuracy of the critical exponents. It should be clear from our studies that \eqref{triple} does indeed capture both the local geometry of the horizon and the behavior of the physical quantities in the critical regime. 

In $D = 10$ we have also explicitly checked that the vev's of the dual scalar field $\mc{S}^2$, $\mc{S}^3$ and  $\mc{S}^4$, follow the same power law behavior near the merger as the other physical quantities. From the point of view of the cone model, this should not be surprising since any observable should exhibit the same behavior near the merger. However,  from the point of view of holography, this is quite interesting. Firstly, the fact that the field theory observables, such as the vev's of the scalar operators, exhibit this scaling behavior indicates how the field theory knows about the topology change in the bulk: none of the field theory observables seem to become singular at the merger point; they simply develop a scaling behavior with exponents determined by the local cone model in the bulk. Secondly, one might naively expect that the vev's, being associated to scalar fields that are coupled to 5-dimensional gravity in AdS, develop a spiraling behavior since in $D = 5$ the critical exponents have a non-zero imaginary part. However, our result shows that this is not the case;  the behavior near criticality is inherited from the full higher-dimensional solution. 

%~~~~~~~~~~~~~~~~~~~~~~~~~~~~~~~~~~~~~~~~~~~~~~~
\section{Discussion and outlook}
\label{disc}
%~~~~~~~~~~~~~~~~~~~~~~~~~~~~~~~~~~~~~~~~~~~~~~\
In this paper we have numerically constructed various families of asymptotically AdS$_p\times S^q$ black holes with a non-trivial field strength; we have considered the specific cases of $(p,q)=(5,5)$ corresponding to IIB supergravity in AdS$_5\times S^5$,  and $(p,q)=(4,7)$ corresponding to 11$D$ supergravity in AdS$_4\times S^7$. These black holes have horizon topology $S^{p-2}\times S^q$ and were called `lumpy' black holes in \cite{Dias:2015pda}. Each family is specified by an integer $\ell$ that labels the harmonic on the internal $S^q$  in the decomposition of the linear perturbations around the Schwarzschild-AdS$_p\times S^q$ black hole. For each harmonic, at a specific value of the horizon radius of the Schwarzschild-AdS$_p\times S^q$ black hole given in \eqref{GLvalues}, there exists a time-independent linear solution (i.e.,~a zero mode) of the Einstein's equations corresponding to a black hole with a slightly deformed horizon on the $S^q$. These linear modes can be uplifted to full non-linear solutions of the Einstein's equations, corresponding to the lumpy black holes. These zero modes signal the onset of the corresponding GL instability of the small Schwarzschild-AdS$_p\times S^q$ black hole. 

Ref.~\cite{Dias:2015pda} was the first one to construct lumpy black holes in AdS$_5\times S^5$; on the other hand, our solutions in the AdS$_4\times S^7$ case are new. Arguably, the solutions of  \cite{Dias:2015pda} are only slightly non-uniform on the $S^5$, with their non-uniformity parameter \eqref{lumpiness} being $\lambda \sim \mathcal{O}(1)$ at most.  In this article we have used similar numerical techniques as in our previous paper \cite{Cardona:2018shd} to extend the various families of lumpy black holes to very large values of $\lambda$, all the way to near the critical regime where they are expected to merge with other families of black holes with different horizon topology. While our results and those of \cite{Dias:2015pda} agree very well near the GL threshold point ($\lambda \ll 1$), we found some inconsistencies further along the branch towards more non-uniform lumpy black holes. We attribute this disagreement to the fact that the resolutions used in \cite{Dias:2015pda} were insufficient to accurately describe their most non-uniform black holes. 

We have constructed the phase diagrams in both the microcanonical and the canonical ensembles all the way to the critical regime. In the microcanonical ensemble, we find that the lumpy black holes are always subdominant with respect to the Schwarzschild-AdS$_p\times S^q$ black hole with the same energy, except for the $\ell=2^-$ family, which dominates in the region near the GL threshold point. In the canonical ensemble, for the range of temperatures that the lumpy black holes exist, thermal AdS$_p\times S^q$ is always the dominant phase whilst the lumpy black holes are always subdominant, even though the $\ell=2^-$  family dominates over the small Schwarzschild-AdS$_p\times S^q$ black hole. The apparent disagreement between these two thermodynamic ensembles in AdS$_5\times S^5$ was highlighted in \cite{Yaffe:2017axl}.

The main goal of this paper was to explore the topology-changing transitions that presumably connect, in the space of static solutions,  AdS$_p\times S^q$ lumpy black holes with other families of black holes with different horizon topologies. These transitions have been well-studied in the past in the context of black holes in KK theory. Kol \cite{Kol:top} predicted that such transitions were of the conifold-type and were governed by a local Ricci flat cone. According to this model, the local cone not only controls the change of topology, but it also determines the near-merger behavior of the physical quantities of the black holes on both sides of the transition. These predictions have been recently numerically verified \cite{Kalisch:2016fkm,Kalisch:2017bin,Cardona:2018shd}. In this paper we have extended these local cone models to black holes in AdS$_5\times S^5$ and AdS$_4\times S^7$ coupled to a (self-dual) five-form and a four-form field strength respectively. By constructing lumpy black holes with very large `lumpiness' parameters (see \eqref{maxlambda}),  we have verified, for the first time, that such local cone models also accurately describe the topology-changing transitions in AdS$_p\times S^q$ with the presence of non-trivial fluxes. Interestingly, we have found that the local cones \eqref{triple} are no longer Ricci flat because of the presence of fluxes in the internal $S^{q-1}$ in this limit. Work towards  understanding the new class of conical geometries that arise when there are fluxes in the internal spheres  is in progress.

For the asymptotically AdS$_5\times S^5$ lumpy black holes, we have extracted the dual field theory observables using the procedure of KK holography \cite{Skenderis:2006kkh}. Our computation fixes some typos and inconsistencies in \cite{Dias:2015pda}. We give the expressions for the vev's of the scalar operators $\mathcal S^2$, $\mathcal S^3$ and $\mathcal S^4$ in \eqref{scalarVEVsLumpy}; on the other hand, the vev's of $\mathcal T^\ell$, $\ell=0,1,2,3,4$, vanish identically in these backgrounds (and in the localized AdS$_5\times S^5$ black holes). The vev's of the dual scalar operators exhibit the same scaling behavior near the merger as the other physical quantities, thus showing how these topology changes are `detected' by the dual CFT. It is important to note that the behavior of the scalar operators near the merger is dictated by the full 10$D$ geometry. The $11D$ solutions can also be dimensionally reduced on the $S^7$. The KK map in this case, to leading and quadratic order, was worked out in \cite{Jang:2016exh}, and \cite{Jang:2018nle} respectively. We will leave the computation of the holographic data for this case for future work. 

The most natural extension of this work is to consider localized black holes in AdS$_5\times S^5$ and in AdS$_4\times S^7$. Localized black holes in AdS$_5\times S^5$ have been previously constructed in \cite{Dias:2016eto}, but their solutions are still quite far from the merger with the lumpy black holes considered in this paper. Using the same numerical techniques as in \cite{Cardona:2018shd}, we should be able to construct localized black holes in the critical regime and verify the predictions of the local cone model from the other `side' of the transition. Work in this direction is in progress.

%%%%%%%%%%%%%%%%%%%%%%%%%%%%%%%%%%%%%%%%%%%%%%%
\acknowledgments\addcontentsline{toc}{section}{Acknowledgements}
%%%%%%%%%%%%%%%%%%%%%%%%%%%%%%%%%%%%%%%%%%%%%%%
B.C.~and P.F.~are financially supported by the European Research Council grant ERC-2014-StG 639022-NewNGR. P.F.~is also supported by a Royal Society University Research Fellowship (Grant No. UF140319 and URF$\backslash$R$\backslash$201026). We would like to thank Jos\'e Juan F.~Melgarejo for useful comments on earlier drafts of this paper. B.C.~would like to thank Universidad de Murcia for warm hospitality during the final stages of this work.

\newpage
\appendix

%%%%%%%%%%%%%%%%%%%%%%%%%%%%%%%%%%%%%%%%%%%%%%%
\section{Kaluza-Klein holography}
\label{app:KKhol}
%%%%%%%%%%%%%%%%%%%%%%%%%%%%%%%%%%%%%%%%%%%%%%%
In this appendix we review the procedure of holographic renormalization for solutions of IIB supergravity that are asymptotically $\text{AdS}_5\times S^5$. This procedure was worked out in \cite{Skenderis:2006kkh} (see also \cite{Skenderis:2006hcv,Skenderis:2007ana} for related work) and named, for obvious reasons, Kaluza-Klein holography, or KK holography for short. 
We review it here to set up our conventions and fix the numerical factors. In Appendix \ref{applyHolo} we apply it to the specific class of solutions considered in this paper. These calculations were previously reported in \cite{Dias:2015pda}. However, since we found some disagreements with their results, we decided to write down our calculations in full detail so that they can be straightforwardly and independently checked, should anyone wish to do so. That said,  we have verified that our calculations pass all consistency checks and hence we are quite certain of their correctness.

Let us begin recalling that in the context of gauge/gravity duality one can compute the vev's of the dual gauge invariant fields in the bulk  by differentiating the (renormalized) on-shell supergravity action once with respect to the sources, a procedure known as holographic renormalization \cite{deHaro:2000hrec} (see also \cite{Bianchi:2001htg,Bianchi:2000hrec}). In this way, one obtains the vev's of gauge invariant operators of the dual field theory in terms of certain coefficients in the asymptotic expansion of the bulk fields. Originally, this procedure was worked out for solutions that are asymptotically AdS. The extension to generic solutions that asymptote to the direct product $\text{AdS}\times X$, where $X$ is some internal (compact) manifold,  requires to first carry out the dimensional reduction on $X$. In general, this dimensional reduction will involve an infinite tower of massive KK fields that come from the deformations of $X$, and which are  dual to an infinite tower of gauge invariant operators in the CFT.  Here we will concentrate on the $\text{AdS}_5\times S^5$ case, which is the relevant one for this paper. 

Given that the information needed is encoded in the asymptotic behavior of the solution near the AdS boundary, one may work out the dimensional reduction analysis as a linear perturbation about the AdS$_5\times S^5$ background and then apply holographic renormalization. This is what the authors of \cite{Skenderis:2006kkh} did in the most general way, using gauge invariant variables. This paper finds the effective 5-dimensional field equations and an explicit non-linear map between the higher-dimensional fields (which is the numerical data we have access to) and the solutions to these 5-dimensional equations. Integrating the latter equations one obtains a 5-dimensional (super)gravity action, which involves certain scalar fields;  from here one can compute the 1-point correlation functions of the dual field theory applying the standard holographic renormalization prescription.

At this point the idea is clear and \cite{Skenderis:2006kkh} provides an algorithm to implement it in practice. However, the expressions that arise are quite lengthly and involved. This is essentially because the dimensional reduction cannot be truncated to the massless sector when the isometries of the internal space are broken. In other words, the mapping between higher- and lower-dimensional fields, $\psi^k$ and $\Psi^k$ respectively, in general is going to be highly non-linear: \beq\label{exp}
\Psi^i = \psi^i - \sum_{j,k}\(A_{jk}^i\psi^j\psi^k+ B_{jk}^iD_\mu\psi^j D^\mu\psi^k +\mc{O}[\psi^3]\),
\enq
where $A_{jk}^i$ and $B_{jk}^i$ are numerical coefficients. Luckily, it turns out that to find the vev of an operator of a given dimension, one only needs the map truncated at certain order in the number of fields \cite{Skenderis:2006kkh}. Let us elaborate a bit more on this. If $\Psi^i$ is dual to an operator of dimension $i$, one would need to expand the field up to order $Z^i$ in the standard Fefferman-Graham (FG) coordinate to extract the operator's vev: \beq
\Psi^i = \Psi^i_{(0)} + \Psi^i_{(1)}Z + \Psi^i_{(2)}Z^2 + \dots + \Psi^i_{(i)}Z^i + \text{irrelevant terms}.
\enq
If only a finite number of terms in the near boundary expansion are needed to compute the vev of \eqref{exp}, then the r.h.s.~of this expression can only involve a finite number of fields. 
Recall that after all, the outcome of this mapping is directly related to a physical observable which must be finite. As we will see in the following sections and in Appendix \ref{applyHolo}, the scalar operators whose vev's we shall compute have at most conformal dimension 4. Hence we will need to compute the near boundary expansion of the corresponding bulk fields up to order $Z^4$ in the FG radial coordinate. This truncates the mapping \eqref{exp} to quadratic order in the number of fields.

In the rest of this appendix we will provide the details of this procedure in the context of this paper. This has been done before in \cite{Skenderis:2006kkh} and in \cite{Dias:2015pda}. Here we repeat the analysis in these references because we use slightly different conventions and we take this opportunity to fix typos. 

%%%%%%%%%%%%%%%%%%%%%%%%%%%%%%%%%%%%%%
\subsection{Kaluza-Klein dimensional reduction on S$^{\text{5}}$}
%%%%%%%%%%%%%%%%%%%%%%%%%%%%%%%%%%%%%%
The first step consists of dimensionally reducing the higher-dimensional solutions on the $S^5$. To this end, we need to expand a general perturbation of AdS$_5\times S^5$ into a complete set of scalar, vector and tensor harmonics. Due to the symmetries of the spacetimes that we consider, we only have to consider scalar harmonics on the $S^5$; from these we can obtain the corresponding scalar derived vector and tensor harmonics.  The details of the basis of harmonics that we use in this paper are summarized in the Appendix \ref{harm}.

We consider solutions that are fluctuations about the AdS$_5\times S^5$ background (denoted by the superscript ``(0)''), \beq
g_{MN} = g_{MN}^{(0)} + h_{MN}, \hs{0.75} F_{(5)} = F_{(5)}^{(0)} + f_{(5)}.
\enq
The indices are split up into $M = (\mu,a)$, with $\mu \in \text{AdS}_5$ and $a \in S^5$. From the point of view of transformations on the $S^5$, the perturbations of the metric components with two legs along the AdS  factor behave as scalars, and the ones with one/two leg/s along the $S^5$, behave as vectors/tensors. Then we may expand the metric as follows: \beq\begin{split}\label{expmetric}
h_{\munu}(z,x) &= \sum_{\ell}h_{\munu}^\ell(z)Y_{\ell}(x), \hs{1.7} h_{\mu a}(z,x) = \sum_{\ell}A_{\mu}^\ell(z)D_aY_{\ell}(x), \\
h_{(ab)}(z,x) &= \sum_{\ell}B^\ell(z)D_{(a}D_{b)}Y_{\ell}(x), \hs{0.75} h_a^a(z,x) = \sum_{\ell}C^\ell(z)Y_{\ell}(x).
\end{split}\enq
Notice that the internal metric perturbation is split into the trace, which behaves as a scalar,  and a traceless symmetric part which behaves as a tensor. Here we use  the  notation $T_{(ab)}$ to denote the symmetric and traceless part of a given 2-tensor $T_{ab}$: $T_{(ab)} =\frac{1}{2} (T_{ab} + T_{ba})-\frac{1}{5}g_{ab}T^a_a$.

The 5-form field strength is expanded in the following way:\footnote{The numerical constants here are introduced to match with the conventions of \cite{Kim:1985spe}.} \beq\begin{split}\label{expform}
f_{\munu\rho\sigma\tau}(z,x) &= \sum_{\ell}5D_{[\mu}b^\ell_{\nu\rho\sigma\tau]}(z)Y_\ell(x), \hs{0.80} f_{a\nu\rho\sigma\tau}(z,x) = \sum_{\ell}b^\ell_{\nu\rho\sigma\tau}(z)D_aY_\ell(x), \\
f_{abcd\mu}(z,x) &= \sum_{\ell}D_\mu b^\ell(z)\omega_{abcd}^{\phantom{abcd}{e}}D_eY_\ell(x), \hs{0.33} f_{abcde}(z,x)
= \sum_{\ell}\omega_{abcde}b^\ell(z)\Lambda_{\ell}Y_\ell(x),
\end{split}\enq
where $\omega_{abcde}$ denotes the volume 5-form of the $S^5$ and $\Lambda_\ell = -\ell(\ell+4)$ is the  harmonic eigenvalue with respect to the scalar Laplacian on the $S^5$. Given the self-duality condition of the 5-form, it turns out that the perturbation functions $b_{\nu\rho\sigma\tau}(z)$ and $b(z)$ are algebraically related, \beq
b_{\nu\rho\sigma\tau} = \omega_{\munu\rho\sigma}^{\phantom{\munu\rho\sigma}\gamma}(A_\gamma^\ell-D_\gamma b^\ell),
\enq
so we may just focus on one of them; $b(z)$ for instance.

A few comments are in order. First, in the expansions \eqref{expform} we have only included fields that couple to scalar-derived vector and tensor harmonics. Although the analysis can be carried out in full generality, including transverse vector and tensor harmonics, the additional terms would break the SO(5) isometry that the class of spacetimes that we consider have. In addition, not all the expansions in (\ref{expmetric}) and (\ref{expform}) start with the lowest harmonic, $\ell = 0$; there are a few exceptions. Given that $\Lambda_0 = D_aY_{0}(x) = D_{(a}D_{b)}Y_{0,1}(x) = 0$, then $A_\mu^0(z), B^0(z), B^1(z)$ and $b^0(z)$ are not defined, and the expansion for the perturbation fields $h_{\mu a}(z,x)$, $h_{(ab)}(z,x)$ and $f_{abcde}(z,x)$ starts at $\ell = 1$, $\ell = 2$ and $\ell = 1$, respectively.

The next step consists of inserting the expansions \eqref{expmetric} and \eqref{expform} into the linearized field equations, work out the mapping between the higher and the lower-dimensional fields, and obtain the 5-dimensional equations of motion for the dimensionally reduced fields. Before that, let us count the number of fields that we have. From the lower-dimensional perspective we have, for each value of $\ell$, a graviton, $h_{\munu}^\ell$, a vector, $A_\mu^\ell$, and three scalars: $B^\ell, C^\ell$ and $b^\ell$. However, some of these fields are pure gauge modes. By finding the gauge transformation rules, we can determine which ones are pure gauge and which ones correspond to the physical modes; a straightforward way to do this is to explicitly fix the gauge. For instance, Refs.~\cite{Sangmin:19983pf,Arutyunov:1999scc,Lee:20044pf,Kim:1985spe} performed the dimensional reduction in the de Donder gauge, which  can be shown to be equivalent to setting to zero the fields $A_\mu^\ell$ and $B^\ell$. In the analysis we do here we follow \cite{Skenderis:2006kkh}, and these are not set to zero but instead they are used to define gauge invariant physical fields. This is much more convenient in practice since, in general, solutions to 10-dimensional supergravity equations will not be in the de Donder gauge.

Consider a gauge transformation: $\delta x^M = -\xi^M$. The shift up to linear terms in the perturbation fields is given by\footnote{Notice that  the background is kept fixed: $\delta g^{(0)}_{MN} = 0 = \delta F^{(0)}$.}
\beq
\begin{split}
\label{deltagf}
\delta h_{MN} &= (D_M\xi_N + D_N\xi_M) + (D_M\xi^Ph_{PN} + D_N\xi^Ph_{PM} + \xi^PD_Ph_{MN}), \\
\delta f_{MNPQR} &= -5D_{[M}\xi^SF^{(0)}_{NPQR]S} - (5D_{[M}\xi^Sf_{NPQR]S} - \xi^SD_Sf_{MNPQR}).
\end{split}
\enq
Here the gauge parameter $\xi_M(z,x)$ is also expanded in harmonics,
\beq
\label{expxi}
\xi_\mu(z,x) = \sum_{\ell} \xi_\mu^\ell(z)Y_\ell(x)\,,\quad \xi_a(z,x) = \sum_{\ell}\xi^\ell(z)D_aY_\ell(x)\,.
\enq
Plugging \eqref{expmetric}, \eqref{expform} and \eqref{expxi} into \eqref{deltagf}, we obtain the gauge transformations of the perturbations at linear order, and from these we can identify the gauge invariant combinations. At linear order these transformations can be directly read off by looking at the coefficients of each harmonic. At quadratic order, one needs to project onto the basis of spherical harmonics to identify the transformation rules.

\subsubsection{Gauge invariance and equations at leading order}
At leading order one obtains the following gauge transformations of the perturbations:
\beq\begin{split}\label{gaugetransflin}
\delta h_{\munu}^\ell &= D_\mu\xi_\nu^\ell + D_\nu\xi_\mu^\ell, \hs{0.75} \ell \geq 0, \\
 \delta A_\mu^\ell &= D_\mu\xi^\ell + \xi_\mu^\ell, \hs{1.32} \ell \geq 1, \\ 
\delta B^\ell &= 2\xi^\ell, \hs{2.55} \ell \geq 2, \\
\delta C^\ell &= 2\Lambda_\ell\xi^\ell, \hs{2.15} \ell \geq 0, \\
\delta b^\ell &= \xi^\ell, \hs{2.8} \ell \geq 1.
\end{split}\enq
Given that the different harmonics $\ell$ are decoupled at linear order, we analyze the cases $\ell = 0,1$ and $\ell \geq 2$, separately. From \eqref{gaugetransflin} one can easily identify the gauge invariant combinations, which we summarize in  the following table:

\begin{center}
\begin{tabular}{|r||r|r|r|}
\cline{2-3}
\multicolumn{1}{r||}{} &   \multirow{2}{*}{Fields} & \multirow{2}{*}{Gauge invariant fields} \\
\multicolumn{1}{r||}{} & & \\ \hline\hline
\multirow{2}{*}{$\ell = 0$} & $h_{\munu}^0$ & \multirow{2}{*}{$\bar{C}^0 \equiv C^0\hs{3.22}$} \\
						   & $C^0$ & \\ \hline\hline
\multirow{4}{*}{$\ell = 1$} & $h_{\munu}^1$ & \multirow{2.5}{*}{$\bar{h}_{\munu}^1 \equiv h_{\munu}^1 - D_\mu \tilde{A}_\nu^1 - D_\nu \tilde{A}_\mu^1$} \\
						   & $C^1$ & \\
						   & $A_\mu^1$ & \multirow{1.5}{*}{$\bar{b}^1 \equiv \frac{1}{4}b^1 - \frac{1}{2\Lambda_1}C^1\hs{1.5}$} \\ 			                                     & $b^1$ & \\ \hline\hline
\multirow{5}{*}{$\ell \geq 2$} & $h_{\munu}^\ell$ & \multirow{2}{*}{$\bar{h}_{\munu}^\ell \equiv h_{\munu}^\ell - D_\mu \tilde{A}_\nu^\ell - D_\nu \tilde{A}_\mu^\ell$} \\
						   & $C^\ell$ & \\
						   & $A_\mu^\ell$ & \multirow{1}{*}{$\bar{C}^\ell \equiv C^\ell - \Lambda_\ell B^\ell\hs{1.8}$}   \\ 			                                    & $b^\ell$ & \multirow{1.5}{*}{$\bar{b}^\ell \equiv \frac{1}{4}b^\ell - \frac{1}{2}B^\ell\hs{1.9}$}  \\
						   & $B^\ell$ & \\ \hline
\end{tabular}
\vs{0.25}
\captionsetup{width=0.9\textwidth}
\captionof{table}{\textsl{Fields and gauge invariant combinations at linear order. For $\ell = 0$ there is only one gauge invariant field and $h_{\munu}^0$ is a deformation of the AdS part of the background metric (since it carries trivial harmonic index). To construct gauge invariant combinations for $\ell > 0$, one needs to introduce the auxiliary field $\tilde{A}_\mu^1 \equiv A_\mu^1 - D_\mu C^1/(2\Lambda_1)$, such that $\delta\tilde{A}^1_\mu = \xi_\mu^1$, and for $\ell > 1$, $\tilde{A}_\mu^\ell \equiv A_\mu^\ell - D_\mu B^\ell/2$ (then $\delta \tilde{A}_\mu^\ell = \xi^\ell_\mu$). }}
\label{tableLin}
\end{center}

Notice that the definition of the field $\bar{b}^\ell$ in Table \ref{tableLin} includes a factor $1/4$ which is not present in  \cite{Skenderis:2006kkh}. This simply corresponds to a different normalization of the field strength: in our equations of motion \eqref{eom} the coupling in front of the field strength is  $1/96$, whereas \cite{Skenderis:2006kkh,Sangmin:19983pf,Arutyunov:1999scc,Kim:1985spe,Lee:20044pf} considered $1/6$. To be consistent with the mapping, we need to rescale our field strength which amounts to the $1/4$ factor in the dimensionally reduced $b^\ell$ field.

To obtain the equations of motion for the gauge invariant  fields, we plug the expansions \eqref{expmetric} and \eqref{expform} into the linearized type IIB supergravity equations of motion and project onto  a given spherical harmonic. The equations for the fields $\bar{h}_{\munu}^\ell$, $\bar{C}^\ell$ nor $\bar{b}^\ell$ are of the Klein-Gordon type,  but these fields are not mass eigenstates.  Diagonalizing,  one finds the following gauge invariant combinations 
\cite{Kim:1985spe,Sangmin:19983pf}: \beq\begin{split}\label{massfields}
\hat{s}^\ell &= \frac{1}{20(\ell+2)}\(\bar{C}^\ell - 10(\ell+4)\bar{b}^\ell\), \hs{3.45} \ell \geq 2, \\
\hat{t}^\ell &= \frac{1}{20(\ell+2)}\(\bar{C}^\ell+10\ell\bar{b}^\ell\), \hs{4.52} \ell \geq 0, \\
\hat{\phi}^\ell_{(\munu)} &= \bar{h}_{(\munu)}^\ell - \frac{1}{(\ell+1)(\ell+3)}D_{(\mu}D_{\nu)}\(\frac{2}{5}\bar{C}^\ell - 12\bar{b}^\ell\), \hs{0.75} \ell \geq 1, \\
\end{split}\enq
satisfying: 
\beq
\begin{split}
\label{equationsSTlin}
\Box\hat{s}^\ell &= \ell(\ell-4)\hat{s}^\ell, \hs{2.32} \ell \geq 2, \\
\Box\hat{t}^\ell & = (\ell+4)(\ell+8)\hat{t}^\ell, \hs{1.36} \ell \geq 0,\\
\Box\hat{\phi}^\ell_{(\munu)} &= \(\ell(\ell+4)-2\)\hat{\phi}^\ell_{(\munu)}, \hs{0.75} \ell \geq 1,
\end{split}
\enq
where the `box' operator denotes the D'Alembertian in AdS$_5$. Now one can easily identify the mass of the different modes (in units of AdS radius $L = 1$); the scalar fields $\hat{s}^\ell$, $\hat{t}^\ell$ and the graviton $\hat{\phi}^\ell_{(\munu)}$ are dual to operators of conformal dimensions $\{\Delta_+,\Delta_- = 4-\Delta_+\} = \{\ell,4-\ell\}$, $\{\Delta_+,\Delta_-\} = \{\ell+8,-\ell-4\}$ and $\{\Delta_+,\Delta_-\} = \{\ell+4,-\ell\}$ respectively.

\subsubsection{Gauge invariance and equations at quadratic order}
At quadratic order the expressions are much more involved;  for this reason we first discuss the scalar sector ($B^\ell, C^\ell$ and $b^\ell$) and then the tensor sector ($h_{\munu}^\ell$ and $A_\mu^\ell$).

As mentioned earlier, at quadratic order the gauge transformations are obtained by projecting onto the basis of spherical harmonics. For instance, the transformation rule for the scalar field $B^\ell$ is given by: \beq
\delta B^{\ell}(z) = \frac{1}{z(\ell)q(\ell)}\int D^{(a}D^{b)}Y_{\ell}\, \delta h_{(ab)}\,\dd\omega_5,
\enq
with $\delta h_{(ab)} = 2D_{(a}\xi^Ph_{b)P} + \xi^PD_Ph_{(ab)}$; $z(\ell)$ and $q(\ell)$ are defined in Appendix \ref{harm}. Let us discuss again different values of $\ell$ separately. For $\ell = 1$ we have two scalars, $C^1$ and $b^1$, whose gauge transformations are the same as those at linear order, equation \eqref{gaugetransflin}, and hence the gauge invariant quantities at quadratic order are the same as those in Table \ref{tableLin}. For $\ell = 0$ we only have one scalar, i.e., $C^0$; its gauge transformation is: 
\beq\begin{split}
\delta C^0 &= \sum_{\ell\geq1}\frac{z(\ell)}{\Omega_5}\(2\xi^\ell B^\ell q(\ell) + \frac{2}{5}\Lambda_\ell \xi^\ell C^\ell + \xi^\mu_\ell D_\mu \bar{C}^\ell - \Lambda_\ell\(\xi^\ell C^\ell + 2\xi^\mu_\ell \tilde{A}_\mu^\ell\)\), 
\end{split}\enq
and the corresponding gauge invariant combination is given by: \beq\begin{split}
\bar{\bar{C}}^0 &= \bar{C}^0 + \sum_{\ell\geq1}\frac{z(\ell)}{\Omega_5}\(\frac{3}{10}\Lambda_{\ell}B^\ell \bar{C}^\ell - \frac{1}{4}\Lambda_\ell\(\Lambda_\ell+8\)(B^\ell)^2 - \tilde{A}^\mu_\ell D_\mu\bar{C}^\ell + \Lambda_\ell \tilde{A}^\mu_\ell\tilde{A}_\mu^\ell\),
\end{split}\enq
where $\bar{C}^\ell$ and $\tilde{A}_\mu^\ell$ are defined in Table \ref{tableLin}. Notice that the sum above starts at $\ell = 1$, but $B^1$ is not defined. We have written it this way just for notational compactness, and from this section onwards whenever we have $B^1$ we actually mean $C^1/\Lambda_1$. Here $\Omega_5$ denotes the surface area of the unit round $S^5$.

For $\ell \geq 2$ we have the whole scalar sector, $B^\ell, C^\ell$ and $b^\ell$. At quadratic order the gauge transformations are given by \beq\begin{split}
\delta B^{\ell_1} &= 2\xi^{\ell_1} + \frac{1}{z_{\ell_1}q_{\ell_1}}\sum_{\ell_2,\ell_3\geq 1}\bigg(2\xi^\mu_{\ell_2}A_\mu^{\ell_3}c_{\ell_1\ell_2\ell_3}+\frac{2}{5}\xi^{\ell_2}C^{\ell_3}d_{\ell_3\ell_1\ell_2}+\xi^\mu_{\ell_2}D_\mu B^{\ell_3}d_{\ell_2\ell_1\ell_3} \\
&\hs{4.2}+\xi^{\ell_2}B^{\ell_3}e_{\ell_1\ell_2\ell_3}\bigg), \\
\delta C^{\ell_1} &= 2\Lambda_{\ell_1}\xi^{\ell_1} + \frac{1}{z_{\ell_1}}\sum_{\ell_2,\ell_3\geq1}\bigg(2B^{\ell_2}\xi^{\ell_3}d_{\ell_1\ell_2\ell_3} + \(\xi^{\ell_2}C^{\ell_3}+2\xi^{\mu}_{\ell_2}A_\mu^{\ell_3}\)b_{\ell_1\ell_2\ell_3} \\
&\hs{4.3}+ \Big(\frac{2}{5}\Lambda_{\ell_2}\xi^{\ell_2}C^{\ell_3} + \xi^\mu_{\ell_2}D_\mu C^{\ell_3}\Big)a_{\ell_1\ell_2\ell_3}\bigg), \\
\delta b^{\ell_1} &= \xi^{\ell_1} + \frac{1}{\Lambda_{\ell_1}z_{\ell_1}}\sum_{\ell_2,\ell_3\geq 1}\(\xi^\mu_{\ell_2}D_\mu b^{\ell_3} + \Lambda_{\ell_2}b^{\ell_2}\xi^{\ell_3}\)\(b_{\ell_1\ell_2\ell_3}+\Lambda_{\ell_3}a_{\ell_1\ell_2\ell_3}\),
\end{split}\enq
with the notation for normalization factors and triple integrals defined in Appendix \ref{harm}: $z_{\ell_1} \equiv z(\ell_1)$, $a_{\ell_1\ell_2\ell_3} \equiv a(\ell_1,\ell_2,\ell_3)$, etc. From these transformations one infers the following gauge invariant fields:\footnote{A few typos have been corrected with respect to the expressions given in \cite{Dias:2015pda,Skenderis:2006kkh}.} 
\beq\begin{split}
\bar{\bar{C}}^{\ell_1} &= \bar{C}^{\ell_1} - \frac{1}{2z_{\ell_1}}\sum_{\ell_2,\ell_3\geq1}\bigg(\frac{2}{5}\Big(\Lambda_{\ell_2}a_{\ell_1\ell_2\ell_3} + \frac{5}{2}b_{\ell_1\ell_2\ell_3}-\frac{\Lambda_{\ell_1}}{q_{\ell_{1}}}d_{\ell_3\ell_1\ell_2}\Big)B^{\ell_2}\bar{C}^{\ell_3} \\
&\hs{3.8} + \Big(d_{\ell_1\ell_2\ell_3} - \frac{\Lambda_{\ell_1}}{2q_{\ell_1}}e_{\ell_1\ell_2\ell_3} \\
&\hs{4.4}+\frac{\Lambda_{\ell_3}}{5}\Big[\Lambda_{\ell_2}a_{\ell_1\ell_2\ell_3} + \frac{5}{2}b_{\ell_1\ell_2\ell_3} - \frac{\Lambda_{\ell_1}}{q_{\ell_1}}d_{\ell_2\ell_1\ell_3}\Big]\Big)B^{\ell_2}B^{\ell_3}\\
&\hs{3.8}+2\tilde{A}_\mu^{\ell_2}\Big(D^\mu\bar{C}^{\ell_3} a_{\ell_1\ell_2\ell_3} + \tilde{A}^\mu_{\ell_3}\Big[b_{\ell_1\ell_2\ell_3}-\frac{\Lambda_{\ell_1}}{q_{\ell_1}}c_{\ell_1\ell_2\ell_3}\Big]\Big)\bigg), \\
\bar{\bar{b}}^{\ell_1} &= \bar{b}^{\ell_1} + \frac{1}{z_{\ell_1}}\sum_{\ell_2,\ell_3\geq1}\bigg(\frac{\Lambda_{\ell_3}}{2\Lambda_{\ell_1}}B^{\ell_2}\bar{b}^{\ell_3}b_{\ell_3\ell_1\ell_2} + \frac{1}{10 q_{\ell_1}}d_{\ell_3\ell_1\ell_2}B^{\ell_2}\bar{C}^{\ell_3} \\
&\hs{3.4}+\frac{1}{8}\Big(\frac{\Lambda_{\ell_3}}{\Lambda_{\ell_1}}b_{\ell_3\ell_1\ell_2} + \frac{2}{5}\frac{\Lambda_{\ell_3}}{q_{\ell_1}}d_{\ell_2\ell_1\ell_3} + \frac{1}{q_{\ell_1}}e_{\ell_1\ell_2\ell_3}\Big)B^{\ell_2}B^{\ell_3} \\
&\hs{3.4}+ \tilde{A}_\mu^{\ell_2}\Big(\frac{1}{2q_{\ell_1}}\tilde{A}_{\ell_3}^{\mu}c_{\ell_1\ell_2\ell_3} + \frac{1}{\Lambda_{\ell_1}}D^\mu\bar{b}^{\ell_3}b_{\ell_2\ell_1\ell_3}\Big)\bigg).
\end{split}\enq

In the tensor sector we are only interested in the massless KK graviton, i.e., $h_{\munu}^0$ (see \cite{Skenderis:2006kkh} for the analysis of massive gravitons). At quadratic order, this field transforms according to: \beq\begin{split}
\delta h_{\munu}^0 &= D_\mu\xi_\nu^0 + D_\nu\xi^0_\mu + \bigg(D_\mu\xi^\alpha_0h_{\alpha\nu}^0 + D_\nu\xi^\alpha_0h_{\alpha\mu}^0 + \xi^\alpha_0D_\alpha h_{\munu}^0 \\
&\hs{0.45}+ \sum_{\ell \geq 1}\frac{z_{\ell}}{\Omega_5}\Big(D_\mu\xi_\ell^\alpha h_{\alpha\nu}^\ell + D_\nu\xi_\ell^\alpha h_{\alpha\mu}^\ell + \xi_\ell^\alpha D_\alpha h_{\munu}^\ell - \Lambda_\ell(\xi^\ell h_{\munu}^\ell + 2D_{(\mu}\xi^\ell A_{\nu)}^\ell)\Big)\bigg).
\end{split}\enq
Then the combination \beq\begin{split}\label{hmunu02ndOrd}
\bar{\bar{h}}_{\munu}^0 &= h_{\munu}^0 + \frac{1}{3}\bar{\bar{C}}^0 g_{\munu}^{(0)} - \sum_{\ell \geq 1}\frac{z_{\ell}}{\Omega_5}\bigg(-\frac{1}{2}\Lambda^\ell\Big(B^\ell\bar{h}^\ell_{\munu} + \frac{1}{2}D_\mu B^\ell D_\nu B^\ell \Big) \\
&\hs{0.45}+ D_\mu \tilde{A}^\alpha_\ell \bar{h}_{\nu\alpha}^\ell + D_\nu \tilde{A}^\alpha_\ell \bar{h}_{\mu\alpha}^\ell + \tilde{A}^\alpha_\ell D_\alpha\bar{h}_{\munu}^\ell + D_\mu \tilde{A}^\alpha_\ell D_\nu\tilde{A}_\alpha^\ell + \tilde{A}^\alpha_\ell\tilde{A}_\alpha^\ell g_{\munu}^{(0)} - \tilde{A}_\mu^\ell\tilde{A}_\nu^\ell \bigg),
\end{split}\enq
where $g_{\munu}^{(0)}$ denotes the AdS$_5$ background metric, is not gauge invariant but provides a correction to the spacetime metric that transforms nicely. Notice that what plays the role of `gauge invariant' field at linear order here is the combination: $\bar{h}_{\munu}^{(0)} \equiv h_{\munu}^0 + \frac{1}{3}\bar{C}^0g_{\munu}^{(0)}$, although it is not gauge invariant either (recall that $\delta \bar{C}^0 = 0$ but $\delta h_{\munu}^0 = D_\mu \xi_\nu^0 + D_\nu\xi_\mu^0$). The shift in $\bar{C}^0$ can be understood as a Weyl transformation that brings the 5-dimensional action to the Einstein frame. Then $\bar{h}_{\munu}^{(0)}$ satisfies the proper linearized Einstein equation \cite{Skenderis:2006kkh}.

From the gauge invariant scalar fields, one can define the same combinations as in (\ref{massfields}), now replacing the `bar' fields by the double `bar' ones, which satisfy the modified equations of motion \cite{Sangmin:19983pf}: \beq\begin{split}\label{scalarq}
\(\Box - \ell(\ell-4)\)\hat{\hat{s}}^\ell &= \frac{1}{2(\ell+2)}\((\ell+4)(\ell+5)Q_1^\ell + Q_2^\ell + (\ell+4)(D_\mu {Q_3^\mu}^\ell + Q_4^\ell)\), \\
\(\Box - (\ell+4)(\ell+8)\)\hat{\hat{t}}^\ell &= \frac{1}{2(\ell+2)}\(\ell(\ell-1)Q_1^\ell + Q_2^\ell - \ell(D_\mu {Q_3^\mu}^\ell+ Q_4^\ell)\), \\
\end{split}\enq
where the double `hat' notation is to emphasize that these are defined in terms of gauge invariant fields at quadratic order (i.e., they are built up from double `bar' fields). Note that the masses are the same as those of the linear fields $\h{s}$ and $\h{t}$, equation \eqref{equationsSTlin}. The box operator is again the D'Alembertian in AdS$_5$ and the $Q$'s that appear in the r.h.s.~of the equations are given by: \beq\begin{split}\label{Qs}
Q_1^{\ell_1} &= \frac{1}{20z_{\ell_1}q_{\ell_1}}\sum_{\ell_2,\ell_3}\Big((c_{\ell_1\ell_2\ell_3} + d_{\ell_2\ell_3\ell_1}+ d_{\ell_3\ell_2\ell_1})T_{\ell_2\ell_3} + 32c_{\ell_1\ell_2\ell_3}D_\mu \hat{s}^{\ell_2}D^\mu\hat{s}^{\ell_3}\Big), \\
Q_2^{\ell_1} &= \frac{1}{20z_{\ell_1}}\sum_{\ell_2,\ell_3}\Big(10S_{\ell_1\ell_2\ell_3}+ (b_{\ell_1\ell_2\ell_3}-2f_{\ell_3}a_{\ell_1\ell_2\ell_3})T_{\ell_2\ell_3} + 32b_{\ell_1\ell_2\ell_3}D_\mu \hat{s}^{\ell_2}D^\mu\hat{s}^{\ell_3}\Big), \\
Q^{\mu\ell_1}_3 &= \frac{1}{z_{\ell_1}f_{\ell_1}}\sum_{\ell_2,\ell_3}\Big(\big(U_{\ell_2}+3V_{\ell_2}\big)\hat{s}^{\ell_2}D^\mu\hat{s}^{\ell_3} +W_{\ell_2}D^{(\mu}D^{\nu)}\hat{s}^{\ell_2}D_\nu\hat{s}^{\ell_3}\Big)b_{\ell_2\ell_1\ell_3}, \\
Q_4^{\ell_1} &= -\frac{1}{4z_{\ell_1}}\sum_{\ell_2,\ell_3}\Big(T_{\ell_2\ell_3} + V_{\ell_2}\big(16f_{\ell_3} - 40V_{\ell_3}\big)\hat{s}^{\ell_2}\hat{s}^{\ell_3}\Big)a_{\ell_1\ell_2\ell_3},
\end{split}\enq
with \beq\begin{split}
T_{\ell_2\ell_3} &= \big(3V_{\ell_2}V_{\ell_3} + 5U_{\ell_2}U_{\ell_3}\big)\hat{s}^{\ell_2}\hat{s}^{\ell_3} + W_{\ell_2}W_{\ell_3}D^{(\mu}D^{\nu)}\hat{s}^{\ell_2}D_{(\mu}D_{\nu)}\hat{s}^{\ell_3}, \\
S_{\ell_1\ell_2\ell_3} &= -V_{\ell_2}V_{\ell_3} b_{\ell_2\ell_1\ell_3} \hat{s}^{\ell_2}\hat{s}^{\ell_3} -8(a_{\ell_1\ell_2\ell_3} f_{\ell_2}f_{\ell_3}\hat{s}^{\ell_2}\hat{s}^{\ell_3} + b_{\ell_1\ell_2\ell_3}D^\mu\hat{s}^{\ell_2}D_\mu\hat{s}^{\ell_3}) \\
&\hs{0.45}+ V_{\ell_3}U_{\ell_2}a_{\ell_1\ell_2\ell_3}D^\mu(\hat{s}^{\ell_2}D_\mu\hat{s}^{\ell_3}) + W_{\ell_2}V_{\ell_3}a_{\ell_1\ell_2\ell_3} D_{\mu}(D^{(\mu}D^{\nu)}\hat{s}^{\ell_2}D_\nu\hat{s}^{\ell_3}) \\
&\hs{0.45} + a_{\ell_1\ell_2\ell_3}V_{\ell_2}(64f_{\ell_3} - 80V_{\ell_3})\hat{s}^{\ell_2}\hat{s}^{\ell_3},
\end{split}\enq
and \beq
V_{\ell} = -\frac{5}{3}U_{\ell} = 2\ell, \hs{0.75} W_{\ell} = \frac{4}{\ell+1}.
\enq

\subsubsection{Non-linear mapping: 5D action}
If we pack together fields and masses: $\hat{\hat{\psi}}^I_{\ell} = \{\h{\h{s}}^{\ell},\h{\h{t}}^{\ell}\}$, $m_{I,\ell}^2 = \{\ell(\ell-4),(\ell+4)(\ell+8)\}$ ($I = 1,2$), the scalar field equations at quadratic order (\ref{scalarq}) may be written as: \beq\label{mastereq}
(\Box - m_{I,{\ell_1}}^2)\hat{\hat{\psi}}^I_{\ell_1} = \sum_{\ell_2,\ell_3\geq1}\(D^I_{\ell_1\ell_2\ell_3}\h{s}^{\ell_2}\h{s}^{\ell_3} + E^I_{\ell_1\ell_2\ell_3}D_\mu\h{s}^{\ell_2}D^\mu\h{s}^{\ell_3}+ F^I_{\ell_1\ell_2\ell_3}D_{(\mu}D_{\nu)}\h{s}^{\ell_2}D^{(\mu}D^{\nu)}\h{s}^{\ell_3}\),
\enq
for some coefficients $D^I_{\ell_1\ell_2\ell_3}$, $E^I_{\ell_1\ell_2\ell_3}$ and $F^I_{\ell_1\ell_2\ell_3}$ that are easily identified using the expressions in (\ref{Qs}) (see Table \ref{tableNumfac}). Notice that the scalar fields  in the l.h.s.~are the gauge invariant combinations at second order whilst the fields in the r.h.s.~are the gauge invariant fields at linear order (given that the r.h.s.~side is quadratic in the fields). In the case  where $\hat{s}^2$ is the  only non-zero field at linear order (as it is for our solutions), the gauge invariant combinations can be written in the form: \beq\label{2fieldsSimp}
\hat{\hat{\psi}}^I_\ell = \hat{\psi}^I_\ell + A^I_{\ell sB}\hat{s}^2 B^2 + A^I_{\ell BB}(B^2)^2 + A^I_{\ell sA}D^\mu\hat{s}^2\tilde{A}^2_\mu + A^I_{\ell AA}(\tilde{A}^2_\mu)^2
\enq
with $\hat{\psi}^I_\ell = \{\h{s}^{\ell},\h{t}^{\ell}\}$. The four coefficients $A^I_{\ell XX}$ are also given in Table \ref{tableNumfac}.

The higher derivative terms on the r.h.s.~of equation (\ref{mastereq}) can be removed via a field redefinition $\hat{\hat{\psi}}^I_{\ell} \riga \Psi^I_{\ell} = \{\mc{S}^{\ell},\mc{T}^{\ell}\}$ of the form: \beq\label{mappingggg}
\Psi^I_{\ell_1} = w^I_{\ell_1}\bigg(\hat{\hat{\psi}}^I_{\ell_1} - \sum_{\ell_2,\ell_3\geq1}\(J_{\ell_1\ell_2\ell_3}^I\h{s}^{\ell_2}\h{s}^{\ell_3} + K_{\ell_1\ell_2\ell_3}^I D_\mu\h{s}^{\ell_2}D^\mu\h{s}^{\ell_3}\)\bigg),
\enq
with \beq\begin{split}\label{JKw}
w^I_{\ell} &= \bigg\{\sqrt{\frac{8\ell(\ell-1)(\ell+2)}{\ell+1}\frac{z_{\ell}}{\Omega_5}},\, \sqrt{\frac{8(\ell+2)(\ell+4)(\ell+5)}{\ell+3}\frac{z_{\ell}}{\Omega_5}}\bigg\}, \\
J_{\ell_1\ell_2\ell_3}^I &= \frac{1}{2}E^I_{\ell_1\ell_2\ell_3} + \frac{1}{4}F^I_{\ell_1\ell_2\ell_3}\(m_{I,\ell_1}^2 - m_{1,\ell_2}^2 - m_{1,\ell_3}^2+8\), \\
K_{\ell_1\ell_2\ell_3}^I &= \frac{1}{2}F^I_{\ell_1\ell_2\ell_3}.
\end{split}\enq

Then the field equations for the fields $\Psi_{\ell}^I$ read \beq\label{eqST}
(\Box - m_{I,{\ell_1}}^2)\Psi_{\ell_1}^I = \sum_{\ell_2,\ell_3\geq1}\lambda^I_{\ell_1\ell_2\ell_3}\mc{S}^{\ell_2}\mc{S}^{\ell_3},
\enq
where \beq\label{coupling}
\lambda_{\ell_1\ell_2\ell_3}^I = \frac{w^I_{\ell_1}}{w^1_{\ell_2}w^1_{\ell_3}}\(D^I_{\ell_1\ell_2\ell_3} + (m_{I,\ell_1}^2 - m_{1,\ell_2}^2 - m_{1,\ell_3}^2)J^I_{\ell_1\ell_2\ell_3} - \frac{2}{5}K^I_{\ell_1\ell_2\ell_3}m_{1,\ell_2}^2m_{1,\ell_3}^2\).
\enq
All quadratic and cubic scalar couplings have been determined and can be found in the literature \cite{Sangmin:19983pf,Arutyunov:1999scc,Lee:20044pf,Kim:1985spe}, but in our case it is sufficient to retain only the quadratic coupling to the field $\h{s}^2$ (or $\mc{S}^2$). The numerical factors up to this quadratic coupling can be found in Table \ref{tableNumfac}.

A very similar approach can be carried out for the massless KK graviton;  the equation of motion for $\bar{\bar{h}}_{\munu}^0$ at quadratic order is sourced by higher derivative interactions \cite{Skenderis:2006kkh} which, just as in the case of  the scalars, can be removed through the field redefinition:
\beq\label{mappinggggG}
G_{\munu} = \bar{\bar{h}}_{\munu}^0 - \frac{1}{12}\bigg(\frac{2}{9}D_\mu D^\rho\hat{s}^2D_{\nu}D_\rho\hat{s}^2 - \frac{10}{3}\hat{s}^2D_\mu D_\nu\hat{s}^2 + \Big(\frac{10}{9}D_\mu\h{s}^ 2D^\mu\h{s}^2- \frac{32}{9}(\h{s}^2)^2\Big)g_{\munu}^{(0)}\bigg).
\enq

\begin{table}[t]
\begin{center}
\begin{tabular}{|r||r|r|r|r|r|}
\cline{2-6}
\multicolumn{1}{r||}{} &  \multicolumn{2}{c|}{\multirow{2}{*}{$\hat{\hat{s}}^\ell$ $(I = 1)$}} & \multicolumn{3}{c|}{\multirow{2}{*}{$\hat{\hat{t}}^\ell$ $(I = 2)$}}  \\
\multicolumn{1}{r||}{} & \multicolumn{2}{r|}{} & \multicolumn{3}{r|}{} \\ \hline
\multicolumn{1}{|r||}{\multirow{2}{*}{$\ell$}} & \multirow{2}{*}{$2$} & \multirow{2}{*}{$4$} & \multirow{2}{*}{$0$} & \multirow{2}{*}{$2$} & \multirow{2}{*}{$4$} \\
 & & & & & \\ \hline\hline
\multirow{2}{*}{$D_{\ell22}^I$} & \multirow{2}{*}{$-16\sqrt{\frac{2}{15}}$} & \multirow{2}{*}{$-\frac{172\sqrt{7}}{25}$} & \multirow{2}{*}{$\frac{229}{75}$} & \multirow{2}{*}{$\frac{304}{25}\sqrt{\frac{6}{5}}$} & \multirow{2}{*}{$\frac{52\sqrt{7}}{5}$} \\
					      & & & & & \\
\multirow{2}{*}{$E_{\ell22}^I$} & \multirow{2}{*}{$\frac{2}{5}\sqrt{\frac{6}{5}}$} & \multirow{2}{*}{$\frac{3\sqrt{7}}{5}$} & \multirow{2}{*}{$-\frac{11}{20}$} & \multirow{2}{*}{$-\frac{6}{5}\sqrt{\frac{6}{5}}$} & \multirow{2}{*}{$-\frac{\sqrt{7}}{5}$} \\
					      & & & & & \\
\multirow{2}{*}{$F_{\ell22}^I$} & \multirow{2}{*}{$\frac{1}{3}\sqrt{\frac{2}{15}}$} & \multirow{2}{*}{$\frac{7\sqrt{7}}{45}$} & \multirow{2}{*}{$\frac{1}{60}$} & \multirow{2}{*}{$\frac{1}{15}\sqrt{\frac{2}{15}}$} & \multirow{2}{*}{$0$} \\
					      & & & & & \\ \hline	  		      
\multirow{2}{*}{$A_{\ell sB}^I$} & \multirow{2}{*}{$\frac{7}{5}\sqrt{\frac{3}{10}}$} & \multirow{2}{*}{$\frac{7\sqrt{7}}{10}$} & \multirow{2}{*}{$-\frac{3}{40}$} & \multirow{2}{*}{$-\frac{7}{5\sqrt{30}}$} & \multirow{2}{*}{$-\frac{\sqrt{7}}{10}$} \\
					      & & & & & \\
\multirow{2}{*}{$A_{\ell BB}^I$} & \multirow{2}{*}{$-\frac{17}{20}\sqrt{\frac{3}{10}}$} & \multirow{2}{*}{$-\frac{\sqrt{7}}{10}$} & \multirow{2}{*}{$-\frac{1}{80}$} & \multirow{2}{*}{$\frac{3}{20}\sqrt{\frac{3}{10}}$} & \multirow{2}{*}{$\frac{\sqrt{7}}{40}$} \\
					      & & & & & \\
\multirow{2}{*}{$A_{\ell sA}^I$} & \multirow{2}{*}{$-\frac{1}{\sqrt{30}}$} & \multirow{2}{*}{$-\frac{3\sqrt{7}}{20}$} & \multirow{2}{*}{$-\frac{1}{48}$} & \multirow{2}{*}{$-\frac{1}{5\sqrt{30}}$} & \multirow{2}{*}{$0$} \\
					      & & & & & \\					      
\multirow{2}{*}{$A_{\ell AA}^I$} & \multirow{2}{*}{$-\frac{1}{4}\sqrt{\frac{3}{10}}$} & \multirow{2}{*}{$-\frac{\sqrt{7}}{20}$} & \multirow{2}{*}{$-\frac{1}{80}$} & \multirow{2}{*}{$-\frac{1}{20}\sqrt{\frac{3}{10}}$} & \multirow{2}{*}{$0$} \\
					      & & & & & \\ \hline 
\multirow{2}{*}{$J_{\ell22}^I$} & \multirow{2}{*}{$\frac{8}{5}\sqrt{\frac{2}{15}}$} & \multirow{2}{*}{$\frac{83\sqrt{7}}{90}$} & \multirow{2}{*}{$-\frac{3}{40}$} & \multirow{2}{*}{$-\frac{8}{15}\sqrt{\frac{2}{15}}$} & \multirow{2}{*}{$-\frac{\sqrt{7}}{10}$} \\
					      & & & & & \\
\multirow{2}{*}{$K_{\ell22}^I$} & \multirow{2}{*}{$\frac{1}{3\sqrt{30}}$} & \multirow{2}{*}{$\frac{7\sqrt{7}}{90}$} & \multirow{2}{*}{$\frac{1}{120}$} & \multirow{2}{*}{$\frac{1}{15\sqrt{30}}$} & \multirow{2}{*}{$0$} \\
					      & & & & & \\ \hline
\multirow{2}{*}{$\lambda_{\ell22}^I$} & \multirow{2}{*}{$-\frac{16}{\sqrt{15}}$} & \multirow{2}{*}{$0$} & \multirow{2}{*}{$0$} & \multirow{2}{*}{$0$} & \multirow{2}{*}{$0$} \\
					      & & & & & \\ \hline
\end{tabular}
\vs{0.25}
\captionsetup{width=0.9\textwidth}
\captionof{table}{\textsl{Numerical coefficients in equations (\ref{mastereq})-(\ref{coupling}). Notice that our numerical coefficients differ with respect to those given in \cite{Skenderis:2006kkh}. More precisely, all the non-zero coefficients for $\ell = 2$ or $\ell = 4$ differ by the same factor, $4\sqrt{2/5}$ and $\sqrt{7/5}$ respectively, compared to the corresponding coefficients in that paper. This is due to the fact that we use a different harmonic representation, which also includes odd values of $\ell$. The coefficients are the same for $\ell = 0$. For $\ell$ odd all coefficients vanish.}}
\label{tableNumfac}
\end{center}
\end{table}

The equations of motion for the scalar fields and the massless graviton after the field redefinitions,  \eqref{mappingggg} and \eqref{mappinggggG}, can be obtained by varying a 5-dimensional action with a  negative cosmological constant.
Therefore these mappings define the desired non-linear KK map from solutions of the 10-dimensional equations $\{\h{\h{s}}^{\ell},\h{\h{t}}^{\ell},\bar{\bar{h}}_{\munu}^0\}$ to solutions of the 5-dimensional equations $\{\mc{S}^\ell,\mc{T}^\ell,G_{\munu}\}$. More precisely, given the normalization factors $w(\psi_\ell^I)$ in (\ref{JKw}), the 5-dimensional theory corresponds to the bosonic sector of $D = 5$, $\mc{N}=8$ gauged supergravity, whose action is given by:
 \beq\label{N8sugra}
S_{5D} = \frac{N^2}{2\pi^2}\int \dd^5x\sqrt{-G}\bigg[\frac{1}{4}R + 3 - \sum_{I,\ell}\Big(\frac{1}{2}(\partial\Psi^I_\ell)^2 + V(\Psi^I_\ell)\Big)\bigg].
\enq
 This yields the equation of motion for $G_{\munu}$: \beq
R_{\munu}[G] = 2\(-2G_{\munu} + T_{\munu} - \frac{1}{3}G_{\munu}T\),
\enq
with \beq
T_{\munu} = \sum_{I,\ell}\(\partial_\mu\Psi^I_\ell \partial_\nu\Psi^I_\ell - G_{\munu}\Big(\frac{1}{2}(\partial\Psi_\ell^I)^2 + V(\Psi^I_\ell)\Big)\),
\enq
and the non-homogeneous Klein-Gordon equations (\ref{eqST}) for the scalar fields.

Let us remark again that if one considered the problem in full generality, the resulting action would involve an infinite tower of KK fields. However, to find 1-point correlation functions we only need to expand the 5-dimensional fields near the AdS boundary up to a certain order. This truncates the infinite KK tower to a finite number of fields. When applying this procedure to our solutions we will see that the only field that contributes to the dual scalar vev's and the dual stress tensor is the field $\mc{S}^2$. This amounts to having  the following non-zero potential in the action \eqref{N8sugra}: 
\beq
V(\mc{S}^2) = \frac{1}{2}m^2_{1,2}(\mc{S}^2)^2 - \frac{16}{3\sqrt{15}}(\mc{S}^2)^3.
\enq

\subsection{Holographic renormalization}
\label{HR}
Having a five-dimensional theory of gravity coupled to scalars in AdS allows us to apply the standard holographic renormalization prescription \cite{deHaro:2000hrec} to calculate the dual field theory observables. 

It is  convenient to introduce the Fefferman-Graham gauge for the $5D$ solution, 
\beq\label{FGexp}
\dd s^2_{5D} = \frac{1}{Z^2}(\dd Z^2 + G_{ij}(Z,X)\dd X^i \dd X^j),
\enq
where the $4$-dimensional AdS boundary $B_4$ is at $Z = 0$, and the additional coordinates $X^k$ run along $B_4$. Solving the bulk equations of motion order by order in a near boundary expansion yields to a schematic form  of the metric $G_{ij}(Z,X)$ and scalar field $\Psi^k(Z,X)$, where $k$ is the dimension of the dual operator, \beq\begin{split}
G_{ij}(Z,X) &= G^{(0)}_{ij}(X) + Z^2G_{ij}^{(2)}(X) + Z^4\(G_{ij}^{(4)}(X) + \log Z^2H_{ij}^{(4)}(X)\) + \dots, \\
\Psi^2(Z,X) &= Z^2\(\log Z^2\Psi_{(0)}^2(X) + \tilde{\Psi}^2_{(0)}(X) + \dots\), \\
\Psi^k(Z,X) &= Z^{4-k}\Psi^k_{(0)}(X) + \dots + Z^k\Psi^k_{(2k-4)} + \dots, \hs{0.75} (k>2).
\end{split}\enq
The leading terms $G_{ij}^{(0)}(X), \Psi_{(0)}^{2}$ and $\Psi_{(0)}^{k}$ parametrize the Dirichlet boundary conditions and  correspond to the dual sources for the CFT stress-energy tensor, and the operators of dimension 2 and $k$ respectively. This data, together with the normalizable modes $G^{(4)}_{ij}, \tilde{\Psi}_{(0)}^2$ and $\Psi^k_{(2k-4)}$,  
is sufficient to determine the series solution completely. Both pieces of data are related to the vev's of the 1-point correlators \cite{Skenderis:2006kkh,Bianchi:2001htg,Bianchi:2000hrec}: \beq\label{1pDS}
\la\mc{O}^2\ra = \frac{N^2}{2\pi^2}\big(2\tilde{\Psi}_{(0)}^2\big), \hs{0.75} \la\mc{O}^k\ra = \frac{N^2}{2\pi^2}\big((2k-4)\Psi_{(2k-4)}^k\big) + \text{lower},
\enq
and 
\beq\begin{split}\label{1pSE}
\la T_{ij}\ra &=  \frac{N^2}{2\pi^2}\bigg(G^{(4)}_{ij} + \frac{1}{8}\(\text{Tr}(G^{(2)})^2 - (\text{Tr}(G^{(2)}))^2\)G^{(0)}_{ij} - \frac{1}{2}(G_{(2)}^2)_{ij} + \frac{1}{4}G_{ij}^{(2)}\text{Tr}(G^{(2)}) \\
&\hs{1.28} + \frac{3}{2}H_{ij}^{(4)} + \frac{1}{3}(\tilde{\Psi}_{(0)}^2)^2G^{(0)}_{ij} + \Big(\frac{2}{3}\Psi_{(0)}^2-\tilde{\Psi}_{(0)}^2\Big)\Psi_{(0)}^2G_{ij}^{(0)} \bigg).
\end{split}\enq

In \eqref{1pDS}, `lower' indicates terms with index smaller than $2k-4$, and they only appear when the weight $k$ of the operator can be written as a sum of weights of other lower operators. Given that in our case the dual operator with smallest conformal dimension is 2, this non-linear term vanishes for $k = 3$ and it is proportional to $\la\mc{O}^2\ra^2$ for $k = 4$. In the latter case, the proportionality factor can be computed for our harmonic representation using the expressions in \cite{Skenderis:2006kkh}; see their equation \eqref{scalarVEVsLumpy}.

%%%%%%%%%%%%%%%%%%%%%%%%%%%%%%%%%%%%%%%%%%%%%%%
\section{Harmonic expansion, stress-energy tensor and scalar vev's}
\label{applyHolo}
%%%%%%%%%%%%%%%%%%%%%%%%%%%%%%%%%%%%%%%%%%%%%%%
In this section we perform the near boundary dimensional reduction for the class of spacetimes that we consider. We first  expand our solutions near the AdS boundary in spherical harmonics on the $S^5$ according to \eqref{expmetric} and \eqref{expform}. We  work with the compactified radial coordinate $y$ in our ansatz \eqref{ansatzlumpy}, and then transform to the FG gauge.

Our ansatz for the metric \eqref{ansatzlumpy} involves products of the variables $Q_2,Q_3$ and $Q_4$, which are not convenient to perform the decomposition in spherical harmonics. For instance, the $(aa)$-component  is not just the polar component of the $S^5$ times the unknown function $Q_4$, but it also includes a shift proportional to the product $Q_2Q_3^2$, coming from  the crossed $y-a$ term. The $Q_3^2$ factor in this term would introduce scalar-derived vector harmonic modes into the component $h_{aa}(y,a)$, which is not compatible with the decompositions in \eqref{expmetric}. We overcome this issue through a simple redefinition of variables: \beq
Q_2 = Q_2', \hs{0.75} Q_3 = \frac{G_p(y)}{4y^2y_0^2(1-y^2)}\frac{Q_3'}{Q_2'}, \hs{0.75} Q_4 = Q_4' - \frac{(1-y^2)(2-a^2)G_p(y)}{16y^2y_0^2}\frac{{Q_3'}^2}{Q_2'}\,,
\enq
so that the new variables $Q_2'$, $Q_3'$ and $Q_4'$ have a transparent decomposition in spherical harmonics. Then the spacetime metric reads 
\beq\begin{aligned}
\label{metricredef}
\dd s^2 =&~ \frac{L^2}{(1-y^2)^2}\bigg(-\frac{G_p(y)}{1-y^2}Q_1\dd t^2 + \frac{4y_0^2y^2(1-y^2)}{G_p(y)}Q_2'\dd y^2 - 2(1-y^2)^2Q_3'\dd y\dd a + y_0^2Q_5\dd\Omega_{(p-2)}^2\bigg)\\
&  + R^2\(\frac{4Q_4'}{2-a^2}\dd a^2 + (1-a^2)^2Q_6\dd\Omega_{(q-1)}^2\).
\end{aligned}\enq
Given the asymptotic boundary conditions that our solutions satisfy, the unknown functions appearing in the metric \eqref{metricredef} have the following near boundary expansion: 
\beq\begin{split}
\label{expansiongen}
Q_k(y,a) &= 1+\sum_{i=1}^\infty(1-y^2)^i\sum_{\ell=0}^i q_k(i)^\ell Y_\ell(a), \hs{0.75} (k=1,2,5,7), \\
Q_3'(y,a) &= \sum_{i=1}^\infty(1-y^2)^i\sum_{\ell=1}^{i+1} q_3(i)^\ell S_1^\ell(a), \\
Q_4'(y,a) &= 1 + \frac{2-a^2}{4}\sum_{i=2}^\infty(1-y^2)^i\sum_{\ell=2}^i q_{S}(i)^\ell S_{aa}^\ell(a) + \frac{1}{5}\sum_{i=1}^\infty(1-y^2)^i\sum_{\ell=0}^{i} q_{T}(i)^\ell Y_\ell(a), \\
Q_6(y,a) &= 1 + \frac{1}{(1-a^2)^2}\sum_{i=2}^\infty(1-y^2)^i\sum_{\ell=2}^i q_{S}(i)^\ell S_{\Omega\Omega}^\ell(a) + \frac{1}{5}\sum_{i=1}^\infty(1-y^2)^i\sum_{\ell=0}^i q_{T}(i)^\ell Y_\ell(a), \\
\end{split}\enq
where the various $q(i)^\ell$'s  are constants that can be determined by solving the equations of motion order by order near the AdS boundary ($y=1$). The factors in the expansions for $Q_4'$ and $Q_6$ have been chosen such that the internal part of the metric can be written as \beq
\dd s^2_{\text{int}} = g_{ab}^{(0)}\dd x^a\dd x^b +  h_{(ab)}(y,a)\dd x^a\dd x^b + \frac{1}{5}\,g_{ab}^{(0)}h(y,a)\, \dd x^a\dd x^b + \mc{O}(h^2), 
\enq
where $h_{(ab)}$ and $h$ are the symmetric-traceless and trace parts respectively. They have an expansion in harmonics as in  \eqref{expmetric}, where now the coefficients of the harmonics are functions of the compact radial coordinate $y$.

At each order in the near  boundary expansion only a finite number of harmonics are necessary to solve the equations of motion. This has already been taken into account in the form of the expansions: notice that at $i$th order in the near boundary expansion, the  sum in  \eqref{expansiongen} over the harmonics only runs up to the $\ell = i$ harmonic.\footnote{For $Q_3'$ this sum runs up to $i+1$; the reason for this is that in the metric ansatz this term appears multiplied by an extra factor of $1-y^2$.} Plugging the expansions \eqref{expansiongen} into the equations of motion yields, at each order and after projecting them into the spherical harmonics, algebraic relations between the different $q(i)^\ell$'s; at the end, only a few $q$'s remain undetermined. Up to order $(1-y)^4$ we obtain: 
\beq
\begin{split}
\label{Q1expanded}
Q_1(y,a) &\simeq 1 - \frac{4}{15}\beta_2Y_2(y-1)^2 - \frac{4}{15}\big(\beta_2 Y_2+30\gamma_3Y_3\big)(y-1)^3 \\
&\hs{0.45}+ \left(16 \delta_0 Y_0 - \frac{1}{15}\Big(1 - \frac{32}{5}\sqrt{\frac{2}{15}}\beta_2\Big)\beta_2Y_2-4\(3\gamma_3 Y_3-4\delta_4Y_4\)\right)(y-1)^4, \\
Q_2(y,a) &\simeq 1 - \frac{4}{15}\beta_2Y_2(y-1)^2 - \frac{4}{15}\big(\beta_2 Y_2+30\gamma_3Y_3\big)(y-1)^3 \\
&\hs{0.45}+ \bigg(\frac{1}{9}\beta_2^2Y_0 - \frac{1}{15}\Big(1 - \frac{96}{25}\sqrt{\frac{6}{5}}\beta_2\Big)\beta_2Y_2-12\gamma_3 Y_3 - 16\Big(\frac{2\sqrt{7}}{375}\beta_2^2 - \delta_4\Big)Y_4\bigg)(y-1)^4, \\
Q_3(y,a) &\simeq \frac{1}{15}\beta_2(S_a)^2(y-1)+\Big(\gamma_1(S_a)^1-\frac{1}{30}\beta_2 (S_a)^2+\gamma_3 (S_a)^3\Big)(y-1)^2 \\
&\hs{0.45}+\bigg(\Big(\frac{1}{30} + \frac{4}{15}\frac{1}{y_0^2} + \frac{4}{225}\beta_2Y_2 - \frac{56}{1125}\sqrt{\frac{2}{15}}\beta_2\Big)\beta_2(S_a)^2 + \frac{4}{3}\Big(\frac{\sqrt{7}}{375}\beta_2^2-\delta_4\Big)(S_a)^4\bigg)(y-1)^3,\\
Q_4(y,a) &\simeq 1+\frac{4}{5}\beta_2Y_2(y-1)^2 + 4\Big(\frac{2}{3}\gamma_1Y_1 + \frac{1}{5}\beta_2Y_2 + 4\gamma_3Y_3\Big)(y-1)^3 \\
&\hs{0.45}+\bigg(-\frac{3}{125}\beta_2^2Y_0 + 4\gamma_1Y_1 + \frac{1}{5}\Big(1 - \frac{992}{375}\sqrt{\frac{2}{15}}\beta_2\Big)\beta_2Y_2 + 24\gamma_3Y_3 + \Big(\frac{544}{5625}\beta_2^2 - \frac{80}{3}\delta_4\Big)Y_4 \\
&\hs{1.2}+ \frac{16}{375}\sqrt{\frac{2}{15}} \beta_2^2(S_{\phantom{\Omega}\Omega}^\Omega)^2 + \frac{8\sqrt{7}}{1125}\beta_2^2(S_{\phantom{\Omega}\Omega}^\Omega)^4 - \frac{16}{225}\beta_2^2(S^a)^2(S^a)^2 \bigg)(y-1)^4, \\
Q_5(y,a) &\simeq 1 - \frac{4}{15}\beta_2Y_2(y-1)^2 - \frac{4}{15}\big(\beta_2 Y_2+30\gamma_3Y_3\big)(y-1)^3 \\
&\hs{0.45}+\bigg(\frac{16}{3}\Big(\frac{1}{900}\beta_2^2-\delta_0\Big)Y_0 - \frac{1}{15}\Big(1-\frac{32}{5}\sqrt{\frac{2}{15}}\beta_2\Big)\beta_2Y_2 - 4\(3\gamma_3 Y_3-4\delta_4 Y_4\)\bigg)(y-1)^4, \\
Q_{6}(y,a) &\simeq 1+\frac{4}{5}\beta_2Y_2(y-1)^2 + 4\Big(\frac{2}{3}\gamma_1Y_1 + \frac{1}{5}\beta_2Y_2 + 4\gamma_3Y_3\Big)(y-1)^3 \\
&\hs{0.45}+\bigg(-\frac{3}{125}\beta_2^2Y_0 + 4\gamma_1 Y_1 + \frac{1}{5}\Big(1-\frac{992}{375}\sqrt{\frac{2}{15}}\beta_2\Big)\beta_2Y_2+ 24\gamma_3 Y_3 \\
&\hs{1.2}+ \Big(\frac{544\sqrt{7}}{5625}\beta_2^2 - \frac{80}{3}\delta_4\Big)Y_4 +\frac{16}{375}\sqrt{\frac{2}{15}} \beta_2^2(S_{\phantom{\Omega}\Omega}^\Omega)^2 + \frac{8\sqrt{7}}{1125}\beta_2^2(S_{\phantom{\Omega}\Omega}^\Omega)^4\bigg)(y-1)^4, \\
Q_7(y,a) &\simeq 1 - \frac{8}{15}\beta_2Y_2(y-1)^2 - \frac{8}{15}\big(\beta_2 Y_2+30\gamma_3Y_3\big)(y-1)^3 \\
&\hs{0.45}+\bigg(16\alpha_0Y_0-\frac{2}{15}\Big(1+\frac{16}{y_0^2}-\frac{512}{75}\sqrt{\frac{2}{15}}\beta_2\Big)\beta_2 Y_2 - 24\gamma_3Y_3 + 32\Big(\frac{\sqrt{7}}{1500}\beta_2^2+\delta_4\big)Y_4\bigg)(y-1)^4.
\end{split}
\enq
The harmonic expansion near the AdS boundary depends on six undetermined constants: $\beta_2$, $\gamma_1$, $\gamma_3$, $\delta_0$, $\delta_4$ and $\alpha_0$. Gauge freedom of both the metric and gauge field allows us to set $\gamma_1$ and $\alpha_0$ to any value and we choose to set them to zero. Note that even if they are not set to zero, observables are gauge invariant and therefore, at the end, they cannot depend on these two constants. The rest of coefficients, $\{\beta_2,\gamma_3,\delta_0,\delta_4\}$, correspond to data that can only be determined from the full bulk solution.

To bring the metric into the FG gauge $(z,\theta)$ in 10$D$ we perform the coordinate transformation $y=y(z,\theta)$, $a=a(z,\theta)$ in a series expansion near the AdS boundary $z = 0$. 
At each order in the FG coordinate $z^i$, we have two functions to solve for and the only requirement is that the AdS part of the metric takes the FG form, i.e.,~$g_{zz} = \frac{1}{z^2}$ (recall that $L = 1$) and $g_{zi} = 0$. This completely determines the form of the coordinate transformation. The first terms are given by: 
\beq\begin{split}
y(z,\theta) & = 1 + \frac{y_0}{2}z - \frac{y_0^2}{8}z^2 + \frac{1}{8}y_0\bigg(1+\frac{1}{2}y_0^2\Big(1+\frac{2}{15}\beta_2Y_2(\theta)\Big)\bigg)z^3 + \dots\\
a(z,\theta) &= \theta + \frac{1}{20}\sqrt{\frac{3}{10}}y_0^2\beta_2\,\theta(2-3\theta^2 + \theta^4)z^2 \\
&\hs{0.45}+ \frac{1}{12}y_0^3\sqrt{2-\theta^2}(1-\theta^2)\bigg(\gamma_1 - \frac{1}{\sqrt{10}}\gamma_3\Big(1-8\theta^2(2-\theta^2)\Big)\bigg)z^3 + \dots
\end{split}
\enq
To obtain the metric up to order $z^4$ we need this change of coordinates up to order $z^7$ for $y(z,\theta)$ and up to order $z^5$ for $a(z,\theta)$. 

Now we can identify the different fields appearing in \eqref{expmetric} and \eqref{expform}.\footnote{Note that because we work with gauge invariant variables, strictly speaking we should not need to put the 10$D$ solution in the FG gauge.} To this end, it is useful to write the background AdS$_5\times S^5$ in FG coordinates since we have to subtract it: \beq\begin{split}
\dd s^2_{(0)} &= \frac{L^2}{z^2}\[\dd z^2 - \Big(1 + \frac{z^2}{2L^2} + \frac{z^4}{16L^4}\Big)\dd t^2+\Big(1 - \frac{z^2}{2L^2} + \frac{z^4}{16L^4}\Big)L^2\dd\Omega_{(3)}\] + L^2\dd\Omega_{(5)}, 
\end{split}\enq
and $F_{(5)}^{(0)}$ is given in (\ref{gaugeSchw}) in a coordinate invariant form. We are now in a position to find the various harmonic contributions to the near boundary expansions of  the fields $h_{MN}$ and $f_{MNPQR}$. For the components of the metric along  the AdS$_5$ factor we find: 
\beq\begin{split}
h_{\munu}^{0}(z) &= -y_0^2\Big(\frac{3h_0}{4}(1+y_0^2)-y_0^2(\frac{h_1}{7200}\beta_2^2+h_0\delta_0)\Big) z^2\eta_{\munu}, \\
h_{\munu}^{2}(z) &= -\frac{1}{10}y_0^2\beta_2\bigg(1+\frac{3}{4}\Big(h_2-\frac{23}{75}\sqrt{\frac{2}{15}}y_0^2\beta_2\Big)z^2\bigg)\eta_{\munu}, \\
h_{\munu}^{3}(z) &= -\frac{4}{3}y_0^3\gamma_3z\eta_{\munu}, \\
h_{\munu}^{4}(z) &= \frac{1}{4}y_0^4\bigg(\frac{13\sqrt{7}}{1500}\beta_2^2 + 5\delta_4\bigg)z^2\eta_{\munu},
\end{split}\enq
with $\{h_0, h_1,h_2\} = \{-1/3,11/3,1/9\}$ when the indices are along the $S^3$ within the AdS$_5$ space, otherwise these constants are all equal to one; $\eta_{\mu\nu}= \text{diag}(0,-1,\sigma_{\h{\imath}\h{\jmath}})$, where $\sigma_{\h{\imath}\h{\jmath}}$ is the standard metric on the unit round $S^3$. Since, by construction, in the FG gauge there are no crossed terms between the AdS$_5$ and $S^5$ factors, we have $A_\mu^\ell(z) = 0$. The scalar fields (from the point of view of AdS$_5$) that come from the components of the 10$D$ metric along the $S^5$ are given by:
\beq\begin{split}\label{BsandCs}
B^2(z) &= \frac{1}{20}y_0^2\beta_2\bigg(1+\frac{5}{8}\Big(1-\frac{1}{25}\sqrt{\frac{2}{15}}y_0^2\beta_2\Big)z^2\bigg)z^2, \hs{1} B^3(z) = \frac{2}{9}y_0^3\gamma_3z^3, \\
B^4(z) &= \frac{1}{48}y_0^4\Big(\frac{\sqrt{7}}{300}\beta_2^2-5\delta_4\Big)z^4, \\
C^0(z) &= \frac{1}{600}y_0^4\beta_2^2z^4, \hs{1.25} C^2(z) = \frac{2}{5}y_0^2\beta_2\bigg(1+\frac{5}{16}\Big(1-\frac{17}{25}\sqrt{\frac{2}{15}}y_0^2\beta_2\Big)z^2\bigg)z^2, \\
C^3(z) &= \frac{16}{3}y_0^3\gamma_3z^3, \hs{1.52} C^4(z) = \frac{1}{300}y_0^4\(\sqrt{7}\beta_2^2-5\delta_4\)z^4.
\end{split}\enq
Finally, the 5-dimensional scalar fields arising from the 5-form are: 
\beq\begin{split}
b^2 &= -\frac{1}{10}y_0^2\beta_2\bigg(1+\frac{3}{8}\Big(1-\frac{31}{75}\sqrt{\frac{2}{15}}y_0^2\beta_2\Big)z^2\bigg)z^2, \hs{1.1} b^3 = -\frac{8}{9}y_0^3\gamma_3z^3, \\
b^4 &= -\frac{1}{8}y_0^4\Big(\frac{7\sqrt{7}}{750}\beta_2^2-5\delta_4\Big)z^4.
\end{split}\enq

From these expressions we can find the gauge invariant combinations at linear order from Table \ref{tableLin}, and then the corresponding gauge invariant mass eigenfields defined in \eqref{massfields}. To leading order in the number of fields, the equations of motion are satisfied up to order $z^2$, since higher order terms  in $z$ can (and will!) receive contributions from the non-linear interactions. 
The leading order field $\h{s}^2$ is given by
\beq
\h{s}^2 = \frac{1}{20}y_0^2\beta_2 z^2 + \mathcal{O}(z^4). 
\enq
The rest of the fields $\h{s}^\ell$, $\ell = 3,4$ are at least $\mc{O}(z^3)$ and $\h{t}^\ell$, $\ell = 0,1,2,3,4$, are $\mc{O}(z^4)$. We do not present the explicit expressions for them in this appendix, but they can be non-trivial at this order and  they are needed to find the dimensionally reduced scalar fields $\mc{S}^\ell$ and $\mc{T}^\ell$.

The relevant fields to the second order are $\h{\h{s}}^2$ and $\bar{\bar{h}}^0_{\munu}$. Using (\ref{2fieldsSimp}) and (\ref{hmunu02ndOrd}) we find: 
\beq\begin{split}
\h{\h{s}}^2 &= \frac{1}{20}y_0^2\beta_2 z^2 + \frac{1}{40}y_0^2\beta_2\bigg(1 - \frac{1}{25}\sqrt{\frac{2}{15}}y_0^2\beta_2\bigg)z^4,\\
\bar{\bar{h}}^0_{zz} &= -\frac{11}{3600} y_0^4\beta_2^2 z^2\,,\\
\bar{\bar{h}}^0_{ij} &= \(\frac{3}{4}h_0' y_0^2\(1+y_0^2\) - \frac{1}{960}h_1' y_0^4\beta_2^2  - h_0' y_0^4 \delta_0\) z^2 \delta_{ij}\,,
\end{split}
\enq
where $\delta_{ij} = \text{diag}(1,\sigma_{\h{\imath}\h{\jmath}})$, and $\{h_0',h_1'\} = \{1/3,-61/45\}$ for indices along the $S^3$ and one otherwise. We are interested in the scalar fields $\mc{S}^\ell$, $\mc{T}^\ell$ given in (\ref{mappingggg}) and the massless KK graviton $G_{\munu}$ (\ref{mappinggggG}). These are 5$D$ fields involving only quadratic contributions in $\h{s}^2$ and its derivatives. After adding the background in the case of the metric, we find: 
\beq\begin{split}
\mc{S}^2 &= \frac{1}{15\sqrt{2}}y_0^2\beta_2z^2 + \frac{1}{30\sqrt{2}}y_0^2\beta_2\bigg(1 - \frac{4}{15}\sqrt{\frac{2}{15}}y_0^2\beta_2\bigg)z^4, \\
G_{zz} &= \frac{1}{z^2} - \frac{13}{10800}y_0^4\beta_2^2 z^2, \\
G_{ij} &= \(\frac{1}{z^2}+\frac{G_0}{2} + \Big(\frac{1}{16} - \frac{3G_0}{4}y_0^2(1+y_0^2)\Big)z^2 - y_0^4\Big(\frac{31G_1}{43200}\beta_2^2 - G_2\delta_0\Big)z^2\)\eta_{ij},
\end{split}
\label{5dfields1}
\enq
with $\eta_{ij} = \text{diag}(-1,\sigma_{\h{\imath}\h{\jmath}})$, $(G_0,G_1,G_2) = (-1,15/31,-1/3)$ for the components along the $S^3$ and  one otherwise. The fields $\mc{T}^\ell$, $\ell = 0,1,2,3,4$ are $\mc{O}(z^5)$. Therefore, up to order $z^4$, there are only two other scalars which can acquire non-trivial vev's: 
\beq
\begin{split}
\mc{S}^3 &= \frac{1}{2\sqrt{3}}y_0^3\gamma_3 z^3,\\
\mc{S}^4 &= -\frac{1}{1200\sqrt{3}}y_0^4(\sqrt{7}\beta_2^2+300\delta_4)z^4.
\end{split}
\enq

Having the 5$D$ fields, we are now in a position to apply the standard holographic renormalization prescription (see \S\ref{HR}). Clearly, the metric in \eqref{5dfields1} is not in FG gauge;  introducing the 5$D$ FG coordinate $Z$ through
\beq
z(Z) = Z+\frac{13}{86400}y_0^4\beta_2^2 Z^5 + \mc{O}(Z^6),
\enq
we can bring it to  the canonical FG form. Finally, identifying the different terms that appear in the FG expansion, we obtain the stress-energy tensor from \eqref{1pSE}: 
\beq\label{thefinalTmunu}
\la T_{ij}\ra = \frac{N^2}{2\pi^2}\bigg(\frac{3}{16}(1+2y_0^2)^2 + y_0^4\Big(\frac{\beta_2^2}{3600}  - \delta_0\Big)\bigg)(1,\tfrac{1}{3}\sigma_{\h{\imath}\h{\jmath}}).
\enq
The boundary geometry $B_4$ is conformally flat and therefore there is no conformal anomaly: the stress-energy tensor is traceless $\la T_i^i\ra = 0$. It is also trivially conserved, $D^i_{(0)}\la T_{ij}\ra = 0$, and in the limit $\beta_2,\delta_0 \riga 0$, it reduces to the stress-energy tensor of the Schwarzschild-AdS black hole, as expected. The energy of the spacetime is simply given by $E = \Omega_3\la T_{00}\ra$, which includes the well-known Casimir energy of the dual $\mc{N} = 4$ SYM on the ESU$_4$. Using the expressions in (\ref{Q1expanded}) one has the relation \beq\label{betad}
\frac{\beta_2^2}{3600}  - \delta_0 = \frac{1}{512}\(\partial_y^4Q_5 - \partial_y^4Q_1\)\Big|_{y = 1},
\enq
which is very useful in order to check the numerics.

Recall that the conformal dimension of operators dual to $\mc{S}^\ell$ is $\Delta_+ = \ell$ ($\Delta_- = 4-\ell$); thus their vev's in our case are given by: \beq\begin{split}\label{scalarVEVsLumpy}
\la \mc{O}_{\mc{S}^2}\ra &= \frac{N^2}{2\pi^2}(2\tilde{S}^2_{(0)}) = \frac{N^2}{2\pi^2}\frac{\sqrt{2}}{15}y_0^2\beta_2, \\
\la \mc{O}_{\mc{S}^3}\ra &= \frac{N^2}{2\pi^2}(2S^3_{(2)}) = \frac{N^2}{2\pi^2}\frac{1}{\sqrt{3}}y_0^3\gamma_3, \\
\la \mc{O}_{\mc{S}^4}\ra &= \frac{N^2}{2\pi^2}\Big(4S^4_{(4)}+ \frac{3\sqrt{21}}{5}(2\tilde{S}^2_{(0)})^2 \Big) = \frac{N^2}{2\pi^2}\frac{1}{\sqrt{3}}y_0^4\Big(\frac{19\sqrt{7}}{1500}\beta_2^2-\delta_4\Big).
\end{split}\enq

%%%%%%%%%%%%%%%%%%%%%%%%%%%%%%%%%%%%%%%%%%%%%%%
\section{Scalar harmonics of S$^{\text{q}}$ with SO(q) symmetry}
\label{harm}
%%%%%%%%%%%%%%%%%%%%%%%%%%%%%%%%%%%%%%%%%%%%%%%
In this appendix we construct the scalar spherical harmonics on $S^q$ that preserve an $S^{q-1}$. These are necessary to perform the dimensional reduction of our solutions and compute the vev's of the dual operators.

The metric of $S^{q}$ in the standard spherical coordinates can be found recursively through: \beq\begin{split}
\dd \Omega_1^2 &= \dd\theta_1^2, \\
\dd \Omega_{q}^2 &= \dd\theta_{q} + \sin^2\theta_{q}\dd\Omega_{q-1}^2, \\
\end{split}\enq
with $\theta_1\in[0,2\pi)$ and $\theta_{k} \in [0,\pi)$, $\forall k = 2,\dots,q$. The corresponding volume forms are: \beq\begin{split}
\dd\omega_1 &= \dd\theta_1, \\
\dd\omega_{q} &= \sin^{q-1}\theta_{q}\dd\theta_{q}\wedge\dd\omega_{q-1},
\end{split}\enq
and the Laplace operators are: \beq\begin{split}
\Delta_1 &= \parcial[^2]{\theta_1^2}, \\
\Delta_{q} &= \frac{1}{\sin^{q-1}\theta_{q}}\parcial[]{\theta_{q}}\(\sin^{q-1}\theta_{q}\parcial[]{\theta_{q}}\) + \frac{1}{\sin^2\theta_{q}}\Delta_{q-1}.
\end{split}\enq

The defining equations for the scalar spherical harmonics are: \beq\begin{split}\label{harmeq}
\(\Delta_1+\ell_1^2\) Y_{\ell_1}(\theta_1) &= 0, \\
\(\Delta_{q}-\Lambda_{\ell_q}\) Y_{\ell_1\dots\ell_{q}}(\theta_1,\dots,\theta_{q}) &= 0,
\end{split}\enq
with $\Lambda_{\ell_n} = -\ell_n(\ell_n+n-1)$, $n = 1,\dots,q$ and $\ell_i = 0,1,2,\dots$. Scalar harmonics in one dimension (on a circle), are given by the Fourier modes: $Y_{\ell_1}(\theta_1) \propto e^{i\ell_1\theta_1}$ (up to a normalization constant). Requiring univaluedness  implies that $\ell_1\in\mathbb{Z}$.

We focus on harmonics that preserve an SO$(q)$ subgroup out of the full SO$(q+1)$ isometry group of the $S^q$. 
Imposing SO$(q)$ symmetry implies that these spherical harmonics can only depend on the polar angle $\theta_{q} \equiv \theta$, and thus they are labeled by a single quantum number $\ell_{q} \equiv \ell$. Therefore, the scalar harmonics that we are interested in satisfy: 
\beq
\(\frac{1}{\sin^{q-1}\theta}\parcial[]{\theta}\(\sin^{q-1}\theta\parcial[]{\theta}\)-\Lambda_{\ell}\) Y_{\ell}(\theta) = 0.
\label{harmonicsSOq}
\enq
Let $Y_{\ell}(\theta) = y(\cos\theta)$ and consider the change of variables $\theta = \arccos(x)$. With the appropriate definitions $2\mu = q-1$ and $\nu = \ell$,  equation \eqref{harmonicsSOq} becomes the Gegenbauer differential equation: 
\beq
(1-x^2)y''(x) - (2\mu+1)xy'(x) + \nu(\nu+2\mu)y(x) = 0.
\enq
The general solution to this equation is \beq
y(x) = (x^2-1)^{(1-2\mu)/4}\(C_1\,P^{1/2-\mu}_{\mu+\nu - 1/2}(x) + C_2\,Q^{1/2-\mu}_{\mu+\nu - 1/2}(x)\),
\enq 
where $C_1$ and $C_2$ are integration constants, and  $P^a_b(x)$ and $Q^a_b(x)$ are the Legendre's polynomials of first and second kind respectively. Finiteness at the poles, $x = \pm1$, requires that $C_2 = 0$, and $\nu$ to be a non-negative integer, i.e.,\ $\ell \geq 0$.

The solution may be written in terms of hypergeometric functions. 
Redefining the integration constant by absorbing the numerical factors that may arise, we can write the regular solution as: \beq
y(x) = \bar{C}_1\,_2F_1\(-\nu,2\mu+\nu;\mu + \frac{1}{2};\frac{1-x}{2}\)\,
\enq
where $\bar{C}_1$ is determined by requiring a proper normalization. To be consistent, we use the same normalization as in the previous literature \cite{Arutyunov:1999scc,Sangmin:19983pf,Lee:20044pf,Kim:1985spe}. The desired scalar harmonics are then given by: \beq\begin{split}
Y_{\ell}(\theta) &= \sqrt{\frac{2^{-\ell-q+2}\pi^{1/2}\Gamma(\ell+q-1)}{\Gamma(\ell+\frac{q-1}{2})\Gamma(q/2)}}\;{}_2F_1\(-\ell,\ell+q-1;\frac{q}{2};\frac{1-\cos\theta}{2}\),
\end{split}\enq
satisfying \beq\label{normalYY}
\int Y_{\ell_1}Y_{\ell_2}\dd\omega_{q} = z(\ell_1)\delta_{\ell_1\ell_2}, \hs{0.75} z(\ell_1) = \frac{\ell_1!(\frac{q-1}{2})!}{2^{\ell_1}(\ell_1+\frac{q-1}{2})!}\Omega_q,
\enq
where $\Omega_q$ is the surface area of the unit $q$-sphere. This normalization ensures that totally symmetric traceless rank $k$  tensors of SO($q+1$), which can be used to represent the scalar harmonics, are normalized to a delta function.

From the scalar harmonics one may define scalar-derived quantities such as vector and tensor harmonics by taking  covariant derivatives on the $S^q$. Letting the metric on the $S^q$ be $\sigma_{ab}$, with compatible covariant derivative $D_a$, the scalar-derived vector and tensor harmonics are given by 
\beq\begin{split}
S_a^{\ell}(\theta) &= D_aY_{\ell}, \\
S_{ab}^{\ell}(\theta) &\equiv D_{(a}D_{b)}Y_{\ell} = D_{a}D_{b}Y_{\ell} - \frac{\Lambda_{\ell}}{q}\sigma_{ab}Y_{\ell}.
\end{split}\enq
With these definitions, these harmonics satisfy: 
\beq\begin{split}
D^aS_a^{\ell} &= \Lambda_{\ell}Y_{\ell}, \hs{4.04} D^aS_{ab}^{\ell} = (q-1)\(1+\frac{\Lambda_{\ell}}{q}\)S_a^{\ell}, \\
D^bD_b S_a^{\ell} &= \(\Lambda_{\ell}+q-1\)S_a^{\ell}, \hs{1.71} D^cD_cS_{ab}^{\ell} = \(\Lambda_{\ell}+2q\)S_{ab}^{\ell}, \\
 \int S_a^{\ell_1}S^{a}_{\ell_2}\,\dd\omega_q &= z(\ell_1)f(\ell_1)\delta^{\ell_1}_{\ell_2}, \hs{1.33} \int S_{ab}^{\ell_1}S^{ab}_{\ell_2}\,\dd\omega_q = z(\ell_1)q(\ell_1)\delta^{\ell_1}_{\ell_2}, \\
\end{split}\enq
where $z(\ell)$ has been defined in (\ref{normalYY}) and \beq
f(\ell) = -\Lambda_\ell, \hs{0.75} q(\ell) = \Lambda_\ell(q-1)\(1+\frac{1}{q}\Lambda_\ell\).
\enq

In order to find the gauge transformations of the lower-dimensional fields in Appendix \ref{app:KKhol}, one also needs the following triple integrals of spherical harmonics: 
\beq\begin{split}
\label{abcde}
a(\ell_1,\ell_2,\ell_3) &= \int Y_{\ell_1} Y_{\ell_2} Y_{\ell_3}\,\dd\omega_q\, \hs{0.9} b(\ell_1,\ell_2,\ell_3) = \int Y_{\ell_1}S_a^{\ell_2} S^{a}_{\ell_3}\,\dd\omega_q,\\
c(\ell_1,\ell_2,\ell_3) &= \int S^{ab}_{\ell_1}S_a^{\ell_2}S_b^{\ell_3}\,\dd\omega_q, \hs{0.75} d(\ell_1,\ell_2,\ell_3) = \int Y_{\ell_1}S^{ab}_{\ell_2}D_aS_b^{\ell_3}\,\dd\omega_q, \\
e(\ell_1,\ell_2,\ell_3) &= \int S^{ab}_{\ell_1}\(2D_a S^{c}_{\ell_2}S_{cb}^{\ell_3}+S^c_{\ell_2}D_cS_{ab}^{\ell_3}\)\,\dd\omega_q.
\end{split}
\enq
The expressions for $b(\ell_1,\ell_2,\ell_3), c(\ell_1,\ell_2,\ell_3), d(\ell_1,\ell_2,\ell_3)$ and $e(\ell_1,\ell_2,\ell_3)$ can be written in terms of $a(\ell_1,\ell_2,\ell_3)$ by partial integrations, e.g., 
\beq
b(\ell_1,\ell_2,\ell_3) = \frac{1}{2}\Big(f(\ell_2)+f(\ell_3)-f(\ell_1)\Big)a(\ell_1,\ell_2,\ell_3).
\enq

%%%%%%%%%%%%%%%%%%%%%%%%%%%%%%%%%%%%%%%%%%%%%%%
\section{Embedding plots in AdS$_\text{5}\times$S$^{\text{5}}$}
\label{appEmb10d}
%%%%%%%%%%%%%%%%%%%%%%%%%%%%%%%%%%%%%%%%%%%%%%%
In  Fig.~\ref{fig:12AdS5} of this Appendix we include the embedding diagrams  for the lumpy black hole families $\ell= 1,2^+$ and $2^-$ in AdS$_5\times S^5$. These figures help to visualize the topology changing transition, in which the lumpy black holes  merge with another family of  black holes with a different horizon topology. In the AdS$_5\times S^5$ case, the values of lumpiness parameter $\lambda$ that we were able to confidently achieve are smaller than in the AdS$_4\times S^7$  case, but yet significantly larger than in  \cite{Dias:2015pda}. The topology changes are qualitatively similar to the corresponding ones in  AdS$_4\times S^7$,  except for the $\ell=2^-$ case; in the AdS$_5\times S^5$ case, the lumpy black holes merge with a black hole family whose horizon topology is $S^8\# S^8$, i.e.,~a double localized black hole.

\newpage

\begin{center}
\begin{center}
{\bf \textsf{$\ell \text{ = 1}$}}
\end{center}
\begin{minipage}{\textwidth}
\begin{center}
\hs{-1}% GNUPLOT: LaTeX picture with Postscript
\begingroup
  \makeatletter
  \providecommand\color[2][]{%
    \GenericError{(gnuplot) \space\space\space\@spaces}{%
      Package color not loaded in conjunction with
      terminal option `colourtext'%
    }{See the gnuplot documentation for explanation.%
    }{Either use 'blacktext' in gnuplot or load the package
      color.sty in LaTeX.}%
    \renewcommand\color[2][]{}%
  }%
  \providecommand\includegraphics[2][]{%
    \GenericError{(gnuplot) \space\space\space\@spaces}{%
      Package graphicx or graphics not loaded%
    }{See the gnuplot documentation for explanation.%
    }{The gnuplot epslatex terminal needs graphicx.sty or graphics.sty.}%
    \renewcommand\includegraphics[2][]{}%
  }%
  \providecommand\rotatebox[2]{#2}%
  \@ifundefined{ifGPcolor}{%
    \newif\ifGPcolor
    \GPcolortrue
  }{}%
  \@ifundefined{ifGPblacktext}{%
    \newif\ifGPblacktext
    \GPblacktexttrue
  }{}%
  % define a \g@addto@macro without @ in the name:
  \let\gplgaddtomacro\g@addto@macro
  % define empty templates for all commands taking text:
  \gdef\gplbacktext{}%
  \gdef\gplfronttext{}%
  \makeatother
  \ifGPblacktext
    % no textcolor at all
    \def\colorrgb#1{}%
    \def\colorgray#1{}%
  \else
    % gray or color?
    \ifGPcolor
      \def\colorrgb#1{\color[rgb]{#1}}%
      \def\colorgray#1{\color[gray]{#1}}%
      \expandafter\def\csname LTw\endcsname{\color{white}}%
      \expandafter\def\csname LTb\endcsname{\color{black}}%
      \expandafter\def\csname LTa\endcsname{\color{black}}%
      \expandafter\def\csname LT0\endcsname{\color[rgb]{1,0,0}}%
      \expandafter\def\csname LT1\endcsname{\color[rgb]{0,1,0}}%
      \expandafter\def\csname LT2\endcsname{\color[rgb]{0,0,1}}%
      \expandafter\def\csname LT3\endcsname{\color[rgb]{1,0,1}}%
      \expandafter\def\csname LT4\endcsname{\color[rgb]{0,1,1}}%
      \expandafter\def\csname LT5\endcsname{\color[rgb]{1,1,0}}%
      \expandafter\def\csname LT6\endcsname{\color[rgb]{0,0,0}}%
      \expandafter\def\csname LT7\endcsname{\color[rgb]{1,0.3,0}}%
      \expandafter\def\csname LT8\endcsname{\color[rgb]{0.5,0.5,0.5}}%
    \else
      % gray
      \def\colorrgb#1{\color{black}}%
      \def\colorgray#1{\color[gray]{#1}}%
      \expandafter\def\csname LTw\endcsname{\color{white}}%
      \expandafter\def\csname LTb\endcsname{\color{black}}%
      \expandafter\def\csname LTa\endcsname{\color{black}}%
      \expandafter\def\csname LT0\endcsname{\color{black}}%
      \expandafter\def\csname LT1\endcsname{\color{black}}%
      \expandafter\def\csname LT2\endcsname{\color{black}}%
      \expandafter\def\csname LT3\endcsname{\color{black}}%
      \expandafter\def\csname LT4\endcsname{\color{black}}%
      \expandafter\def\csname LT5\endcsname{\color{black}}%
      \expandafter\def\csname LT6\endcsname{\color{black}}%
      \expandafter\def\csname LT7\endcsname{\color{black}}%
      \expandafter\def\csname LT8\endcsname{\color{black}}%
    \fi
  \fi
    \setlength{\unitlength}{0.0500bp}%
    \ifx\gptboxheight\undefined%
      \newlength{\gptboxheight}%
      \newlength{\gptboxwidth}%
      \newsavebox{\gptboxtext}%
    \fi%
    \setlength{\fboxrule}{0.5pt}%
    \setlength{\fboxsep}{1pt}%
\begin{picture}(9920.00,3258.00)%
    \gplgaddtomacro\gplbacktext{%
      \csname LTb\endcsname%%
      \put(1067,704){\makebox(0,0)[r]{\strut{}$-1.5$}}%
      \csname LTb\endcsname%%
      \put(1067,1335){\makebox(0,0)[r]{\strut{}$-0.5$}}%
      \csname LTb\endcsname%%
      \put(1067,1966){\makebox(0,0)[r]{\strut{}$0.5$}}%
      \csname LTb\endcsname%%
      \put(1067,2597){\makebox(0,0)[r]{\strut{}$1.5$}}%
      \csname LTb\endcsname%%
      \put(1515,484){\makebox(0,0){\strut{}$-6$}}%
      \csname LTb\endcsname%%
      \put(2777,484){\makebox(0,0){\strut{}$-4$}}%
      \csname LTb\endcsname%%
      \put(4039,484){\makebox(0,0){\strut{}$-2$}}%
      \csname LTb\endcsname%%
      \put(5301,484){\makebox(0,0){\strut{}$0$}}%
      \csname LTb\endcsname%%
      \put(6563,484){\makebox(0,0){\strut{}$2$}}%
      \csname LTb\endcsname%%
      \put(7825,484){\makebox(0,0){\strut{}$4$}}%
      \csname LTb\endcsname%%
      \put(9087,484){\makebox(0,0){\strut{}$6$}}%
    }%
    \gplgaddtomacro\gplfronttext{%
      \csname LTb\endcsname%%
      \put(319,1650){\rotatebox{-270}{\makebox(0,0){\strut{}$X/L$}}}%
      \put(5300,154){\makebox(0,0){\strut{}$Y/L$}}%
      \put(5300,2927){\makebox(0,0){\textsf{Fixed S$^{\text{3}}$ coordinates}}}%
    }%
    \gplbacktext
    \put(0,0){\includegraphics{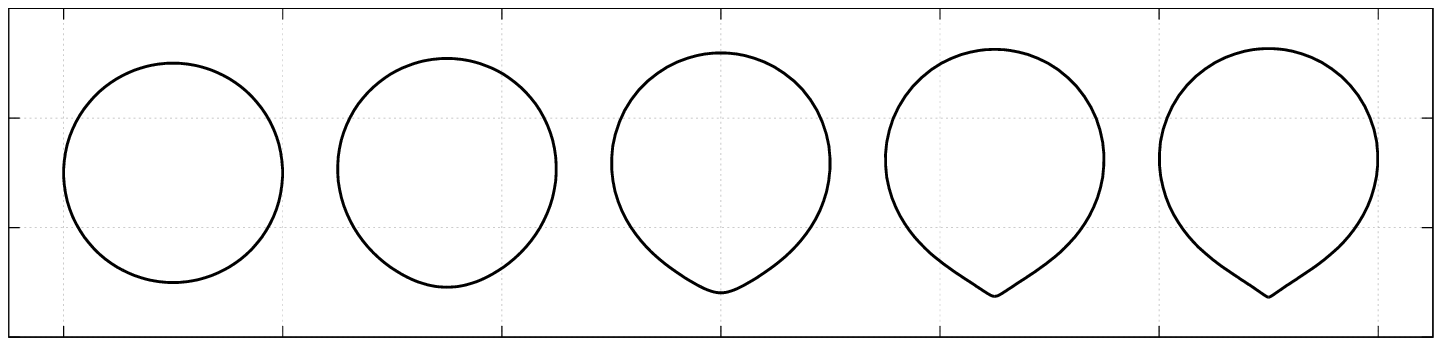}}%
    \gplfronttext
  \end{picture}%
\endgroup

\hs{-0.75}% GNUPLOT: LaTeX picture with Postscript
\begingroup
  \makeatletter
  \providecommand\color[2][]{%
    \GenericError{(gnuplot) \space\space\space\@spaces}{%
      Package color not loaded in conjunction with
      terminal option `colourtext'%
    }{See the gnuplot documentation for explanation.%
    }{Either use 'blacktext' in gnuplot or load the package
      color.sty in LaTeX.}%
    \renewcommand\color[2][]{}%
  }%
  \providecommand\includegraphics[2][]{%
    \GenericError{(gnuplot) \space\space\space\@spaces}{%
      Package graphicx or graphics not loaded%
    }{See the gnuplot documentation for explanation.%
    }{The gnuplot epslatex terminal needs graphicx.sty or graphics.sty.}%
    \renewcommand\includegraphics[2][]{}%
  }%
  \providecommand\rotatebox[2]{#2}%
  \@ifundefined{ifGPcolor}{%
    \newif\ifGPcolor
    \GPcolortrue
  }{}%
  \@ifundefined{ifGPblacktext}{%
    \newif\ifGPblacktext
    \GPblacktexttrue
  }{}%
  % define a \g@addto@macro without @ in the name:
  \let\gplgaddtomacro\g@addto@macro
  % define empty templates for all commands taking text:
  \gdef\gplbacktext{}%
  \gdef\gplfronttext{}%
  \makeatother
  \ifGPblacktext
    % no textcolor at all
    \def\colorrgb#1{}%
    \def\colorgray#1{}%
  \else
    % gray or color?
    \ifGPcolor
      \def\colorrgb#1{\color[rgb]{#1}}%
      \def\colorgray#1{\color[gray]{#1}}%
      \expandafter\def\csname LTw\endcsname{\color{white}}%
      \expandafter\def\csname LTb\endcsname{\color{black}}%
      \expandafter\def\csname LTa\endcsname{\color{black}}%
      \expandafter\def\csname LT0\endcsname{\color[rgb]{1,0,0}}%
      \expandafter\def\csname LT1\endcsname{\color[rgb]{0,1,0}}%
      \expandafter\def\csname LT2\endcsname{\color[rgb]{0,0,1}}%
      \expandafter\def\csname LT3\endcsname{\color[rgb]{1,0,1}}%
      \expandafter\def\csname LT4\endcsname{\color[rgb]{0,1,1}}%
      \expandafter\def\csname LT5\endcsname{\color[rgb]{1,1,0}}%
      \expandafter\def\csname LT6\endcsname{\color[rgb]{0,0,0}}%
      \expandafter\def\csname LT7\endcsname{\color[rgb]{1,0.3,0}}%
      \expandafter\def\csname LT8\endcsname{\color[rgb]{0.5,0.5,0.5}}%
    \else
      % gray
      \def\colorrgb#1{\color{black}}%
      \def\colorgray#1{\color[gray]{#1}}%
      \expandafter\def\csname LTw\endcsname{\color{white}}%
      \expandafter\def\csname LTb\endcsname{\color{black}}%
      \expandafter\def\csname LTa\endcsname{\color{black}}%
      \expandafter\def\csname LT0\endcsname{\color{black}}%
      \expandafter\def\csname LT1\endcsname{\color{black}}%
      \expandafter\def\csname LT2\endcsname{\color{black}}%
      \expandafter\def\csname LT3\endcsname{\color{black}}%
      \expandafter\def\csname LT4\endcsname{\color{black}}%
      \expandafter\def\csname LT5\endcsname{\color{black}}%
      \expandafter\def\csname LT6\endcsname{\color{black}}%
      \expandafter\def\csname LT7\endcsname{\color{black}}%
      \expandafter\def\csname LT8\endcsname{\color{black}}%
    \fi
  \fi
    \setlength{\unitlength}{0.0500bp}%
    \ifx\gptboxheight\undefined%
      \newlength{\gptboxheight}%
      \newlength{\gptboxwidth}%
      \newsavebox{\gptboxtext}%
    \fi%
    \setlength{\fboxrule}{0.5pt}%
    \setlength{\fboxsep}{1pt}%
\begin{picture}(10204.00,3400.00)%
    \gplgaddtomacro\gplbacktext{%
      \csname LTb\endcsname%%
      \put(1495,704){\makebox(0,0)[r]{\strut{}$-2.5$}}%
      \csname LTb\endcsname%%
      \put(1495,1111){\makebox(0,0)[r]{\strut{}$-1.5$}}%
      \csname LTb\endcsname%%
      \put(1495,1518){\makebox(0,0)[r]{\strut{}$-0.5$}}%
      \csname LTb\endcsname%%
      \put(1495,1925){\makebox(0,0)[r]{\strut{}$0.5$}}%
      \csname LTb\endcsname%%
      \put(1495,2332){\makebox(0,0)[r]{\strut{}$1.5$}}%
      \csname LTb\endcsname%%
      \put(1495,2739){\makebox(0,0)[r]{\strut{}$2.5$}}%
      \csname LTb\endcsname%%
      \put(2390,484){\makebox(0,0){\strut{}$-3$}}%
      \csname LTb\endcsname%%
      \put(3408,484){\makebox(0,0){\strut{}$-2$}}%
      \csname LTb\endcsname%%
      \put(4425,484){\makebox(0,0){\strut{}$-1$}}%
      \csname LTb\endcsname%%
      \put(5443,484){\makebox(0,0){\strut{}$0$}}%
      \csname LTb\endcsname%%
      \put(6460,484){\makebox(0,0){\strut{}$1$}}%
      \csname LTb\endcsname%%
      \put(7477,484){\makebox(0,0){\strut{}$2$}}%
      \csname LTb\endcsname%%
      \put(8495,484){\makebox(0,0){\strut{}$3$}}%
    }%
    \gplgaddtomacro\gplfronttext{%
      \csname LTb\endcsname%%
      \put(747,1721){\rotatebox{-270}{\makebox(0,0){\strut{}$X/L$}}}%
      \put(5442,154){\makebox(0,0){\strut{}$Y/L$}}%
      \put(5442,3069){\makebox(0,0){\textsf{Fixed S$^\text{4}$ coordinates}}}%
    }%
    \gplbacktext
    \put(0,0){\includegraphics{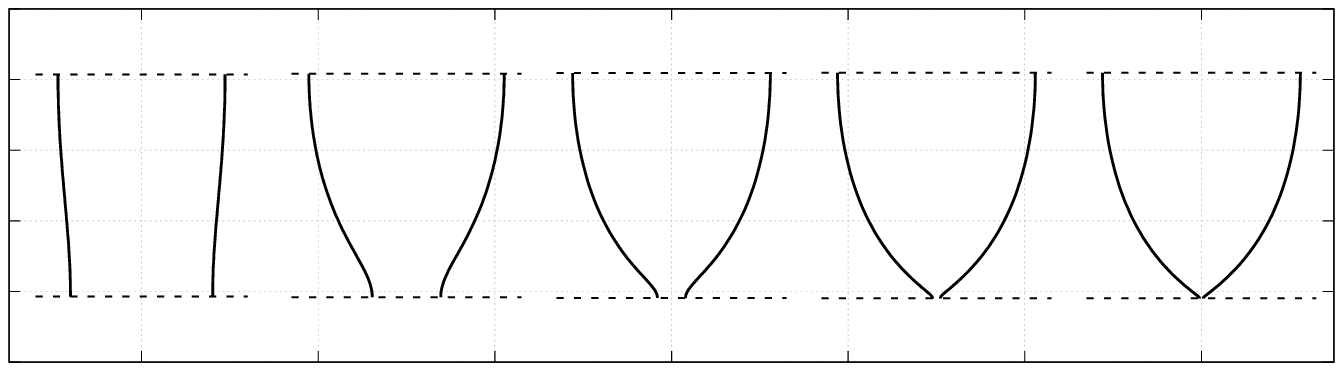}}%
    \gplfronttext
  \end{picture}%
\endgroup

\end{center}
\end{minipage}
\end{center}

\begin{center}
\begin{center}
{\bf \textsf{$\ell \text{ = 2$^\mathbf{+}$}$}}
\end{center}
\begin{minipage}{\textwidth}
\begin{center}
\hs{-1}% GNUPLOT: LaTeX picture with Postscript
\begingroup
  \makeatletter
  \providecommand\color[2][]{%
    \GenericError{(gnuplot) \space\space\space\@spaces}{%
      Package color not loaded in conjunction with
      terminal option `colourtext'%
    }{See the gnuplot documentation for explanation.%
    }{Either use 'blacktext' in gnuplot or load the package
      color.sty in LaTeX.}%
    \renewcommand\color[2][]{}%
  }%
  \providecommand\includegraphics[2][]{%
    \GenericError{(gnuplot) \space\space\space\@spaces}{%
      Package graphicx or graphics not loaded%
    }{See the gnuplot documentation for explanation.%
    }{The gnuplot epslatex terminal needs graphicx.sty or graphics.sty.}%
    \renewcommand\includegraphics[2][]{}%
  }%
  \providecommand\rotatebox[2]{#2}%
  \@ifundefined{ifGPcolor}{%
    \newif\ifGPcolor
    \GPcolortrue
  }{}%
  \@ifundefined{ifGPblacktext}{%
    \newif\ifGPblacktext
    \GPblacktexttrue
  }{}%
  % define a \g@addto@macro without @ in the name:
  \let\gplgaddtomacro\g@addto@macro
  % define empty templates for all commands taking text:
  \gdef\gplbacktext{}%
  \gdef\gplfronttext{}%
  \makeatother
  \ifGPblacktext
    % no textcolor at all
    \def\colorrgb#1{}%
    \def\colorgray#1{}%
  \else
    % gray or color?
    \ifGPcolor
      \def\colorrgb#1{\color[rgb]{#1}}%
      \def\colorgray#1{\color[gray]{#1}}%
      \expandafter\def\csname LTw\endcsname{\color{white}}%
      \expandafter\def\csname LTb\endcsname{\color{black}}%
      \expandafter\def\csname LTa\endcsname{\color{black}}%
      \expandafter\def\csname LT0\endcsname{\color[rgb]{1,0,0}}%
      \expandafter\def\csname LT1\endcsname{\color[rgb]{0,1,0}}%
      \expandafter\def\csname LT2\endcsname{\color[rgb]{0,0,1}}%
      \expandafter\def\csname LT3\endcsname{\color[rgb]{1,0,1}}%
      \expandafter\def\csname LT4\endcsname{\color[rgb]{0,1,1}}%
      \expandafter\def\csname LT5\endcsname{\color[rgb]{1,1,0}}%
      \expandafter\def\csname LT6\endcsname{\color[rgb]{0,0,0}}%
      \expandafter\def\csname LT7\endcsname{\color[rgb]{1,0.3,0}}%
      \expandafter\def\csname LT8\endcsname{\color[rgb]{0.5,0.5,0.5}}%
    \else
      % gray
      \def\colorrgb#1{\color{black}}%
      \def\colorgray#1{\color[gray]{#1}}%
      \expandafter\def\csname LTw\endcsname{\color{white}}%
      \expandafter\def\csname LTb\endcsname{\color{black}}%
      \expandafter\def\csname LTa\endcsname{\color{black}}%
      \expandafter\def\csname LT0\endcsname{\color{black}}%
      \expandafter\def\csname LT1\endcsname{\color{black}}%
      \expandafter\def\csname LT2\endcsname{\color{black}}%
      \expandafter\def\csname LT3\endcsname{\color{black}}%
      \expandafter\def\csname LT4\endcsname{\color{black}}%
      \expandafter\def\csname LT5\endcsname{\color{black}}%
      \expandafter\def\csname LT6\endcsname{\color{black}}%
      \expandafter\def\csname LT7\endcsname{\color{black}}%
      \expandafter\def\csname LT8\endcsname{\color{black}}%
    \fi
  \fi
    \setlength{\unitlength}{0.0500bp}%
    \ifx\gptboxheight\undefined%
      \newlength{\gptboxheight}%
      \newlength{\gptboxwidth}%
      \newsavebox{\gptboxtext}%
    \fi%
    \setlength{\fboxrule}{0.5pt}%
    \setlength{\fboxsep}{1pt}%
\begin{picture}(9920.00,3258.00)%
    \gplgaddtomacro\gplbacktext{%
      \csname LTb\endcsname%%
      \put(1067,704){\makebox(0,0)[r]{\strut{}$-1.5$}}%
      \csname LTb\endcsname%%
      \put(1067,1335){\makebox(0,0)[r]{\strut{}$-0.5$}}%
      \csname LTb\endcsname%%
      \put(1067,1966){\makebox(0,0)[r]{\strut{}$0.5$}}%
      \csname LTb\endcsname%%
      \put(1067,2597){\makebox(0,0)[r]{\strut{}$1.5$}}%
      \csname LTb\endcsname%%
      \put(1515,484){\makebox(0,0){\strut{}$-6$}}%
      \csname LTb\endcsname%%
      \put(2777,484){\makebox(0,0){\strut{}$-4$}}%
      \csname LTb\endcsname%%
      \put(4039,484){\makebox(0,0){\strut{}$-2$}}%
      \csname LTb\endcsname%%
      \put(5301,484){\makebox(0,0){\strut{}$0$}}%
      \csname LTb\endcsname%%
      \put(6563,484){\makebox(0,0){\strut{}$2$}}%
      \csname LTb\endcsname%%
      \put(7825,484){\makebox(0,0){\strut{}$4$}}%
      \csname LTb\endcsname%%
      \put(9087,484){\makebox(0,0){\strut{}$6$}}%
    }%
    \gplgaddtomacro\gplfronttext{%
      \csname LTb\endcsname%%
      \put(319,1650){\rotatebox{-270}{\makebox(0,0){\strut{}$X/L$}}}%
      \put(5300,154){\makebox(0,0){\strut{}$Y/L$}}%
      \put(5300,2927){\makebox(0,0){\textsf{Fixed S$^{\text{3}}$ coordinates}}}%
    }%
    \gplbacktext
    \put(0,0){\includegraphics{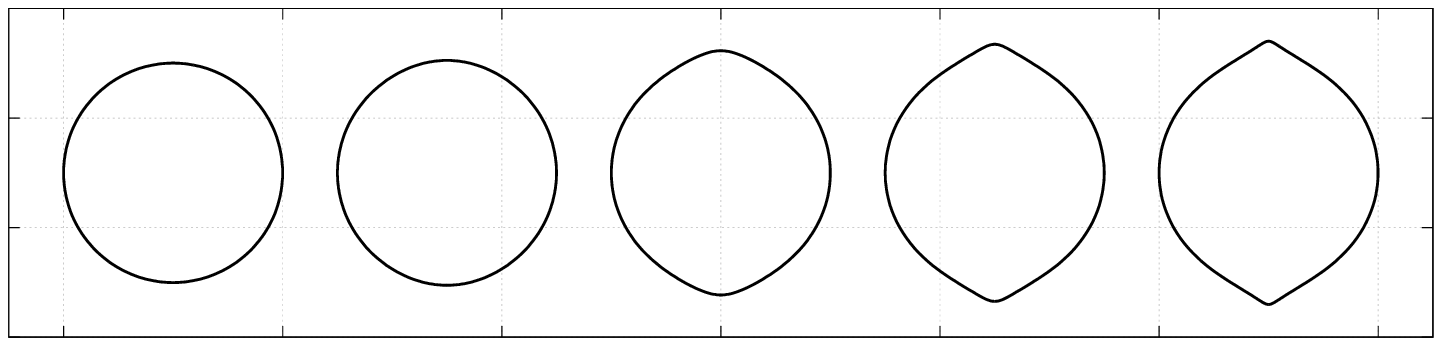}}%
    \gplfronttext
  \end{picture}%
\endgroup

\end{center}
\end{minipage}

\begin{minipage}{\textwidth}
\begin{center}
\hs{-0.75}% GNUPLOT: LaTeX picture with Postscript
\begingroup
  \makeatletter
  \providecommand\color[2][]{%
    \GenericError{(gnuplot) \space\space\space\@spaces}{%
      Package color not loaded in conjunction with
      terminal option `colourtext'%
    }{See the gnuplot documentation for explanation.%
    }{Either use 'blacktext' in gnuplot or load the package
      color.sty in LaTeX.}%
    \renewcommand\color[2][]{}%
  }%
  \providecommand\includegraphics[2][]{%
    \GenericError{(gnuplot) \space\space\space\@spaces}{%
      Package graphicx or graphics not loaded%
    }{See the gnuplot documentation for explanation.%
    }{The gnuplot epslatex terminal needs graphicx.sty or graphics.sty.}%
    \renewcommand\includegraphics[2][]{}%
  }%
  \providecommand\rotatebox[2]{#2}%
  \@ifundefined{ifGPcolor}{%
    \newif\ifGPcolor
    \GPcolortrue
  }{}%
  \@ifundefined{ifGPblacktext}{%
    \newif\ifGPblacktext
    \GPblacktexttrue
  }{}%
  % define a \g@addto@macro without @ in the name:
  \let\gplgaddtomacro\g@addto@macro
  % define empty templates for all commands taking text:
  \gdef\gplbacktext{}%
  \gdef\gplfronttext{}%
  \makeatother
  \ifGPblacktext
    % no textcolor at all
    \def\colorrgb#1{}%
    \def\colorgray#1{}%
  \else
    % gray or color?
    \ifGPcolor
      \def\colorrgb#1{\color[rgb]{#1}}%
      \def\colorgray#1{\color[gray]{#1}}%
      \expandafter\def\csname LTw\endcsname{\color{white}}%
      \expandafter\def\csname LTb\endcsname{\color{black}}%
      \expandafter\def\csname LTa\endcsname{\color{black}}%
      \expandafter\def\csname LT0\endcsname{\color[rgb]{1,0,0}}%
      \expandafter\def\csname LT1\endcsname{\color[rgb]{0,1,0}}%
      \expandafter\def\csname LT2\endcsname{\color[rgb]{0,0,1}}%
      \expandafter\def\csname LT3\endcsname{\color[rgb]{1,0,1}}%
      \expandafter\def\csname LT4\endcsname{\color[rgb]{0,1,1}}%
      \expandafter\def\csname LT5\endcsname{\color[rgb]{1,1,0}}%
      \expandafter\def\csname LT6\endcsname{\color[rgb]{0,0,0}}%
      \expandafter\def\csname LT7\endcsname{\color[rgb]{1,0.3,0}}%
      \expandafter\def\csname LT8\endcsname{\color[rgb]{0.5,0.5,0.5}}%
    \else
      % gray
      \def\colorrgb#1{\color{black}}%
      \def\colorgray#1{\color[gray]{#1}}%
      \expandafter\def\csname LTw\endcsname{\color{white}}%
      \expandafter\def\csname LTb\endcsname{\color{black}}%
      \expandafter\def\csname LTa\endcsname{\color{black}}%
      \expandafter\def\csname LT0\endcsname{\color{black}}%
      \expandafter\def\csname LT1\endcsname{\color{black}}%
      \expandafter\def\csname LT2\endcsname{\color{black}}%
      \expandafter\def\csname LT3\endcsname{\color{black}}%
      \expandafter\def\csname LT4\endcsname{\color{black}}%
      \expandafter\def\csname LT5\endcsname{\color{black}}%
      \expandafter\def\csname LT6\endcsname{\color{black}}%
      \expandafter\def\csname LT7\endcsname{\color{black}}%
      \expandafter\def\csname LT8\endcsname{\color{black}}%
    \fi
  \fi
    \setlength{\unitlength}{0.0500bp}%
    \ifx\gptboxheight\undefined%
      \newlength{\gptboxheight}%
      \newlength{\gptboxwidth}%
      \newsavebox{\gptboxtext}%
    \fi%
    \setlength{\fboxrule}{0.5pt}%
    \setlength{\fboxsep}{1pt}%
\begin{picture}(10204.00,3400.00)%
    \gplgaddtomacro\gplbacktext{%
      \csname LTb\endcsname%%
      \put(1495,704){\makebox(0,0)[r]{\strut{}$-2.5$}}%
      \csname LTb\endcsname%%
      \put(1495,1111){\makebox(0,0)[r]{\strut{}$-1.5$}}%
      \csname LTb\endcsname%%
      \put(1495,1518){\makebox(0,0)[r]{\strut{}$-0.5$}}%
      \csname LTb\endcsname%%
      \put(1495,1925){\makebox(0,0)[r]{\strut{}$0.5$}}%
      \csname LTb\endcsname%%
      \put(1495,2332){\makebox(0,0)[r]{\strut{}$1.5$}}%
      \csname LTb\endcsname%%
      \put(1495,2739){\makebox(0,0)[r]{\strut{}$2.5$}}%
      \csname LTb\endcsname%%
      \put(2390,484){\makebox(0,0){\strut{}$-3$}}%
      \csname LTb\endcsname%%
      \put(3408,484){\makebox(0,0){\strut{}$-2$}}%
      \csname LTb\endcsname%%
      \put(4425,484){\makebox(0,0){\strut{}$-1$}}%
      \csname LTb\endcsname%%
      \put(5443,484){\makebox(0,0){\strut{}$0$}}%
      \csname LTb\endcsname%%
      \put(6460,484){\makebox(0,0){\strut{}$1$}}%
      \csname LTb\endcsname%%
      \put(7477,484){\makebox(0,0){\strut{}$2$}}%
      \csname LTb\endcsname%%
      \put(8495,484){\makebox(0,0){\strut{}$3$}}%
    }%
    \gplgaddtomacro\gplfronttext{%
      \csname LTb\endcsname%%
      \put(747,1721){\rotatebox{-270}{\makebox(0,0){\strut{}$X/L$}}}%
      \put(5442,154){\makebox(0,0){\strut{}$Y/L$}}%
      \put(5442,3069){\makebox(0,0){\textsf{Fixed S$^{\text{4}}$ coordinates}}}%
    }%
    \gplbacktext
    \put(0,0){\includegraphics{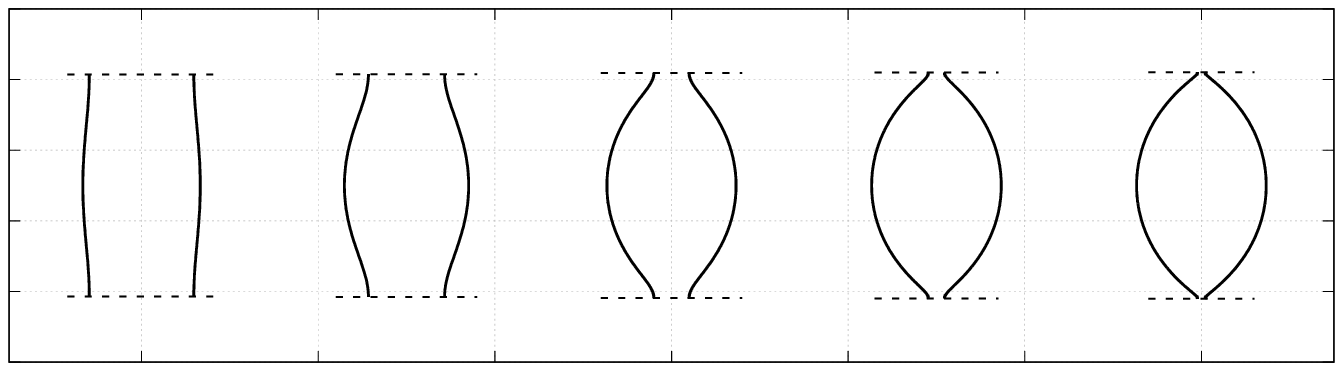}}%
    \gplfronttext
  \end{picture}%
\endgroup

\end{center}
\end{minipage}
\end{center}

\begin{center}
\begin{center}
{\bf \textsf{$\ell \text{ = 2$^\mathbf{-}$}$}}
\end{center}
\begin{minipage}{\textwidth}
\begin{center}
\hs{-1}% GNUPLOT: LaTeX picture with Postscript
\begingroup
  \makeatletter
  \providecommand\color[2][]{%
    \GenericError{(gnuplot) \space\space\space\@spaces}{%
      Package color not loaded in conjunction with
      terminal option `colourtext'%
    }{See the gnuplot documentation for explanation.%
    }{Either use 'blacktext' in gnuplot or load the package
      color.sty in LaTeX.}%
    \renewcommand\color[2][]{}%
  }%
  \providecommand\includegraphics[2][]{%
    \GenericError{(gnuplot) \space\space\space\@spaces}{%
      Package graphicx or graphics not loaded%
    }{See the gnuplot documentation for explanation.%
    }{The gnuplot epslatex terminal needs graphicx.sty or graphics.sty.}%
    \renewcommand\includegraphics[2][]{}%
  }%
  \providecommand\rotatebox[2]{#2}%
  \@ifundefined{ifGPcolor}{%
    \newif\ifGPcolor
    \GPcolortrue
  }{}%
  \@ifundefined{ifGPblacktext}{%
    \newif\ifGPblacktext
    \GPblacktexttrue
  }{}%
  % define a \g@addto@macro without @ in the name:
  \let\gplgaddtomacro\g@addto@macro
  % define empty templates for all commands taking text:
  \gdef\gplbacktext{}%
  \gdef\gplfronttext{}%
  \makeatother
  \ifGPblacktext
    % no textcolor at all
    \def\colorrgb#1{}%
    \def\colorgray#1{}%
  \else
    % gray or color?
    \ifGPcolor
      \def\colorrgb#1{\color[rgb]{#1}}%
      \def\colorgray#1{\color[gray]{#1}}%
      \expandafter\def\csname LTw\endcsname{\color{white}}%
      \expandafter\def\csname LTb\endcsname{\color{black}}%
      \expandafter\def\csname LTa\endcsname{\color{black}}%
      \expandafter\def\csname LT0\endcsname{\color[rgb]{1,0,0}}%
      \expandafter\def\csname LT1\endcsname{\color[rgb]{0,1,0}}%
      \expandafter\def\csname LT2\endcsname{\color[rgb]{0,0,1}}%
      \expandafter\def\csname LT3\endcsname{\color[rgb]{1,0,1}}%
      \expandafter\def\csname LT4\endcsname{\color[rgb]{0,1,1}}%
      \expandafter\def\csname LT5\endcsname{\color[rgb]{1,1,0}}%
      \expandafter\def\csname LT6\endcsname{\color[rgb]{0,0,0}}%
      \expandafter\def\csname LT7\endcsname{\color[rgb]{1,0.3,0}}%
      \expandafter\def\csname LT8\endcsname{\color[rgb]{0.5,0.5,0.5}}%
    \else
      % gray
      \def\colorrgb#1{\color{black}}%
      \def\colorgray#1{\color[gray]{#1}}%
      \expandafter\def\csname LTw\endcsname{\color{white}}%
      \expandafter\def\csname LTb\endcsname{\color{black}}%
      \expandafter\def\csname LTa\endcsname{\color{black}}%
      \expandafter\def\csname LT0\endcsname{\color{black}}%
      \expandafter\def\csname LT1\endcsname{\color{black}}%
      \expandafter\def\csname LT2\endcsname{\color{black}}%
      \expandafter\def\csname LT3\endcsname{\color{black}}%
      \expandafter\def\csname LT4\endcsname{\color{black}}%
      \expandafter\def\csname LT5\endcsname{\color{black}}%
      \expandafter\def\csname LT6\endcsname{\color{black}}%
      \expandafter\def\csname LT7\endcsname{\color{black}}%
      \expandafter\def\csname LT8\endcsname{\color{black}}%
    \fi
  \fi
    \setlength{\unitlength}{0.0500bp}%
    \ifx\gptboxheight\undefined%
      \newlength{\gptboxheight}%
      \newlength{\gptboxwidth}%
      \newsavebox{\gptboxtext}%
    \fi%
    \setlength{\fboxrule}{0.5pt}%
    \setlength{\fboxsep}{1pt}%
\begin{picture}(9920.00,3258.00)%
    \gplgaddtomacro\gplbacktext{%
      \csname LTb\endcsname%%
      \put(1067,704){\makebox(0,0)[r]{\strut{}$-1.5$}}%
      \csname LTb\endcsname%%
      \put(1067,1335){\makebox(0,0)[r]{\strut{}$-0.5$}}%
      \csname LTb\endcsname%%
      \put(1067,1966){\makebox(0,0)[r]{\strut{}$0.5$}}%
      \csname LTb\endcsname%%
      \put(1067,2597){\makebox(0,0)[r]{\strut{}$1.5$}}%
      \csname LTb\endcsname%%
      \put(1515,484){\makebox(0,0){\strut{}$-6$}}%
      \csname LTb\endcsname%%
      \put(2777,484){\makebox(0,0){\strut{}$-4$}}%
      \csname LTb\endcsname%%
      \put(4039,484){\makebox(0,0){\strut{}$-2$}}%
      \csname LTb\endcsname%%
      \put(5301,484){\makebox(0,0){\strut{}$0$}}%
      \csname LTb\endcsname%%
      \put(6563,484){\makebox(0,0){\strut{}$2$}}%
      \csname LTb\endcsname%%
      \put(7825,484){\makebox(0,0){\strut{}$4$}}%
      \csname LTb\endcsname%%
      \put(9087,484){\makebox(0,0){\strut{}$6$}}%
    }%
    \gplgaddtomacro\gplfronttext{%
      \csname LTb\endcsname%%
      \put(319,1650){\rotatebox{-270}{\makebox(0,0){\strut{}$X/L$}}}%
      \put(5300,154){\makebox(0,0){\strut{}$Y/L$}}%
      \put(5300,2927){\makebox(0,0){\textsf{Fixed S$^{\text{3}}$ coordinates}}}%
    }%
    \gplbacktext
    \put(0,0){\includegraphics{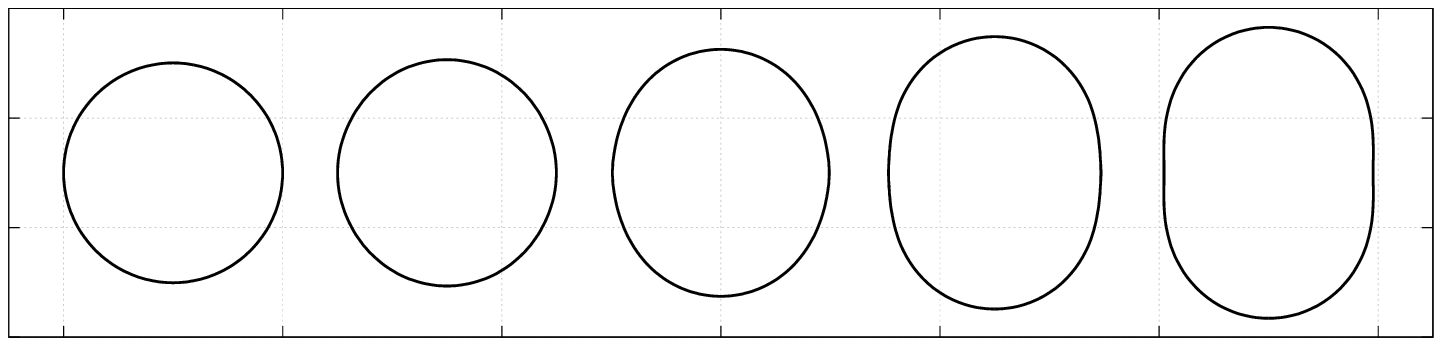}}%
    \gplfronttext
  \end{picture}%
\endgroup

\hs{-0.75}% GNUPLOT: LaTeX picture with Postscript
\begingroup
  \makeatletter
  \providecommand\color[2][]{%
    \GenericError{(gnuplot) \space\space\space\@spaces}{%
      Package color not loaded in conjunction with
      terminal option `colourtext'%
    }{See the gnuplot documentation for explanation.%
    }{Either use 'blacktext' in gnuplot or load the package
      color.sty in LaTeX.}%
    \renewcommand\color[2][]{}%
  }%
  \providecommand\includegraphics[2][]{%
    \GenericError{(gnuplot) \space\space\space\@spaces}{%
      Package graphicx or graphics not loaded%
    }{See the gnuplot documentation for explanation.%
    }{The gnuplot epslatex terminal needs graphicx.sty or graphics.sty.}%
    \renewcommand\includegraphics[2][]{}%
  }%
  \providecommand\rotatebox[2]{#2}%
  \@ifundefined{ifGPcolor}{%
    \newif\ifGPcolor
    \GPcolortrue
  }{}%
  \@ifundefined{ifGPblacktext}{%
    \newif\ifGPblacktext
    \GPblacktexttrue
  }{}%
  % define a \g@addto@macro without @ in the name:
  \let\gplgaddtomacro\g@addto@macro
  % define empty templates for all commands taking text:
  \gdef\gplbacktext{}%
  \gdef\gplfronttext{}%
  \makeatother
  \ifGPblacktext
    % no textcolor at all
    \def\colorrgb#1{}%
    \def\colorgray#1{}%
  \else
    % gray or color?
    \ifGPcolor
      \def\colorrgb#1{\color[rgb]{#1}}%
      \def\colorgray#1{\color[gray]{#1}}%
      \expandafter\def\csname LTw\endcsname{\color{white}}%
      \expandafter\def\csname LTb\endcsname{\color{black}}%
      \expandafter\def\csname LTa\endcsname{\color{black}}%
      \expandafter\def\csname LT0\endcsname{\color[rgb]{1,0,0}}%
      \expandafter\def\csname LT1\endcsname{\color[rgb]{0,1,0}}%
      \expandafter\def\csname LT2\endcsname{\color[rgb]{0,0,1}}%
      \expandafter\def\csname LT3\endcsname{\color[rgb]{1,0,1}}%
      \expandafter\def\csname LT4\endcsname{\color[rgb]{0,1,1}}%
      \expandafter\def\csname LT5\endcsname{\color[rgb]{1,1,0}}%
      \expandafter\def\csname LT6\endcsname{\color[rgb]{0,0,0}}%
      \expandafter\def\csname LT7\endcsname{\color[rgb]{1,0.3,0}}%
      \expandafter\def\csname LT8\endcsname{\color[rgb]{0.5,0.5,0.5}}%
    \else
      % gray
      \def\colorrgb#1{\color{black}}%
      \def\colorgray#1{\color[gray]{#1}}%
      \expandafter\def\csname LTw\endcsname{\color{white}}%
      \expandafter\def\csname LTb\endcsname{\color{black}}%
      \expandafter\def\csname LTa\endcsname{\color{black}}%
      \expandafter\def\csname LT0\endcsname{\color{black}}%
      \expandafter\def\csname LT1\endcsname{\color{black}}%
      \expandafter\def\csname LT2\endcsname{\color{black}}%
      \expandafter\def\csname LT3\endcsname{\color{black}}%
      \expandafter\def\csname LT4\endcsname{\color{black}}%
      \expandafter\def\csname LT5\endcsname{\color{black}}%
      \expandafter\def\csname LT6\endcsname{\color{black}}%
      \expandafter\def\csname LT7\endcsname{\color{black}}%
      \expandafter\def\csname LT8\endcsname{\color{black}}%
    \fi
  \fi
    \setlength{\unitlength}{0.0500bp}%
    \ifx\gptboxheight\undefined%
      \newlength{\gptboxheight}%
      \newlength{\gptboxwidth}%
      \newsavebox{\gptboxtext}%
    \fi%
    \setlength{\fboxrule}{0.5pt}%
    \setlength{\fboxsep}{1pt}%
\begin{picture}(10204.00,3400.00)%
    \gplgaddtomacro\gplbacktext{%
      \csname LTb\endcsname%%
      \put(1444,704){\makebox(0,0)[r]{\strut{}$-2.5$}}%
      \csname LTb\endcsname%%
      \put(1444,1111){\makebox(0,0)[r]{\strut{}$-1.5$}}%
      \csname LTb\endcsname%%
      \put(1444,1518){\makebox(0,0)[r]{\strut{}$-0.5$}}%
      \csname LTb\endcsname%%
      \put(1444,1925){\makebox(0,0)[r]{\strut{}$0.5$}}%
      \csname LTb\endcsname%%
      \put(1444,2332){\makebox(0,0)[r]{\strut{}$1.5$}}%
      \csname LTb\endcsname%%
      \put(1444,2739){\makebox(0,0)[r]{\strut{}$2.5$}}%
      \csname LTb\endcsname%%
      \put(2390,484){\makebox(0,0){\strut{}$-3$}}%
      \csname LTb\endcsname%%
      \put(3408,484){\makebox(0,0){\strut{}$-2$}}%
      \csname LTb\endcsname%%
      \put(4425,484){\makebox(0,0){\strut{}$-1$}}%
      \csname LTb\endcsname%%
      \put(5443,484){\makebox(0,0){\strut{}$0$}}%
      \csname LTb\endcsname%%
      \put(6460,484){\makebox(0,0){\strut{}$1$}}%
      \csname LTb\endcsname%%
      \put(7478,484){\makebox(0,0){\strut{}$2$}}%
      \csname LTb\endcsname%%
      \put(8495,484){\makebox(0,0){\strut{}$3$}}%
    }%
    \gplgaddtomacro\gplfronttext{%
      \csname LTb\endcsname%%
      \put(696,1721){\rotatebox{-270}{\makebox(0,0){\strut{}$X/L$}}}%
      \put(5442,154){\makebox(0,0){\strut{}$Y/L$}}%
      \put(5442,3069){\makebox(0,0){\textsf{Fixed S$^{\text{4}}$ coordinates}}}%
    }%
    \gplbacktext
    \put(0,0){\includegraphics{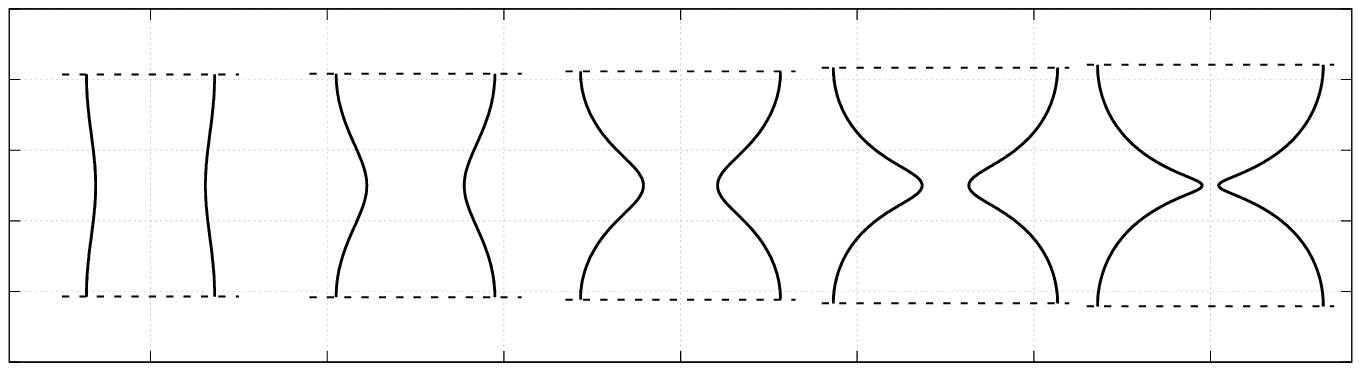}}%
    \gplfronttext
  \end{picture}%
\endgroup

\end{center}
\end{minipage}
\captionsetup{width=0.9\textwidth}
\captionof{figure}{\textsl{Embeddings plots for fixed $S^3/S^4$ coordinates of lumpy black holes in AdS$_5 \times S^5$ with $\ell = 1, 2^+$ and $2^-$. The corresponding plots for $\ell = 3$ are similar to those for $\ell = 1$, but the pinch-off appears at the opposite pole.}}
\label{fig:12AdS5}
\end{center}

\begin{center}
% GNUPLOT: LaTeX picture with Postscript
\begingroup
  \makeatletter
  \providecommand\color[2][]{%
    \GenericError{(gnuplot) \space\space\space\@spaces}{%
      Package color not loaded in conjunction with
      terminal option `colourtext'%
    }{See the gnuplot documentation for explanation.%
    }{Either use 'blacktext' in gnuplot or load the package
      color.sty in LaTeX.}%
    \renewcommand\color[2][]{}%
  }%
  \providecommand\includegraphics[2][]{%
    \GenericError{(gnuplot) \space\space\space\@spaces}{%
      Package graphicx or graphics not loaded%
    }{See the gnuplot documentation for explanation.%
    }{The gnuplot epslatex terminal needs graphicx.sty or graphics.sty.}%
    \renewcommand\includegraphics[2][]{}%
  }%
  \providecommand\rotatebox[2]{#2}%
  \@ifundefined{ifGPcolor}{%
    \newif\ifGPcolor
    \GPcolortrue
  }{}%
  \@ifundefined{ifGPblacktext}{%
    \newif\ifGPblacktext
    \GPblacktexttrue
  }{}%
  % define a \g@addto@macro without @ in the name:
  \let\gplgaddtomacro\g@addto@macro
  % define empty templates for all commands taking text:
  \gdef\gplbacktext{}%
  \gdef\gplfronttext{}%
  \makeatother
  \ifGPblacktext
    % no textcolor at all
    \def\colorrgb#1{}%
    \def\colorgray#1{}%
  \else
    % gray or color?
    \ifGPcolor
      \def\colorrgb#1{\color[rgb]{#1}}%
      \def\colorgray#1{\color[gray]{#1}}%
      \expandafter\def\csname LTw\endcsname{\color{white}}%
      \expandafter\def\csname LTb\endcsname{\color{black}}%
      \expandafter\def\csname LTa\endcsname{\color{black}}%
      \expandafter\def\csname LT0\endcsname{\color[rgb]{1,0,0}}%
      \expandafter\def\csname LT1\endcsname{\color[rgb]{0,1,0}}%
      \expandafter\def\csname LT2\endcsname{\color[rgb]{0,0,1}}%
      \expandafter\def\csname LT3\endcsname{\color[rgb]{1,0,1}}%
      \expandafter\def\csname LT4\endcsname{\color[rgb]{0,1,1}}%
      \expandafter\def\csname LT5\endcsname{\color[rgb]{1,1,0}}%
      \expandafter\def\csname LT6\endcsname{\color[rgb]{0,0,0}}%
      \expandafter\def\csname LT7\endcsname{\color[rgb]{1,0.3,0}}%
      \expandafter\def\csname LT8\endcsname{\color[rgb]{0.5,0.5,0.5}}%
    \else
      % gray
      \def\colorrgb#1{\color{black}}%
      \def\colorgray#1{\color[gray]{#1}}%
      \expandafter\def\csname LTw\endcsname{\color{white}}%
      \expandafter\def\csname LTb\endcsname{\color{black}}%
      \expandafter\def\csname LTa\endcsname{\color{black}}%
      \expandafter\def\csname LT0\endcsname{\color{black}}%
      \expandafter\def\csname LT1\endcsname{\color{black}}%
      \expandafter\def\csname LT2\endcsname{\color{black}}%
      \expandafter\def\csname LT3\endcsname{\color{black}}%
      \expandafter\def\csname LT4\endcsname{\color{black}}%
      \expandafter\def\csname LT5\endcsname{\color{black}}%
      \expandafter\def\csname LT6\endcsname{\color{black}}%
      \expandafter\def\csname LT7\endcsname{\color{black}}%
      \expandafter\def\csname LT8\endcsname{\color{black}}%
    \fi
  \fi
    \setlength{\unitlength}{0.0500bp}%
    \ifx\gptboxheight\undefined%
      \newlength{\gptboxheight}%
      \newlength{\gptboxwidth}%
      \newsavebox{\gptboxtext}%
    \fi%
    \setlength{\fboxrule}{0.5pt}%
    \setlength{\fboxsep}{1pt}%
\begin{picture}(3968.00,3968.00)%
    \gplgaddtomacro\gplbacktext{%
      \csname LTb\endcsname%%
      \put(1078,704){\makebox(0,0)[r]{\strut{}$-1.9$}}%
      \csname LTb\endcsname%%
      \put(1078,1355){\makebox(0,0)[r]{\strut{}$-0.95$}}%
      \csname LTb\endcsname%%
      \put(1078,2006){\makebox(0,0)[r]{\strut{}$0$}}%
      \csname LTb\endcsname%%
      \put(1078,2656){\makebox(0,0)[r]{\strut{}$0.95$}}%
      \csname LTb\endcsname%%
      \put(1078,3307){\makebox(0,0)[r]{\strut{}$1.9$}}%
      \csname LTb\endcsname%%
      \put(1210,484){\makebox(0,0){\strut{}$-0.8$}}%
      \csname LTb\endcsname%%
      \put(1800,484){\makebox(0,0){\strut{}$-0.4$}}%
      \csname LTb\endcsname%%
      \put(2391,484){\makebox(0,0){\strut{}$0$}}%
      \csname LTb\endcsname%%
      \put(2981,484){\makebox(0,0){\strut{}$0.4$}}%
      \csname LTb\endcsname%%
      \put(3571,484){\makebox(0,0){\strut{}$0.8$}}%
    }%
    \gplgaddtomacro\gplfronttext{%
      \csname LTb\endcsname%%
      \put(198,2005){\rotatebox{-270}{\makebox(0,0){\strut{}$X/L$}}}%
      \put(2390,154){\makebox(0,0){\strut{}$Y/L$}}%
      \put(2390,3637){\makebox(0,0){\textsf{Fixed S$^{\textsf{4}}$ coordinates}}}%
    }%
    \gplbacktext
    \put(0,0){\includegraphics{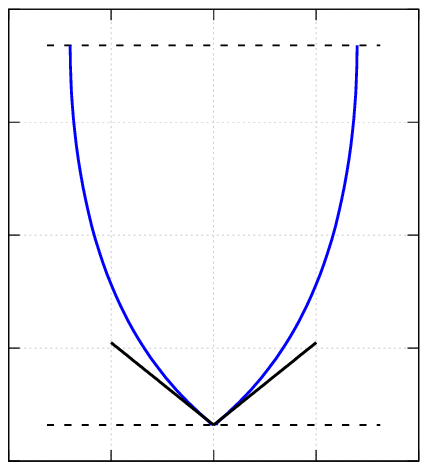}}%
    \gplfronttext
  \end{picture}%
\endgroup

% GNUPLOT: LaTeX picture with Postscript
\begingroup
  \makeatletter
  \providecommand\color[2][]{%
    \GenericError{(gnuplot) \space\space\space\@spaces}{%
      Package color not loaded in conjunction with
      terminal option `colourtext'%
    }{See the gnuplot documentation for explanation.%
    }{Either use 'blacktext' in gnuplot or load the package
      color.sty in LaTeX.}%
    \renewcommand\color[2][]{}%
  }%
  \providecommand\includegraphics[2][]{%
    \GenericError{(gnuplot) \space\space\space\@spaces}{%
      Package graphicx or graphics not loaded%
    }{See the gnuplot documentation for explanation.%
    }{The gnuplot epslatex terminal needs graphicx.sty or graphics.sty.}%
    \renewcommand\includegraphics[2][]{}%
  }%
  \providecommand\rotatebox[2]{#2}%
  \@ifundefined{ifGPcolor}{%
    \newif\ifGPcolor
    \GPcolortrue
  }{}%
  \@ifundefined{ifGPblacktext}{%
    \newif\ifGPblacktext
    \GPblacktexttrue
  }{}%
  % define a \g@addto@macro without @ in the name:
  \let\gplgaddtomacro\g@addto@macro
  % define empty templates for all commands taking text:
  \gdef\gplbacktext{}%
  \gdef\gplfronttext{}%
  \makeatother
  \ifGPblacktext
    % no textcolor at all
    \def\colorrgb#1{}%
    \def\colorgray#1{}%
  \else
    % gray or color?
    \ifGPcolor
      \def\colorrgb#1{\color[rgb]{#1}}%
      \def\colorgray#1{\color[gray]{#1}}%
      \expandafter\def\csname LTw\endcsname{\color{white}}%
      \expandafter\def\csname LTb\endcsname{\color{black}}%
      \expandafter\def\csname LTa\endcsname{\color{black}}%
      \expandafter\def\csname LT0\endcsname{\color[rgb]{1,0,0}}%
      \expandafter\def\csname LT1\endcsname{\color[rgb]{0,1,0}}%
      \expandafter\def\csname LT2\endcsname{\color[rgb]{0,0,1}}%
      \expandafter\def\csname LT3\endcsname{\color[rgb]{1,0,1}}%
      \expandafter\def\csname LT4\endcsname{\color[rgb]{0,1,1}}%
      \expandafter\def\csname LT5\endcsname{\color[rgb]{1,1,0}}%
      \expandafter\def\csname LT6\endcsname{\color[rgb]{0,0,0}}%
      \expandafter\def\csname LT7\endcsname{\color[rgb]{1,0.3,0}}%
      \expandafter\def\csname LT8\endcsname{\color[rgb]{0.5,0.5,0.5}}%
    \else
      % gray
      \def\colorrgb#1{\color{black}}%
      \def\colorgray#1{\color[gray]{#1}}%
      \expandafter\def\csname LTw\endcsname{\color{white}}%
      \expandafter\def\csname LTb\endcsname{\color{black}}%
      \expandafter\def\csname LTa\endcsname{\color{black}}%
      \expandafter\def\csname LT0\endcsname{\color{black}}%
      \expandafter\def\csname LT1\endcsname{\color{black}}%
      \expandafter\def\csname LT2\endcsname{\color{black}}%
      \expandafter\def\csname LT3\endcsname{\color{black}}%
      \expandafter\def\csname LT4\endcsname{\color{black}}%
      \expandafter\def\csname LT5\endcsname{\color{black}}%
      \expandafter\def\csname LT6\endcsname{\color{black}}%
      \expandafter\def\csname LT7\endcsname{\color{black}}%
      \expandafter\def\csname LT8\endcsname{\color{black}}%
    \fi
  \fi
    \setlength{\unitlength}{0.0500bp}%
    \ifx\gptboxheight\undefined%
      \newlength{\gptboxheight}%
      \newlength{\gptboxwidth}%
      \newsavebox{\gptboxtext}%
    \fi%
    \setlength{\fboxrule}{0.5pt}%
    \setlength{\fboxsep}{1pt}%
\begin{picture}(3968.00,3968.00)%
    \gplgaddtomacro\gplbacktext{%
      \csname LTb\endcsname%%
      \put(1078,704){\makebox(0,0)[r]{\strut{}$-1.3$}}%
      \csname LTb\endcsname%%
      \put(1078,1355){\makebox(0,0)[r]{\strut{}$-0.65$}}%
      \csname LTb\endcsname%%
      \put(1078,2006){\makebox(0,0)[r]{\strut{}$0$}}%
      \csname LTb\endcsname%%
      \put(1078,2656){\makebox(0,0)[r]{\strut{}$0.65$}}%
      \csname LTb\endcsname%%
      \put(1078,3307){\makebox(0,0)[r]{\strut{}$1.3$}}%
      \csname LTb\endcsname%%
      \put(1210,484){\makebox(0,0){\strut{}$-1.3$}}%
      \csname LTb\endcsname%%
      \put(1800,484){\makebox(0,0){\strut{}$-0.65$}}%
      \csname LTb\endcsname%%
      \put(2391,484){\makebox(0,0){\strut{}$0$}}%
      \csname LTb\endcsname%%
      \put(2981,484){\makebox(0,0){\strut{}$0.65$}}%
      \csname LTb\endcsname%%
      \put(3571,484){\makebox(0,0){\strut{}$1.3$}}%
    }%
    \gplgaddtomacro\gplfronttext{%
      \csname LTb\endcsname%%
      \put(198,2005){\rotatebox{-270}{\makebox(0,0){\strut{}$X/L$}}}%
      \put(2390,154){\makebox(0,0){\strut{}$Y/L$}}%
      \put(2390,3637){\makebox(0,0){\textsf{Fixed S$^{\textsf{3}}$ coordinates}}}%
    }%
    \gplbacktext
    \put(0,0){\includegraphics{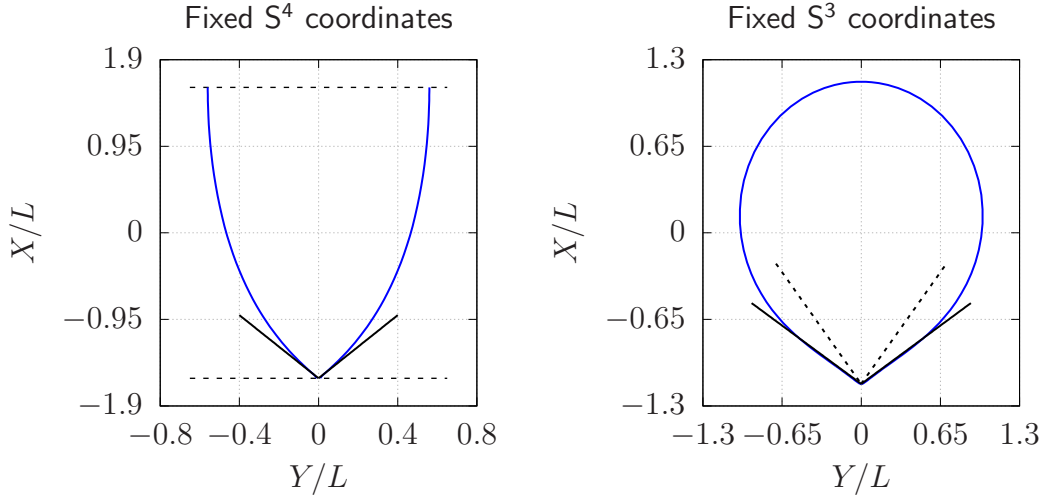}}%
    \gplfronttext
  \end{picture}%
\endgroup

\captionsetup{width=0.9\textwidth}
\captionof{figure}{\textsl{Embedding diagrams of $\ell = 1$ lumpy black hole (blue line) in AdS$_5\times S^5$, for the largest lumpiness parameter we have reached with fixed $S^4$ coordinates (left) and fixed $S^3$ coordinates (right). The black lines correspond to the cone embeddings proposed to mediate the topology change transition. For fixed $S^3$ coordinates, the embedding dictated by the Ricci-flat cone is indicated with dashed lines.}}
\label{fig:embcone2}
\end{center}

In Fig.~\ref{fig:embcone2} we show the embedding plots of the most critical solution we have found along the $\ell = 1$ branch, and compare them with the embeddings of the local cone. The embedding along the internal $S^4$  indicates that the local cone must be sourced by some flux field along the internal directions, and it is consistent with \eqref{alphflux}. 

%%%%%%%%%%%%%%%%%%%%%%%%%%%%%%%%%%%%%%%%%%%%%%%
\section{Numerics and convergence tests}
\label{app:conve}
%%%%%%%%%%%%%%%%%%%%%%%%%%%%%%%%%%%%%%%%%%%%%%%
In this appendix we provide further details of our numerical construction of the lumpy black hole solutions and present some convergence tests.

Our numerical approach to construct critical solutions relies on the redefinition of certain unknown functions to ensure that the new functions have  bounded values at the merger. However, steep gradients are unavoidable and therefore, very close to critical limit, these redefinitions are useless if they are not accompanied by a numerical grid that can properly resolve these gradients. To this end, it is convenient to divide the original grid into other small subregions in order to be able to increase the grid resolution just where it is necessary. Of course, it is not desirable to consider many subdomains as it may result in a loss of computational efficiency. A moderate number of patches together with mesh-refinement overcomes the efficiency issue while redistributing the density of points in a useful manner.

\begin{figure}[t]
\begin{center}
\begin{minipage}{\textwidth}
\begin{minipage}[h!]{0.5\textwidth}
\begin{center}
{\bf \textsf{No mesh-refinement}}
\end{center}
\end{minipage}
\begin{minipage}[h!]{0.5\textwidth}
\begin{center}
{\bf \textsf{With mesh-refinement}}
\end{center}
\end{minipage}
\begin{minipage}[h!]{0.5\textwidth}
\hs{-0.5}\input{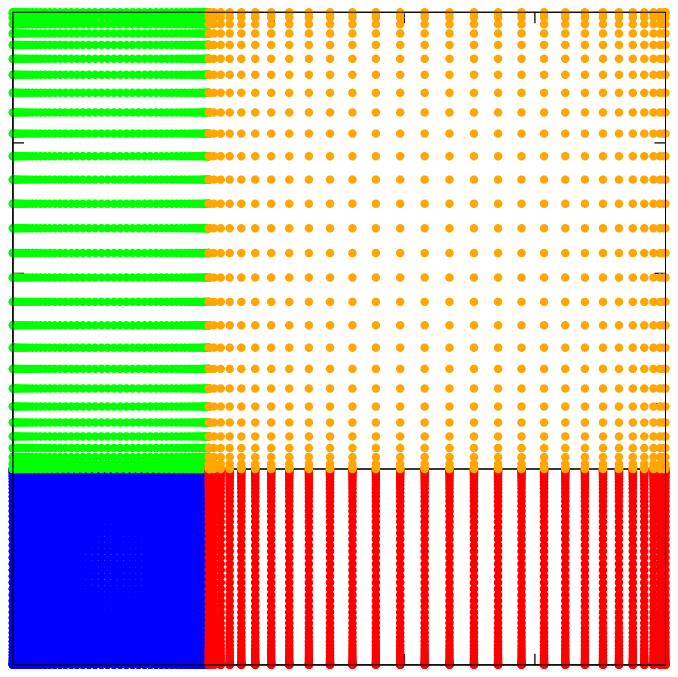}
\end{minipage}
\hfill
\begin{minipage}[h!]{0.5\textwidth}
\hs{-0.5}\input{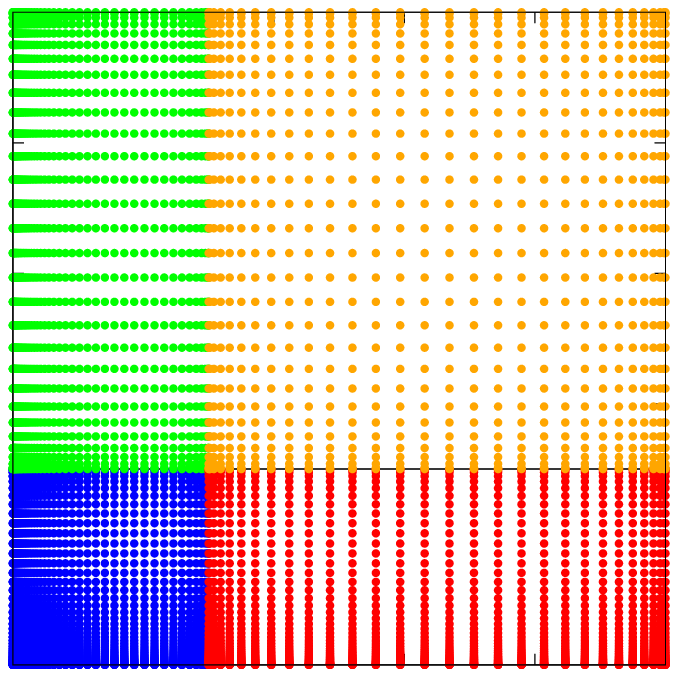}
\end{minipage}
\end{minipage}
\captionsetup{width=0.9\textwidth}
\captionof{figure}{\textsl{Physical grid to construct solutions with $\ell$ even, with grid parameters $x^* = 0.3 = y^*$. Left: simple Chebyshev grid points for the various patches. Right: mesh-refined grid points $\tilde{x} = \text{mesh}(x; 0, x^*,\chi_x)$, $\tilde{y} = \text{mesh}(y; 0, y^*,\chi_y)$, with $\chi_{x,y} = 4$, in the blue patch. In the green and red patches only $x$- and $y$-directions need to be refined respectively.}}
\label{gridss}
\end{center}
\end{figure}

In this paper we have used up to 4 subdomains and a mesh-refinement function as in \cite{Cardona:2018shd}
\beq
\text{mesh}(X;A,B,C) = A + \frac{B-A}{\sinh C}\,\sinh\left(C\,\frac{X-A}{B-A}\right)\,,
\enq
in the patch where the singular behavior appears, which is typically at a pole in the near horizon region. Therefore, where needed, we redefine the coordinates as $\tilde x = \text{mesh}(x;0,x^*,\chi_x)$ and $\tilde y = \text{mesh}(y;0,y^*,\chi_y)$. In total, this introduces four new parameters: $x^*$ and $y^*$, corresponding to the coordinate values where the different patches meet, and $\chi_x$ and $\chi_y$, which control the `strength' of the density of the new grid points along each coordinate direction. Typical values that we used are $\chi_x, \chi_y = 4$. At each shared boundary one must impose continuity of the functions and their first normal derivatives. An example of such grids is shown in Fig.~\ref{gridss}. As for the resolution used, for all families we began with a single patch with $60\times 60$ grid points, and then we increased the number of patches up to four as we moved along the branch of solutions towards the critical regime. The smallest patch where the peaks of the functions typically appear has $50\times50$ resolution, whereas for the other patches  lower resolutions suffice, e.g.,~$\sim30\times 50$ or $30\times 30$.

To check whether the solution on a given grid is good enough, we monitor the numerical error. If this is greater than a few percent, then the resolution needs to be increased. To this end, we use the squared norm of the DeTurck vector $\xi^2$ and its non-zero components $\xi_x$ and $\xi_y$ since they should all vanish in the continuum limit and hence they are good measures of the error. In the case of using a pseudo-spectral numerical approach, as in this article, the error should be exponentially suppressed in the continuum limit, i.e.,~with increasing the grid size. Insufficient resolution can also reflect itself in some unphysical behavior of the physical parameters in the phase diagram of the family of black holes under consideration. For this reason, we also monitor the physical quantities along the branch.

In Fig.~\ref{fig:convergence} we compare the maximum values of $\xi^2$ and $\xi_y$ for a reference solution (with $\lambda\sim 0.2$) at different resolutions in each of the branch of solutions we have constructed.  To produce this figure we picked up a reference solution of each branch, we interpolated it at different resolutions and then we filtered through the Newton-Raphson loop. Solutions with other values of $\lambda$ behave in a qualitatively similar manner. Of course, the larger the $\lambda$ is the larger the error is with the same grid structure and resolution. 

\begin{center}
\begin{minipage}{\textwidth}
\begin{minipage}[h!]{0.5\textwidth}
\begin{center}
\hs{1.25}{\bf \textsf{AdS$_{\textbf{5}}\times$S$^{\textbf{5}}$}}
\end{center}
\end{minipage}
\begin{minipage}[h!]{0.5\textwidth}
\begin{center}
\hs{1.25}{\bf \textsf{AdS$_{\textbf{4}}\times$S$^{\textbf{7}}$}}
\end{center}
\end{minipage}
\begin{minipage}[h!]{0.5\textwidth}
% GNUPLOT: LaTeX picture with Postscript
\begingroup
  \makeatletter
  \providecommand\color[2][]{%
    \GenericError{(gnuplot) \space\space\space\@spaces}{%
      Package color not loaded in conjunction with
      terminal option `colourtext'%
    }{See the gnuplot documentation for explanation.%
    }{Either use 'blacktext' in gnuplot or load the package
      color.sty in LaTeX.}%
    \renewcommand\color[2][]{}%
  }%
  \providecommand\includegraphics[2][]{%
    \GenericError{(gnuplot) \space\space\space\@spaces}{%
      Package graphicx or graphics not loaded%
    }{See the gnuplot documentation for explanation.%
    }{The gnuplot epslatex terminal needs graphicx.sty or graphics.sty.}%
    \renewcommand\includegraphics[2][]{}%
  }%
  \providecommand\rotatebox[2]{#2}%
  \@ifundefined{ifGPcolor}{%
    \newif\ifGPcolor
    \GPcolortrue
  }{}%
  \@ifundefined{ifGPblacktext}{%
    \newif\ifGPblacktext
    \GPblacktexttrue
  }{}%
  % define a \g@addto@macro without @ in the name:
  \let\gplgaddtomacro\g@addto@macro
  % define empty templates for all commands taking text:
  \gdef\gplbacktext{}%
  \gdef\gplfronttext{}%
  \makeatother
  \ifGPblacktext
    % no textcolor at all
    \def\colorrgb#1{}%
    \def\colorgray#1{}%
  \else
    % gray or color?
    \ifGPcolor
      \def\colorrgb#1{\color[rgb]{#1}}%
      \def\colorgray#1{\color[gray]{#1}}%
      \expandafter\def\csname LTw\endcsname{\color{white}}%
      \expandafter\def\csname LTb\endcsname{\color{black}}%
      \expandafter\def\csname LTa\endcsname{\color{black}}%
      \expandafter\def\csname LT0\endcsname{\color[rgb]{1,0,0}}%
      \expandafter\def\csname LT1\endcsname{\color[rgb]{0,1,0}}%
      \expandafter\def\csname LT2\endcsname{\color[rgb]{0,0,1}}%
      \expandafter\def\csname LT3\endcsname{\color[rgb]{1,0,1}}%
      \expandafter\def\csname LT4\endcsname{\color[rgb]{0,1,1}}%
      \expandafter\def\csname LT5\endcsname{\color[rgb]{1,1,0}}%
      \expandafter\def\csname LT6\endcsname{\color[rgb]{0,0,0}}%
      \expandafter\def\csname LT7\endcsname{\color[rgb]{1,0.3,0}}%
      \expandafter\def\csname LT8\endcsname{\color[rgb]{0.5,0.5,0.5}}%
    \else
      % gray
      \def\colorrgb#1{\color{black}}%
      \def\colorgray#1{\color[gray]{#1}}%
      \expandafter\def\csname LTw\endcsname{\color{white}}%
      \expandafter\def\csname LTb\endcsname{\color{black}}%
      \expandafter\def\csname LTa\endcsname{\color{black}}%
      \expandafter\def\csname LT0\endcsname{\color{black}}%
      \expandafter\def\csname LT1\endcsname{\color{black}}%
      \expandafter\def\csname LT2\endcsname{\color{black}}%
      \expandafter\def\csname LT3\endcsname{\color{black}}%
      \expandafter\def\csname LT4\endcsname{\color{black}}%
      \expandafter\def\csname LT5\endcsname{\color{black}}%
      \expandafter\def\csname LT6\endcsname{\color{black}}%
      \expandafter\def\csname LT7\endcsname{\color{black}}%
      \expandafter\def\csname LT8\endcsname{\color{black}}%
    \fi
  \fi
    \setlength{\unitlength}{0.0500bp}%
    \ifx\gptboxheight\undefined%
      \newlength{\gptboxheight}%
      \newlength{\gptboxwidth}%
      \newsavebox{\gptboxtext}%
    \fi%
    \setlength{\fboxrule}{0.5pt}%
    \setlength{\fboxsep}{1pt}%
\begin{picture}(4534.00,4534.00)%
    \gplgaddtomacro\gplbacktext{%
      \csname LTb\endcsname%%
      \put(1078,916){\makebox(0,0)[r]{\strut{}$10^{-18}$}}%
      \csname LTb\endcsname%%
      \put(1078,1341){\makebox(0,0)[r]{\strut{}$10^{-16}$}}%
      \csname LTb\endcsname%%
      \put(1078,1765){\makebox(0,0)[r]{\strut{}$10^{-14}$}}%
      \csname LTb\endcsname%%
      \put(1078,2190){\makebox(0,0)[r]{\strut{}$10^{-12}$}}%
      \csname LTb\endcsname%%
      \put(1078,2615){\makebox(0,0)[r]{\strut{}$10^{-10}$}}%
      \csname LTb\endcsname%%
      \put(1078,3039){\makebox(0,0)[r]{\strut{}$10^{-8}$}}%
      \csname LTb\endcsname%%
      \put(1078,3464){\makebox(0,0)[r]{\strut{}$10^{-6}$}}%
      \csname LTb\endcsname%%
      \put(1078,3888){\makebox(0,0)[r]{\strut{}$10^{-4}$}}%
      \csname LTb\endcsname%%
      \put(1078,4313){\makebox(0,0)[r]{\strut{}$10^{-2}$}}%
      \csname LTb\endcsname%%
      \put(1210,484){\makebox(0,0){\strut{}$0$}}%
      \csname LTb\endcsname%%
      \put(1535,484){\makebox(0,0){\strut{}$0.5$}}%
      \csname LTb\endcsname%%
      \put(1860,484){\makebox(0,0){\strut{}$1$}}%
      \csname LTb\endcsname%%
      \put(2186,484){\makebox(0,0){\strut{}$1.5$}}%
      \csname LTb\endcsname%%
      \put(2511,484){\makebox(0,0){\strut{}$2$}}%
      \csname LTb\endcsname%%
      \put(2836,484){\makebox(0,0){\strut{}$2.5$}}%
      \csname LTb\endcsname%%
      \put(3161,484){\makebox(0,0){\strut{}$3$}}%
      \csname LTb\endcsname%%
      \put(3487,484){\makebox(0,0){\strut{}$3.5$}}%
      \csname LTb\endcsname%%
      \put(3812,484){\makebox(0,0){\strut{}$4$}}%
      \csname LTb\endcsname%%
      \put(4137,484){\makebox(0,0){\strut{}$4.5$}}%
    }%
    \gplgaddtomacro\gplfronttext{%
      \csname LTb\endcsname%%
      \put(198,2508){\rotatebox{-270}{\makebox(0,0){\strut{}$\text{max}\[\xi^2\]$}}}%
      \put(2673,154){\makebox(0,0){\strut{}$N\times 10^3$}}%
      \csname LTb\endcsname%%
      \put(3546,4085){\makebox(0,0)[r]{\strut{}$\ell=1$}}%
      \csname LTb\endcsname%%
      \put(3546,3755){\makebox(0,0)[r]{\strut{}$\ell=3$}}%
      \csname LTb\endcsname%%
      \put(3546,3425){\makebox(0,0)[r]{\strut{}$\ell=2^+$}}%
      \csname LTb\endcsname%%
      \put(3546,3095){\makebox(0,0)[r]{\strut{}$\ell=2^-$}}%
    }%
    \gplbacktext
    \put(0,0){\includegraphics{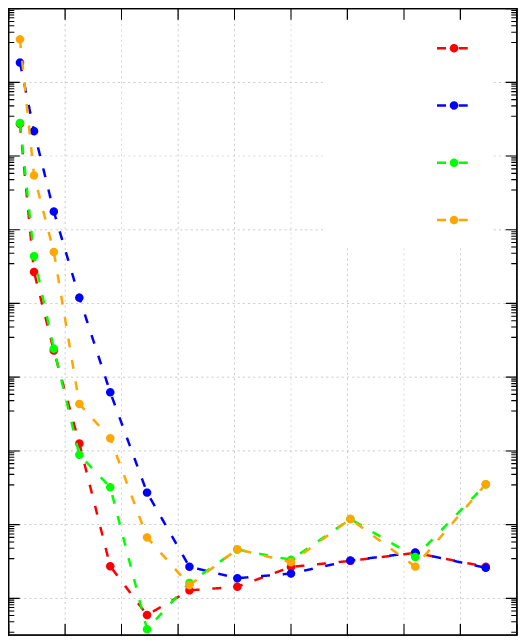}}%
    \gplfronttext
  \end{picture}%
\endgroup

\end{minipage}
\hfill
\begin{minipage}[h!]{0.5\textwidth}
\hs{0.15}% GNUPLOT: LaTeX picture with Postscript
\begingroup
  \makeatletter
  \providecommand\color[2][]{%
    \GenericError{(gnuplot) \space\space\space\@spaces}{%
      Package color not loaded in conjunction with
      terminal option `colourtext'%
    }{See the gnuplot documentation for explanation.%
    }{Either use 'blacktext' in gnuplot or load the package
      color.sty in LaTeX.}%
    \renewcommand\color[2][]{}%
  }%
  \providecommand\includegraphics[2][]{%
    \GenericError{(gnuplot) \space\space\space\@spaces}{%
      Package graphicx or graphics not loaded%
    }{See the gnuplot documentation for explanation.%
    }{The gnuplot epslatex terminal needs graphicx.sty or graphics.sty.}%
    \renewcommand\includegraphics[2][]{}%
  }%
  \providecommand\rotatebox[2]{#2}%
  \@ifundefined{ifGPcolor}{%
    \newif\ifGPcolor
    \GPcolortrue
  }{}%
  \@ifundefined{ifGPblacktext}{%
    \newif\ifGPblacktext
    \GPblacktexttrue
  }{}%
  % define a \g@addto@macro without @ in the name:
  \let\gplgaddtomacro\g@addto@macro
  % define empty templates for all commands taking text:
  \gdef\gplbacktext{}%
  \gdef\gplfronttext{}%
  \makeatother
  \ifGPblacktext
    % no textcolor at all
    \def\colorrgb#1{}%
    \def\colorgray#1{}%
  \else
    % gray or color?
    \ifGPcolor
      \def\colorrgb#1{\color[rgb]{#1}}%
      \def\colorgray#1{\color[gray]{#1}}%
      \expandafter\def\csname LTw\endcsname{\color{white}}%
      \expandafter\def\csname LTb\endcsname{\color{black}}%
      \expandafter\def\csname LTa\endcsname{\color{black}}%
      \expandafter\def\csname LT0\endcsname{\color[rgb]{1,0,0}}%
      \expandafter\def\csname LT1\endcsname{\color[rgb]{0,1,0}}%
      \expandafter\def\csname LT2\endcsname{\color[rgb]{0,0,1}}%
      \expandafter\def\csname LT3\endcsname{\color[rgb]{1,0,1}}%
      \expandafter\def\csname LT4\endcsname{\color[rgb]{0,1,1}}%
      \expandafter\def\csname LT5\endcsname{\color[rgb]{1,1,0}}%
      \expandafter\def\csname LT6\endcsname{\color[rgb]{0,0,0}}%
      \expandafter\def\csname LT7\endcsname{\color[rgb]{1,0.3,0}}%
      \expandafter\def\csname LT8\endcsname{\color[rgb]{0.5,0.5,0.5}}%
    \else
      % gray
      \def\colorrgb#1{\color{black}}%
      \def\colorgray#1{\color[gray]{#1}}%
      \expandafter\def\csname LTw\endcsname{\color{white}}%
      \expandafter\def\csname LTb\endcsname{\color{black}}%
      \expandafter\def\csname LTa\endcsname{\color{black}}%
      \expandafter\def\csname LT0\endcsname{\color{black}}%
      \expandafter\def\csname LT1\endcsname{\color{black}}%
      \expandafter\def\csname LT2\endcsname{\color{black}}%
      \expandafter\def\csname LT3\endcsname{\color{black}}%
      \expandafter\def\csname LT4\endcsname{\color{black}}%
      \expandafter\def\csname LT5\endcsname{\color{black}}%
      \expandafter\def\csname LT6\endcsname{\color{black}}%
      \expandafter\def\csname LT7\endcsname{\color{black}}%
      \expandafter\def\csname LT8\endcsname{\color{black}}%
    \fi
  \fi
    \setlength{\unitlength}{0.0500bp}%
    \ifx\gptboxheight\undefined%
      \newlength{\gptboxheight}%
      \newlength{\gptboxwidth}%
      \newsavebox{\gptboxtext}%
    \fi%
    \setlength{\fboxrule}{0.5pt}%
    \setlength{\fboxsep}{1pt}%
\begin{picture}(4534.00,4534.00)%
    \gplgaddtomacro\gplbacktext{%
      \csname LTb\endcsname%%
      \put(1078,704){\makebox(0,0)[r]{\strut{}$10^{-20}$}}%
      \csname LTb\endcsname%%
      \put(1078,1606){\makebox(0,0)[r]{\strut{}$10^{-15}$}}%
      \csname LTb\endcsname%%
      \put(1078,2509){\makebox(0,0)[r]{\strut{}$10^{-10}$}}%
      \csname LTb\endcsname%%
      \put(1078,3411){\makebox(0,0)[r]{\strut{}$10^{-5}$}}%
      \csname LTb\endcsname%%
      \put(1078,4313){\makebox(0,0)[r]{\strut{}$10^{0}$}}%
      \csname LTb\endcsname%%
      \put(1210,484){\makebox(0,0){\strut{}$0$}}%
      \csname LTb\endcsname%%
      \put(1535,484){\makebox(0,0){\strut{}$0.5$}}%
      \csname LTb\endcsname%%
      \put(1860,484){\makebox(0,0){\strut{}$1$}}%
      \csname LTb\endcsname%%
      \put(2186,484){\makebox(0,0){\strut{}$1.5$}}%
      \csname LTb\endcsname%%
      \put(2511,484){\makebox(0,0){\strut{}$2$}}%
      \csname LTb\endcsname%%
      \put(2836,484){\makebox(0,0){\strut{}$2.5$}}%
      \csname LTb\endcsname%%
      \put(3161,484){\makebox(0,0){\strut{}$3$}}%
      \csname LTb\endcsname%%
      \put(3487,484){\makebox(0,0){\strut{}$3.5$}}%
      \csname LTb\endcsname%%
      \put(3812,484){\makebox(0,0){\strut{}$4$}}%
      \csname LTb\endcsname%%
      \put(4137,484){\makebox(0,0){\strut{}$4.5$}}%
    }%
    \gplgaddtomacro\gplfronttext{%
      \csname LTb\endcsname%%
      \put(198,2508){\rotatebox{-270}{\makebox(0,0){\strut{}$\text{max}\[\xi^2\]$}}}%
      \put(2673,154){\makebox(0,0){\strut{}$N\times 10^3$}}%
      \csname LTb\endcsname%%
      \put(3546,4085){\makebox(0,0)[r]{\strut{}$\ell=1$}}%
      \csname LTb\endcsname%%
      \put(3546,3755){\makebox(0,0)[r]{\strut{}$\ell=3$}}%
      \csname LTb\endcsname%%
      \put(3546,3425){\makebox(0,0)[r]{\strut{}$\ell=2^+$}}%
      \csname LTb\endcsname%%
      \put(3546,3095){\makebox(0,0)[r]{\strut{}$\ell=2^-$}}%
    }%
    \gplbacktext
    \put(0,0){\includegraphics{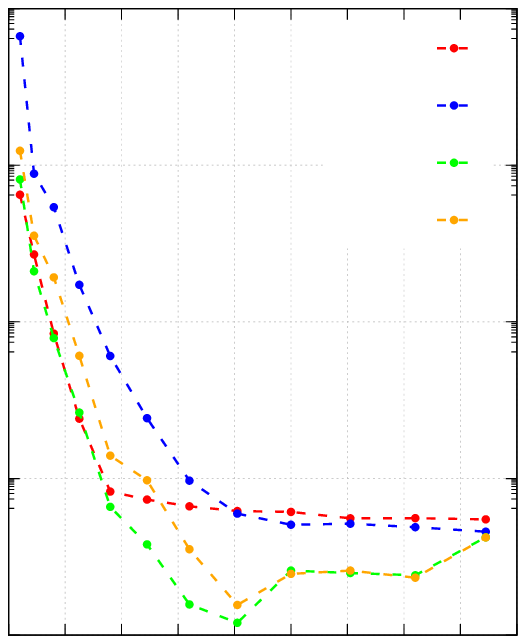}}%
    \gplfronttext
  \end{picture}%
\endgroup

\end{minipage}

\hfill

\begin{minipage}[h!]{0.5\textwidth}
% GNUPLOT: LaTeX picture with Postscript
\begingroup
  \makeatletter
  \providecommand\color[2][]{%
    \GenericError{(gnuplot) \space\space\space\@spaces}{%
      Package color not loaded in conjunction with
      terminal option `colourtext'%
    }{See the gnuplot documentation for explanation.%
    }{Either use 'blacktext' in gnuplot or load the package
      color.sty in LaTeX.}%
    \renewcommand\color[2][]{}%
  }%
  \providecommand\includegraphics[2][]{%
    \GenericError{(gnuplot) \space\space\space\@spaces}{%
      Package graphicx or graphics not loaded%
    }{See the gnuplot documentation for explanation.%
    }{The gnuplot epslatex terminal needs graphicx.sty or graphics.sty.}%
    \renewcommand\includegraphics[2][]{}%
  }%
  \providecommand\rotatebox[2]{#2}%
  \@ifundefined{ifGPcolor}{%
    \newif\ifGPcolor
    \GPcolortrue
  }{}%
  \@ifundefined{ifGPblacktext}{%
    \newif\ifGPblacktext
    \GPblacktexttrue
  }{}%
  % define a \g@addto@macro without @ in the name:
  \let\gplgaddtomacro\g@addto@macro
  % define empty templates for all commands taking text:
  \gdef\gplbacktext{}%
  \gdef\gplfronttext{}%
  \makeatother
  \ifGPblacktext
    % no textcolor at all
    \def\colorrgb#1{}%
    \def\colorgray#1{}%
  \else
    % gray or color?
    \ifGPcolor
      \def\colorrgb#1{\color[rgb]{#1}}%
      \def\colorgray#1{\color[gray]{#1}}%
      \expandafter\def\csname LTw\endcsname{\color{white}}%
      \expandafter\def\csname LTb\endcsname{\color{black}}%
      \expandafter\def\csname LTa\endcsname{\color{black}}%
      \expandafter\def\csname LT0\endcsname{\color[rgb]{1,0,0}}%
      \expandafter\def\csname LT1\endcsname{\color[rgb]{0,1,0}}%
      \expandafter\def\csname LT2\endcsname{\color[rgb]{0,0,1}}%
      \expandafter\def\csname LT3\endcsname{\color[rgb]{1,0,1}}%
      \expandafter\def\csname LT4\endcsname{\color[rgb]{0,1,1}}%
      \expandafter\def\csname LT5\endcsname{\color[rgb]{1,1,0}}%
      \expandafter\def\csname LT6\endcsname{\color[rgb]{0,0,0}}%
      \expandafter\def\csname LT7\endcsname{\color[rgb]{1,0.3,0}}%
      \expandafter\def\csname LT8\endcsname{\color[rgb]{0.5,0.5,0.5}}%
    \else
      % gray
      \def\colorrgb#1{\color{black}}%
      \def\colorgray#1{\color[gray]{#1}}%
      \expandafter\def\csname LTw\endcsname{\color{white}}%
      \expandafter\def\csname LTb\endcsname{\color{black}}%
      \expandafter\def\csname LTa\endcsname{\color{black}}%
      \expandafter\def\csname LT0\endcsname{\color{black}}%
      \expandafter\def\csname LT1\endcsname{\color{black}}%
      \expandafter\def\csname LT2\endcsname{\color{black}}%
      \expandafter\def\csname LT3\endcsname{\color{black}}%
      \expandafter\def\csname LT4\endcsname{\color{black}}%
      \expandafter\def\csname LT5\endcsname{\color{black}}%
      \expandafter\def\csname LT6\endcsname{\color{black}}%
      \expandafter\def\csname LT7\endcsname{\color{black}}%
      \expandafter\def\csname LT8\endcsname{\color{black}}%
    \fi
  \fi
    \setlength{\unitlength}{0.0500bp}%
    \ifx\gptboxheight\undefined%
      \newlength{\gptboxheight}%
      \newlength{\gptboxwidth}%
      \newsavebox{\gptboxtext}%
    \fi%
    \setlength{\fboxrule}{0.5pt}%
    \setlength{\fboxsep}{1pt}%
\begin{picture}(4534.00,4534.00)%
    \gplgaddtomacro\gplbacktext{%
      \csname LTb\endcsname%%
      \put(1078,704){\makebox(0,0)[r]{\strut{}$10^{-10}$}}%
      \csname LTb\endcsname%%
      \put(1078,1105){\makebox(0,0)[r]{\strut{}$10^{-9}$}}%
      \csname LTb\endcsname%%
      \put(1078,1506){\makebox(0,0)[r]{\strut{}$10^{-8}$}}%
      \csname LTb\endcsname%%
      \put(1078,1907){\makebox(0,0)[r]{\strut{}$10^{-7}$}}%
      \csname LTb\endcsname%%
      \put(1078,2308){\makebox(0,0)[r]{\strut{}$10^{-6}$}}%
      \csname LTb\endcsname%%
      \put(1078,2709){\makebox(0,0)[r]{\strut{}$10^{-5}$}}%
      \csname LTb\endcsname%%
      \put(1078,3110){\makebox(0,0)[r]{\strut{}$10^{-4}$}}%
      \csname LTb\endcsname%%
      \put(1078,3511){\makebox(0,0)[r]{\strut{}$10^{-3}$}}%
      \csname LTb\endcsname%%
      \put(1078,3912){\makebox(0,0)[r]{\strut{}$10^{-2}$}}%
      \csname LTb\endcsname%%
      \put(1078,4313){\makebox(0,0)[r]{\strut{}$10^{-1}$}}%
      \csname LTb\endcsname%%
      \put(1210,484){\makebox(0,0){\strut{}$0$}}%
      \csname LTb\endcsname%%
      \put(1535,484){\makebox(0,0){\strut{}$0.5$}}%
      \csname LTb\endcsname%%
      \put(1860,484){\makebox(0,0){\strut{}$1$}}%
      \csname LTb\endcsname%%
      \put(2186,484){\makebox(0,0){\strut{}$1.5$}}%
      \csname LTb\endcsname%%
      \put(2511,484){\makebox(0,0){\strut{}$2$}}%
      \csname LTb\endcsname%%
      \put(2836,484){\makebox(0,0){\strut{}$2.5$}}%
      \csname LTb\endcsname%%
      \put(3161,484){\makebox(0,0){\strut{}$3$}}%
      \csname LTb\endcsname%%
      \put(3487,484){\makebox(0,0){\strut{}$3.5$}}%
      \csname LTb\endcsname%%
      \put(3812,484){\makebox(0,0){\strut{}$4$}}%
      \csname LTb\endcsname%%
      \put(4137,484){\makebox(0,0){\strut{}$4.5$}}%
    }%
    \gplgaddtomacro\gplfronttext{%
      \csname LTb\endcsname%%
      \put(198,2508){\rotatebox{-270}{\makebox(0,0){\strut{}$\text{max}\[\xi_y\]$}}}%
      \put(2673,154){\makebox(0,0){\strut{}$N\times10^3$}}%
      \csname LTb\endcsname%%
      \put(3546,4085){\makebox(0,0)[r]{\strut{}$\ell=1$}}%
      \csname LTb\endcsname%%
      \put(3546,3755){\makebox(0,0)[r]{\strut{}$\ell=3$}}%
      \csname LTb\endcsname%%
      \put(3546,3425){\makebox(0,0)[r]{\strut{}$\ell=2^+$}}%
      \csname LTb\endcsname%%
      \put(3546,3095){\makebox(0,0)[r]{\strut{}$\ell=2^-$}}%
    }%
    \gplbacktext
    \put(0,0){\includegraphics{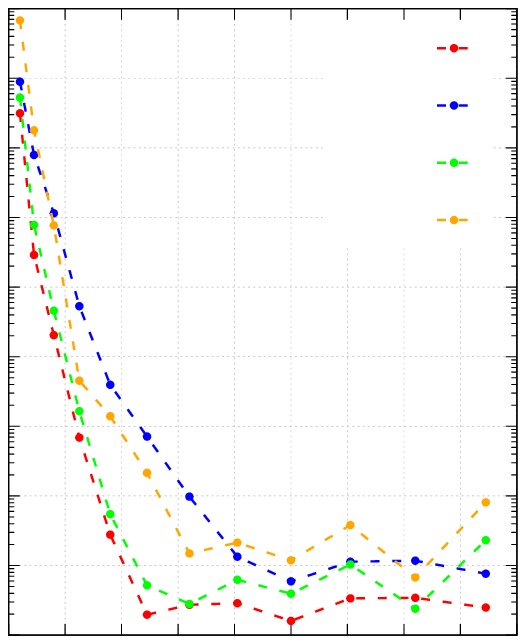}}%
    \gplfronttext
  \end{picture}%
\endgroup

\end{minipage}
\hfill
\begin{minipage}[h!]{0.5\textwidth}
\hs{0.15}% GNUPLOT: LaTeX picture with Postscript
\begingroup
  \makeatletter
  \providecommand\color[2][]{%
    \GenericError{(gnuplot) \space\space\space\@spaces}{%
      Package color not loaded in conjunction with
      terminal option `colourtext'%
    }{See the gnuplot documentation for explanation.%
    }{Either use 'blacktext' in gnuplot or load the package
      color.sty in LaTeX.}%
    \renewcommand\color[2][]{}%
  }%
  \providecommand\includegraphics[2][]{%
    \GenericError{(gnuplot) \space\space\space\@spaces}{%
      Package graphicx or graphics not loaded%
    }{See the gnuplot documentation for explanation.%
    }{The gnuplot epslatex terminal needs graphicx.sty or graphics.sty.}%
    \renewcommand\includegraphics[2][]{}%
  }%
  \providecommand\rotatebox[2]{#2}%
  \@ifundefined{ifGPcolor}{%
    \newif\ifGPcolor
    \GPcolortrue
  }{}%
  \@ifundefined{ifGPblacktext}{%
    \newif\ifGPblacktext
    \GPblacktexttrue
  }{}%
  % define a \g@addto@macro without @ in the name:
  \let\gplgaddtomacro\g@addto@macro
  % define empty templates for all commands taking text:
  \gdef\gplbacktext{}%
  \gdef\gplfronttext{}%
  \makeatother
  \ifGPblacktext
    % no textcolor at all
    \def\colorrgb#1{}%
    \def\colorgray#1{}%
  \else
    % gray or color?
    \ifGPcolor
      \def\colorrgb#1{\color[rgb]{#1}}%
      \def\colorgray#1{\color[gray]{#1}}%
      \expandafter\def\csname LTw\endcsname{\color{white}}%
      \expandafter\def\csname LTb\endcsname{\color{black}}%
      \expandafter\def\csname LTa\endcsname{\color{black}}%
      \expandafter\def\csname LT0\endcsname{\color[rgb]{1,0,0}}%
      \expandafter\def\csname LT1\endcsname{\color[rgb]{0,1,0}}%
      \expandafter\def\csname LT2\endcsname{\color[rgb]{0,0,1}}%
      \expandafter\def\csname LT3\endcsname{\color[rgb]{1,0,1}}%
      \expandafter\def\csname LT4\endcsname{\color[rgb]{0,1,1}}%
      \expandafter\def\csname LT5\endcsname{\color[rgb]{1,1,0}}%
      \expandafter\def\csname LT6\endcsname{\color[rgb]{0,0,0}}%
      \expandafter\def\csname LT7\endcsname{\color[rgb]{1,0.3,0}}%
      \expandafter\def\csname LT8\endcsname{\color[rgb]{0.5,0.5,0.5}}%
    \else
      % gray
      \def\colorrgb#1{\color{black}}%
      \def\colorgray#1{\color[gray]{#1}}%
      \expandafter\def\csname LTw\endcsname{\color{white}}%
      \expandafter\def\csname LTb\endcsname{\color{black}}%
      \expandafter\def\csname LTa\endcsname{\color{black}}%
      \expandafter\def\csname LT0\endcsname{\color{black}}%
      \expandafter\def\csname LT1\endcsname{\color{black}}%
      \expandafter\def\csname LT2\endcsname{\color{black}}%
      \expandafter\def\csname LT3\endcsname{\color{black}}%
      \expandafter\def\csname LT4\endcsname{\color{black}}%
      \expandafter\def\csname LT5\endcsname{\color{black}}%
      \expandafter\def\csname LT6\endcsname{\color{black}}%
      \expandafter\def\csname LT7\endcsname{\color{black}}%
      \expandafter\def\csname LT8\endcsname{\color{black}}%
    \fi
  \fi
    \setlength{\unitlength}{0.0500bp}%
    \ifx\gptboxheight\undefined%
      \newlength{\gptboxheight}%
      \newlength{\gptboxwidth}%
      \newsavebox{\gptboxtext}%
    \fi%
    \setlength{\fboxrule}{0.5pt}%
    \setlength{\fboxsep}{1pt}%
\begin{picture}(4534.00,4534.00)%
    \gplgaddtomacro\gplbacktext{%
      \csname LTb\endcsname%%
      \put(1078,1065){\makebox(0,0)[r]{\strut{}$10^{-10}$}}%
      \csname LTb\endcsname%%
      \put(1078,1787){\makebox(0,0)[r]{\strut{}$10^{-8}$}}%
      \csname LTb\endcsname%%
      \put(1078,2509){\makebox(0,0)[r]{\strut{}$10^{-6}$}}%
      \csname LTb\endcsname%%
      \put(1078,3230){\makebox(0,0)[r]{\strut{}$10^{-4}$}}%
      \csname LTb\endcsname%%
      \put(1078,3952){\makebox(0,0)[r]{\strut{}$10^{-2}$}}%
      \csname LTb\endcsname%%
      \put(1210,484){\makebox(0,0){\strut{}$0$}}%
      \csname LTb\endcsname%%
      \put(1535,484){\makebox(0,0){\strut{}$0.5$}}%
      \csname LTb\endcsname%%
      \put(1860,484){\makebox(0,0){\strut{}$1$}}%
      \csname LTb\endcsname%%
      \put(2186,484){\makebox(0,0){\strut{}$1.5$}}%
      \csname LTb\endcsname%%
      \put(2511,484){\makebox(0,0){\strut{}$2$}}%
      \csname LTb\endcsname%%
      \put(2836,484){\makebox(0,0){\strut{}$2.5$}}%
      \csname LTb\endcsname%%
      \put(3161,484){\makebox(0,0){\strut{}$3$}}%
      \csname LTb\endcsname%%
      \put(3487,484){\makebox(0,0){\strut{}$3.5$}}%
      \csname LTb\endcsname%%
      \put(3812,484){\makebox(0,0){\strut{}$4$}}%
      \csname LTb\endcsname%%
      \put(4137,484){\makebox(0,0){\strut{}$4.5$}}%
    }%
    \gplgaddtomacro\gplfronttext{%
      \csname LTb\endcsname%%
      \put(198,2508){\rotatebox{-270}{\makebox(0,0){\strut{}$\text{max}\[\xi_y\]$}}}%
      \put(2673,154){\makebox(0,0){\strut{}$N\times10^3$}}%
      \csname LTb\endcsname%%
      \put(3546,4085){\makebox(0,0)[r]{\strut{}$\ell=1$}}%
      \csname LTb\endcsname%%
      \put(3546,3755){\makebox(0,0)[r]{\strut{}$\ell=3$}}%
      \csname LTb\endcsname%%
      \put(3546,3425){\makebox(0,0)[r]{\strut{}$\ell=2^+$}}%
      \csname LTb\endcsname%%
      \put(3546,3095){\makebox(0,0)[r]{\strut{}$\ell=2^-$}}%
    }%
    \gplbacktext
    \put(0,0){\includegraphics{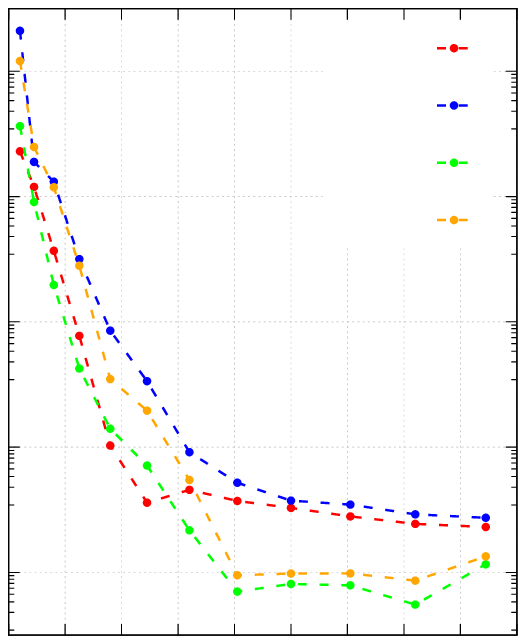}}%
    \gplfronttext
  \end{picture}%
\endgroup

\end{minipage}
\end{minipage}

\captionsetup{width=0.9\textwidth}
\captionof{figure}{\textsl{Logarithmic plot of the maximum values of $\xi^2$ and $\xi_y$ on the whole integration domain, as a function of the grid size. In both cases the error decays exponentially, as expected. The plot for the DeTurck vector component along the $x$-direction is very similar to that for the $y$-component, so we do not include it here.}}
\label{fig:convergence}
\end{center}

Indeed, in all cases shown we observe that those quantities vanish in the continuum limit. %This ensures that our numerical solutions provide a good approximation to the actual solution of the continuous problem.
Note that for the component $\xi_y$, the values themselves, being coordinate dependent, are not physical. However, the fact that they become zero for large $N$ confirms that we are finding an Einstein metric rather than a Ricci  soliton. Small deviations from exponential convergence to the continuum in some cases can be explained by the interpolation errors induced at the boundaries of the patches and potential non-smoothness due to the deTurck gauge fixing.\footnote{The $L_\infty$ norm used here to measure the errors over-penalises local errors, such as those induced at the boundaries between patches.}  Notice also that beyond a certain resolution the different curves flatten out due to the finite machine precision.

\newpage

%-------------------------------------------------------
% Bibliography
\addcontentsline{toc}{section}{References}
%-------------------------------------------------------

\bibliography{refs}

\providecommand{\href}[2]{#2}\begingroup\raggedright\begin{thebibliography}{10}

%%%%%%%%%%%% INTRO %%%%%%%%%%%%

\bibitem{Maldacena:97}
J. M. Maldacena, \emph{{The Large $N$ Limit of Superconformal Field Theories and Supergravity}}, \href{https://link.springer.com/article/10.1023\%2FA\%3A1026654312961}{\emph{Adv. Theor. Math. Phys.}{\bfseries~2} (1998) 231} [\href{http://arxiv.org/abs/hep-th/9711200}{{\ttfamily hep-th/9711200}}].

\bibitem{Hooft:74}
G.~`t Hooft, \emph{{A planar diagram theory for strong interactions}}, \href{https://www.sciencedirect.com/science/article/pii/0550321374901540?via\%3Dihub}{\emph{Nucl. Phys. B}{\bfseries~72} (1974) 461}.

\bibitem{Maldacenaetal:00}
O. Aharony, S. S. Gubser, J. M. Maldacena, H. Ooguri and Y. Oz, \emph{{Large $N$ Field Theories, String Theory and Gravity}}, \href{https://www.sciencedirect.com/science/article/abs/pii/S0370157399000836?via\%3Dihub}{\emph{Phys. Rept.}{\bfseries~323} (2000) 183} [\href{http://arxiv.org/abs/hep-th/9905111}{{\ttfamily hep-th/9905111}}].

\bibitem{ammonbook}
 M.~Ammon and J.~Erdmenger, ``Gauge/Gravity Duality: Foundations and Applications,'' {\em Cambridge University Press} (2015).

\bibitem{marolf:13}
D. Marolf, M. Rangamani and T. Wiseman, \emph{{Holographic thermal field theory on curved spacetimes}}, \href{https://iopscience.iop.org/article/10.1088/0264-9381/31/6/063001}{\emph{Class. Quant. Grav.}{\bfseries~31} (2014) 063001} [\href{https://arxiv.org/abs/1312.0612}{{\ttfamily hep-th/1312.0612}}].

\bibitem{Hawking:83}
S. Hawking and D. N. Page, \emph{{Thermodynamics of Black Holes in anti-De Sitter Space}}, \href{https://link.springer.com/article/10.1007\%2FBF01208266}{\emph{Commun.~Math.~Phys.}{\bfseries~87} (1983) 577}.

\bibitem{witten:98}
E.~Witten, \emph{{Anti-de Sitter space and holography}}, \href{https://www.intlpress.com/site/pub/pages/journals/items/atmp/content/vols/0002/0002/a002/}{\emph{Adv.~Theor.~Math.~Phys.}{\bfseries~2} (1998) 253} [\href{https://arxiv.org/abs/hep-th/9802150}{{\ttfamily hep-th/9802150}}].

\bibitem{witten:981}
E. Witten, \emph{{Anti-de Sitter space, thermal phase transition, and confinement in gauge theories}}, \href{https://www.intlpress.com/site/pub/pages/journals/items/atmp/content/vols/0002/0003/a003/}{\emph{Adv.~Theor.~Math.~Phys.} {\bfseries 2} (1998) 505} [\href{https://arxiv.org/abs/hep-th/9803131}{{\ttfamily hep-th/9803131}}].

\bibitem{Gregory:1993vy}
R.~Gregory and R.~Laflamme, \emph{{Black strings and p-branes are unstable}}, \href{https://doi.org/10.1103/PhysRevLett.70.2837}{\emph{Phys. Rev. Lett.}{\bfseries~70} (1993) 2837} [\href{https://arxiv.org/abs/hep-th/9301052}{{\ttfamily hep-th/9301052}}].

\bibitem{Lehner:2010}
L.~Lehner, F.~Pretorius, \emph{{Black Strings, Low Viscosity Fluids, and Violation of Cosmic Censorship}}, \href{https://journals.aps.org/prl/abstract/10.1103/PhysRevLett.105.101102}{\emph{Phys.~Rev.~Lett.}{\bfseries~105} (2010) 101102} [\href{https://arxiv.org/abs/1006.5960}{{\ttfamily hep-th/1006.5960}}].

\bibitem{Banks:98}
T. Banks, M.R. Douglas, G.T. Horowitz and E. Martinec, \emph{{AdS Dynamics from Conformal Field Theory}} (1998) [\href{https://arxiv.org/abs/hep-th/9808016}{{\ttfamily hep-th/9808016}}].

\bibitem{Peet:98}
A.W. Peet and S.F. Ross, \emph{{Microcanonical Phases of String Theory on AdS$_m \times S^n$}}, \href{https://iopscience.iop.org/article/10.1088/1126-6708/1998/12/020}{\emph{JHEP}{\bfseries~12} (1998) 020} [\href{https://arxiv.org/abs/hep-th/9810200}{{\ttfamily hep-th/9810200}}].

\bibitem{Hubeny:02}
V.E. Hubeny and M. Rangamani, \emph{{Unstable Horizons}}, \href{https://iopscience.iop.org/article/10.1088/1126-6708/2002/05/027}{\emph{JHEP} {\bfseries 0205} (2002) 027} [\href{https://arxiv.org/abs/hep-th/0202189}{{\ttfamily hep-th/0202189}}].

\bibitem{Prestidge:1999uq}
T.~Prestidge, \emph{{Dynamic and thermodynamic stability and negative modes in Schwarzschild-anti-de Sitter}}, \href{https://doi.org/10.1103/PhysRevD.61.084002}{\emph{Phys. Rev.} {\bfseries D61} (2000) 084002} [\href{https://arxiv.org/abs/hep-th/9907163}{{\ttfamily hep-th/9907163}}].

\bibitem{Wiseman:2002zc}
T.~Wiseman, \emph{{Static axisymmetric vacuum solutions and nonuniform black strings}}, \href{https://doi.org/10.1088/0264-9381/20/6/308}{\emph{Class.~Quant.~Grav.} {\bfseries 20} (2003) 1137} [\href{https://arxiv.org/abs/hep-th/0209051}{{\ttfamily hep-th/0209051}}].

\bibitem{Gubser:2001ac}
S.~S. Gubser, \emph{{On nonuniform black branes}}, \href{https://doi.org/10.1088/0264-9381/19/19/303}{\emph{Class. Quant. Grav.} {\bfseries 19} (2002) 4825} [\href{https://arxiv.org/abs/hep-th/0110193}{{\ttfamily hep-th/0110193}}].

\bibitem{Dias:2015pda}
\'O.~J.~C. Dias, J.~E. Santos and B.~Way, \emph{{Lumpy AdS$_{5}\times S^{5}$ black holes and black belts}}, \href{https://doi.org/10.1007/JHEP04(2015)060}{\emph{JHEP} {\bfseries 04} (2015) 060} [\href{https://arxiv.org/abs/1501.06574}{{\ttfamily hep-th/1501.06574}}].

%\bibitem{Gregory:00}
%R. Gregory, \emph{{Black string instabilities in anti-de Sitter space}}, \href{https://iopscience.iop.org/article/10.1088/0264-9381/17/18/103}{\emph{Class. Quant. Grav.}{\bfseries~17} (2000) L125} [\href{https://arxiv.org/abs/hep-th/0004101}{{\ttfamily hep-th/0004101}}].

%\bibitem{Horowitz:11}
%G.T. Horowitz and T. Wiseman, \emph{{General black holes in Kaluza-Klein theory}} (2011) [\href{https://arxiv.org/abs/1107.5563}{{\ttfamily gr-qc/1107.5563}}].

\bibitem{Sorkin:2003ka}
E.~Sorkin, B.~Kol and T.~Piran, \emph{{Caged black holes: Black holes in compactified spacetimes. II: 5d numerical implementation}}, \href{https://journals.aps.org/prd/abstract/10.1103/PhysRevD.69.064032}{\emph{Phys.~Rev.~D }{\bfseries 69} (2004) 064032} [\href{https://arxiv.org/abs/hep-th/0310096}{{\ttfamily hep-th/0310096}}].

\bibitem{Dias:2016eto}
\'O.~J.~C. Dias, J.~E. Santos and B.~Way, \emph{{Localised $AdS_5\times S^5$ Black Holes}}, \href{https://doi.org/10.1103/PhysRevLett.117.151101}{\emph{Phys. Rev. Lett.} {\bfseries 117} (2016) 151101} [\href{https://arxiv.org/abs/1605.04911}{{\ttfamily hep-th/1605.04911}}].

\bibitem{Kol:top}
B. Kol, \emph{{Topology change in General Relativity and the black-hole black-string transition}}, \href{https://iopscience.iop.org/article/10.1088/1126-6708/2005/10/049}{\emph{JHEP} {\bfseries 0510} (2005) 049} [\href{https://arxiv.org/abs/hep-th/0206220}{{\ttfamily hep-th/0206220}}].

\bibitem{Kol:2003ja}
B.~Kol and T.~Wiseman, \emph{{Evidence that highly non-uniform black strings have a conical waist}}, \href{https://iopscience.iop.org/article/10.1088/0264-9381/20/15/315}{\emph{Class.~Quant.~Grav.} {\bfseries 20} (2003) 3493} [\href{https://arxiv.org/abs/hep-th/0304070}{{\ttfamily hep-th/0304070}}].

\bibitem{Kalisch:2016fkm}
M.~Kalisch and M.~Ansorg, \emph{{Pseudo-spectral construction of non-uniform black string solutions in five and six spacetime dimensions}}, \href{https://doi.org/10.1088/0264-9381/33/21/215005}{\emph{Class. Quant. Grav.} {\bfseries 33} (2016) 215005} [\href{https://arxiv.org/abs/1607.03099}{{\ttfamily gr-qc/1607.03099}}].

\bibitem{Kalisch:2017bin}
M.~Kalisch, S.~Moeckel and M.~Ammon, \emph{{Critical behavior of the black hole/black string transition}}, \href{https://doi.org/10.1007/JHEP08(2017)049}{\emph{JHEP} {\bfseries 08} (2017) 049} [\href{https://arxiv.org/abs/1706.02323} {{\ttfamily gr-qc/1706.02323}}].

\bibitem{Cardona:2018shd}
B.~Cardona and P.~Figueras, \emph{{Critical Kaluza-Klein black holes and black strings in D = 10}}, \href{https://link.springer.com/article/10.1007/JHEP11(2018)120}{\emph{JHEP} {\bfseries 11} (2018) 120} [\href{https://arxiv.org/abs/1806.11129} {{\ttfamily hep-th/1806.11129}}].

\bibitem{Ammon:2018sin}
M.~Ammon, M.~Kalisch and S.~Moeckel, \emph{{Notes on ten-dimensional localized black holes and deconfined states in two-dimensional SYM}}, \href{https://link.springer.com/article/10.1007\%2FJHEP11\%282018\%29090}{\emph{JHEP} {\bfseries 11} (2018) 090} [\href{https://arxiv.org/abs/1806.11174} {{\ttfamily hep-th/1806.11174}}].

\bibitem{Emparan:top2}
R. Emparan and R. Suzuki, \emph{{Topology-changing horizons at large D as Ricci flows}}, \href{https://link.springer.com/article/10.1007\%2FJHEP07\%282019\%29094}{\emph{JHEP} {\bfseries 07} (2019) 094} [\href{https://arxiv.org/abs/1905.01062}{{\ttfamily hep-th/1905.01062}}].

\bibitem{Emparan:top}
R. Emparan and N. Haddad, \emph{{Self-similar critical geometries at horizon intersections and mergers}}, \href{https://link.springer.com/article/10.1007\%2FJHEP10\%282011\%29064}{\emph{JHEP} {\bfseries 1110} (2011) 064} [\href{https://arxiv.org/abs/1109.1983}{{\ttfamily hep-th/1109.1983}}].

\bibitem{Figueras:bumpy}
R. Emparan, P. Figueras, M. Martinez, \emph{{Bumpy black holes}}, \href{https://link.springer.com/article/10.1007\%2FJHEP12\%282014\%29072}{\emph{JHEP} {\bfseries 12} (2014) 072} [\href{https://arxiv.org/abs/1410.4764}{{\ttfamily hep-th/1410.4764}}].

\bibitem{Skenderis:2006kkh}
K.~Skenderis and M.~Taylor, \emph{{Kaluza-Klein Holography}},
  \href{https://iopscience.iop.org/article/10.1088/1126-6708/2006/05/057}{\emph{JHEP}
  {\bfseries 05} (2006) 057}
  [\href{https://arxiv.org/abs/hep-th/0603016}{{\ttfamily hep-th/0603016}}].

\bibitem{Aharony:ABJM}
O. Aharony, O. Bergman, D.L. Jafferis and J. Maldacena, \emph{{$\mc{N}=6$ superconformal Chern-Simons-matter theories, M2-branes and their gravity duals}}, \href{https://iopscience.iop.org/article/10.1088/1126-6708/2008/10/091}{\emph{JHEP} {\bfseries 10} (2008) 091} [\href{https://arxiv.org/abs/0806.1218}{{\ttfamily hep-th/0806.1218}}].

\bibitem{freedmanbook}
D.Z. Freedman and Van Proeyen, ``Supergravity,'' {\em Cambridge University Press} (2012).

\bibitem{Headrick:10}
M. Headrick, S. Kitchen, and T. Wiseman, \emph{{A new approach to static numerical relativity, and its application to Kaluza-Klein black holes}}, \href{https://iopscience.iop.org/article/10.1088/0264-9381/27/3/035002}{\emph{Class. Quant. Grav.} {\bfseries 27} (2010) 035002} [\href{https://arxiv.org/abs/0905.1822}{{\ttfamily gr-qc/0905.1822}}].

\bibitem{Figueras:Riccisol}
P. Figueras, J. Lucietti and T. Wiseman, \emph{{Ricci solitons, Ricci flow, and strongly coupled CFT in the Schwarzschild Unruh or Boulware vacua}}, \href{https://iopscience.iop.org/article/10.1088/0264-9381/28/21/215018}{\emph{Class. Quant. Grav.} {\bfseries 28} (2011) 215018} [\href{https://arxiv.org/abs/1104.4489}{{\ttfamily hep-th/1104.4489}}].

\bibitem{Wiseman:2011by}
T.~Wiseman, \emph{{Numerical construction of static and stationary black holes}} (2011) [\href{https://arxiv.org/abs/1107.5513}{{\ttfamily gr-qc/1107.5513}}].

\bibitem{Dias:revst}
\'O.J.C. Dias, J.E. Santos and B. Way, \emph{{Numerical Methods for Finding Stationary Gravitational Solutions}}, \href{https://iopscience.iop.org/article/10.1088/0264-9381/33/13/133001}{\emph{Class. Quant. Grav.} {\bfseries 33} (2016) 133001} [\href{https://arxiv.org/abs/1510.02804}{{\ttfamily hep-th/1510.02804}}].

\bibitem{Balasubramanian:1999re}
V.~Balasubramanian and P.~Kraus, \emph{{A Stress tensor for Anti-de Sitter gravity}}, \href{https://doi.org/10.1007/s002200050764}{\emph{Commun. Math. Phys.} {\bfseries 208} (1999) 413} [\href{https://arxiv.org/abs/hep-th/9902121}{{\ttfamily hep-th/9902121}}].

  \bibitem{deHaro:2000hrec}
S.~de Haro, K.~Skenderis and S.N.~Solodukhin, \emph{{Holographic Reconstruction of Spacetime and Renormalization in the AdS/CFT Correspondence}},
  \href{https://link.springer.com/article/10.1007/s002200100381}{\emph{Commun.~Math.~Phys}
  {\bfseries 217} (2001) 595}
  [\href{https://arxiv.org/abs/hep-th/0002230}{{\ttfamily hep-th/0002230}}].

\bibitem{Yaffe:2017axl}
L.~G.~Yaffe, \emph{{Large $N$ phase transitions and the fate of small Schwarzschild-AdS black holes}},
  \href{https://journals.aps.org/prd/abstract/10.1103/PhysRevD.97.026010}{\emph{Phys. Rev. D}
  {\bfseries 97} (2018) 026010} [\href{https://arxiv.org/abs/1710.06455}{{\ttfamily hep-th/1710.06455}}].
  
  \bibitem{Jang:2016exh}
D.~Jang, Y.~Kim, O.K.~Kwon and D.D.~Tolla, \emph{{Exact Holography of the Mass-deformed M2-brane Theory}},
  \href{https://link.springer.com/article/10.1140/epjc/s10052-017-4909-3}{\emph{Eur.~Phys.~J.~C}
  {\bfseries 77} (2017) 342}
  [\href{https://arxiv.org/abs/1610.01490}{{\ttfamily hep-th/1610.01490}}], \emph{{Mass-deformed ABJM Theory and LLM Geometries: Exact Holography}},
  \href{https://link.springer.com/article/10.1007%2FJHEP04%282017%29104}{\emph{JHEP}
  {\bfseries 04} (2017) 104}
  [\href{https://arxiv.org/abs/1612.05066}{{\ttfamily hep-th/1612.05066}}].
  
\bibitem{Jang:2018nle}
D.~Jang, Y.~Kim, O.K.~Kwon and D.D.~Tolla, \emph{{Holography of Massive M2-brane Theory: Non-linear Extension}},
  [\href{https://arxiv.org/abs/1803.10660}{{\ttfamily hep-th/1803.10660}}].
  
  \bibitem{Kol:2005vy}
B.~Kol, \emph{{Choptuik scaling and the merger transition}}, \href{https://doi.org/10.1088/1126-6708/2006/10/017}{\emph{JHEP} {\bfseries 10} (2006) 017} [\href{https://arxiv.org/abs/hep-th/0502033}{{\ttfamily hep-th/0502033}}].

%%%%%%%%%%%% KALUZA-KLEIN HOLOGRAPHY %%%%%%%%%%%%
   
\bibitem{Skenderis:2006hcv}
K.~Skenderis and M.~Taylor, \emph{{Holographic Coulomb branch vevs}},
  \href{https://iopscience.iop.org/article/10.1088/1126-6708/2006/08/001}{\emph{JHEP} {\bfseries 08} (2006) 001}
  [\href{https://arxiv.org/abs/hep-th/0604169}{{\ttfamily hep-th/0604169}}].
  
\bibitem{Skenderis:2007ana}
K.~Skenderis and M.~Taylor, \emph{{Anatomy of bubbling solutions}},
  \href{https://iopscience.iop.org/article/10.1088/1126-6708/2007/09/019}{\emph{JHEP} {\bfseries 09} (2007) 019}
  [\href{https://arxiv.org/abs/0706.0216}{{\ttfamily hep-th/0706.0216}}].
  
\bibitem{Bianchi:2001htg}
M.~Bianchi, D.Z.~Freedman, K.~Skenderis, \emph{{How to go with an RG Flow}},
  \href{https://iopscience.iop.org/article/10.1088/1126-6708/2001/08/041}{\emph{JHEP}
  {\bfseries 08} (2001) 041}
  [\href{https://arxiv.org/abs/hep-th/0105276}{{\ttfamily hep-th/0105276}}].

\bibitem{Bianchi:2000hrec}
M.~Bianchi, D.Z.~Freedman, K.~Skenderis, \emph{{Holographic Renormalization}},
  \href{https://www.sciencedirect.com/science/article/pii/S0550321302001797?via%3Dihub}{\emph{Nucl.~Phys.~B}
  {\bfseries 631} (2002) 159}
  [\href{https://arxiv.org/abs/hep-th/0112119}{{\ttfamily hep-th/0112119}}].
  
  \bibitem{Kim:1985spe}
H.J.~Kim, L.J.~Romans, and P.~van Nieuwenhuizen, \emph{{Mass spectrum of chiral ten-dimensional $\mc{N}=2$ supergravity on $S^5$}},
  \href{https://journals.aps.org/prd/abstract/10.1103/PhysRevD.32.389}{\emph{Phys.~Rev.~D}
  {\bfseries 32} (1985) 389}.  

\bibitem{Sangmin:19983pf}
S.~Lee, S.~Minwalla, M.~Rangamani and N.~Seiberg, \emph{{Three-Point Functions of Chiral Operators in $D=4$, $\mc{N}=4$ SYM at Large $N$}},
  \href{https://dx.doi.org/10.4310/ATMP.1998.v2.n4.a1}{\emph{Adv.~Theor.~Math.~Phys.}
  {\bfseries 2} (1998) 697}
  [\href{https://arxiv.org/abs/hep-th/9806074}{{\ttfamily hep-th/9806074}}].
  
\bibitem{Arutyunov:1999scc}
G.~Arutyunov and S.~Frolov, \emph{{Some Cubic Couplings in Type IIB Supergravity on $AdS_5\times S^5$ and Three-point Functions in $SYM_4$ at Large $N$}},
  \href{https://journals.aps.org/prd/abstract/10.1103/PhysRevD.61.064009}{\emph{Phys.~Rev.~D}
  {\bfseries 61} (2000) 064009}
  [\href{https://arxiv.org/abs/hep-th/9907085}{{\ttfamily hep-th/9907085}}].
  
\bibitem{Lee:20044pf}
S.Lee, \emph{{$AdS_5/CFT_4$ Four-point Functions of Chiral Primary Operators: Cubic Vertices}},
  \href{https://www.sciencedirect.com/science/article/pii/S0550321399006148?via%3Dihub}{\emph{Nucl.~Phys.~B}
  {\bfseries 563} (1999) 349}
  [\href{https://arxiv.org/abs/hep-th/9907108}{{\ttfamily hep-th/9907108}}].
  
%\bibitem{Biran:1984kk11d}
%B.~Biran, A.~Casher, A.~Englert, M.~Rooman and P.~Spindel, \emph{{The fluctuating seven-sphere in eleven-dimensional supergravity}}, \href{https://www.sciencedirect.com/science/article/abs/pii/037026938490666X?via%3Dihub}{\emph{Phys.~Lett.~B} {\bfseries 134} (1984) 179}; A.~Casher, F.~Englert, H.~Nicolai, and M.~Rooman, \emph{{The mass spectrum of supergravity on the round seven-sphere}}, \href{https://www.sciencedirect.com/science/article/pii/0550321384903924?via%3Dihub}{\emph{Nucl.~Phys.~B} {\bfseries 243} (1984) 173}.
  
\end{thebibliography}\endgroup
\bibliographystyle{jhep}

\end{document}